\documentclass{aa}  

\usepackage{graphicx}
\usepackage{txfonts}
\usepackage{diagbox}
\usepackage{subcaption}
\usepackage{amssymb}
\newcommand{\NOTll}{\hskip 0.4mm \not \hskip -0.4mm \ll}

\begin{document}

   \title{Post-Outburst Chemistry in a Very Low-Luminosity Object}

   \subtitle{Peculiar High Abundance of Nitric Oxide}

   \author{B. M. Kulterer
          \inst{1,2}
          \and
          S. F. Wampfler
          \inst{2}
          \and
          N. F. W. Ligterink
          \inst{3,4}
          \and
          N. Murillo
          \inst{5,6}
          \and
          T.-H. Hsieh
          \inst{7}
          \and
          M. K. McClure
          \inst{8}
          \and
          A. Boogert
          \inst{9}
          \and
          K. Kipfer
          \inst{3}
          \and
          P. Bjerkeli
          \inst{10}
          \and 
          M. N. Drozdovskaya
          \inst{2,11}
          }

   \institute{Center for Astrophysics | Harvard \& Smithsonian, Cambridge, MA 02138, USA \\
               \email{beatrice.kulterer@cfa.harvard.edu}
        \and 
              Center for Space and Habitability, Universität Bern, Gesellschaftsstrasse 6, 3012 Bern, Switzerland 
        \and
            Space Research \& Planetary Sciences, Physics Institute, University of Bern, 3012 Bern, Switzerland 
        \and
            Faculty of Aerospace Engineering, Delft University of Technology, Delft, The Netherlands
        \and 
            Instituto de Astronomía, Universidad Nacional Autónoma de México, AP106, Ensenada CP 22830, B. C., México
        \and 
            Star and Planet Formation Laboratory, RIKEN Cluster for Pioneering Research, Wako, Saitama 351-0198, Japan
        \and 
            Max Planck Institut für Extraterrestrische Physik (MPE), Giessenbachstrasse 1, 85748 Garching, Germany 
        \and
            Leiden Observatory, Leiden University, PO Box 9513, NL–2300 RA Leiden, The Netherlands 
        \and 
            Institute for Astronomy, University of Hawai’i at Manoa, 2680 Woodlawn Drive, Honolulu, HI 96822, USA 
        \and 
            Chalmers University of Technology, Department of Space, Earth and Environment, 412 96 Gothenburg, Sweden 
        \and 
            Physikalisch-Meteorologisches Observatorium Davos und Weltstrahlungszentrum (PMOD/WRC), Dorfstrasse 33, CH-7260, Davos Dorf, Switzerland
             }

   \date{Received -; accepted -}

 
  \abstract
   {Very Low Luminosity Objects (VeLLOs) are deeply embedded, and extremely faint objects (L$_{\rm int}$~$<$~0.1~L$_{\odot}$), and are thought to be in the quiescent phase of the episodic accretion process. They fill an important gap in our understanding of star formation.}
   {The VeLLO in the isolated DC3272+18 cloud has undergone an outburst in the past $\sim$~10$^4$~yr, and is thus an ideal target for investigating the chemical inventory in the gas phase of an object of its type. The aim of this study is to investigate the direct impact of the outburst on the chemical processes in the object and identify molecules that can act as tracers of past heating events.}
   {Observations with the Atacama Pathfinder EXperiment (APEX) in four spectral windows in the frequency range of 213.6--272.4~GHz have been carried out to identify molecules that can be directly linked to the past outburst, utilize the line fluxes, column densities, and the abundance ratios of the detected species to characterize the different physical components of the VeLLO, and probe for the presence of complex organic molecules.}
   {Nitric oxide (NO) is detected for the first time in a source of this type, and its formation could be induced by the sublimation of grain-surface species during the outburst. In addition, the observations securely detect CH$_3$OH, H$_2$CO, D$_2$CO, SO, SO$_2$, CO, $^{13}$CO, C$^{18}$O, N$_2$D$^+$, HCO$^+$, DCO$^+$, HCN, DCN, HNC, c-C$_3$H$_2$, and C$_2$D. The upper state energies of the securely detected lines and their derived line intensity ratios indicate that most of the probed material stems from regions of cold gas in the envelope enshrouding the VeLLO in the DC3272+18 cloud with a temperature of $\sim$~10~K. In addition, c-C$_3$H$_2$ traces a second, warmer gas reservoir with a temperature of $\sim$~35~K. The high D/H ratio derived from D$_2$CO points towards its origin from the prestellar stage, while deuteration of the gas-phase species DCO$^+$, DCN, and C$_2$D could still be on-going in the gas in the envelope.}
   {The gas probed by the observations has already cooled down after the past heating event caused by the outburst, but it still has lasting effects on the chemistry in the envelope of the VeLLO. CH$_3$OH, H$_2$CO, SO, SO$_2$, and CO have sublimated from grains during the outburst and did not fully freeze out yet, which indicates that the outburst took place $<$~10$^4$~yr ago. A pathway to form NO directly in the gas phase is from the photodissociation products created after the sublimation of H$_2$O and NH$_3$ from the ices. While the present time water snowline has likely retreated to pre-outburst small radius, the volatile NO species is still extensively present in the gas phase, as evident by its high column density relative to methanol in the observations. This suggests that NO could be potentially used to trace the water snowline in outbursting sources. In order to rule out non-thermal desorption processes that could also have led to the formation of NO, this proposition has to be verified with future observations at higher spatial resolution, and by searching for NO in additional targets.}

   \keywords{Astrochemistry - ISM: abundances - ISM: individual objects: DC3272+18 - Stars: formation - Stars: low-mass
               }

   \maketitle
%

\section{Introduction}

Very Low Luminosity Objects (VeLLOs) have first been identified by the "Cores to Disk" (c2D) legacy survey of the \textit{Spitzer} Space Telescope (\citealt{Evans03,Young04}), and have been classified as protostellar objects that are deeply embedded in molecular clouds with an internal luminosity of $\leq$~0.1~L$_{\odot}$ (\citealt{Young04,diFrancesco07}). The nature and the future evolution of VeLLOs are currently under debate, as only a handful of sources have been studied in detail, because their low luminosity makes observations challenging. Their low internal luminosity suggests that they could be either young Class 0 sources (e.g., IRAM 04191: \citet{Andre99,Dunham06}; L1521F: \citet{Bourke06}; Cha-MMS1: \citet{Belloche06,Vaisala14};  IRAS 16253: \citet{Hsieh15,Hsieh18}), or even extremely low-mass protostars or proto-brown dwarf candidates (e.g., L328: \citet{Lee09,Lee13} ; L1148: \citet{Kauffmann11}; IC 348-SMM2E: \citet{Palau14}). One hypothesis is that VeLLOs are objects in the quiescent phase of the episodic accretion process. Studies as carried out for L673-7 (\citealt{Dunham10a}), and for L1521F (\citealt{Takahashi13}) have derived average mass accretion rates from the outflows of these objects that are a few times higher than their current internal luminosity allows for. Moreover, the aforementioned studied objects have bolometric temperatures and spectral energy distributions that are similar to protostars. However, their luminosities are significantly fainter than the luminosity that would be expected for the least massive protostars in standard star formation models with a constant mass accretion rate (\citealt{Shu87,Dunham06}). Episodic accretion could bypass this issue, if outbursts boost the accretion rate for short periods of time, before the objects return back to a more quiescent phase which is accompanied by less efficient mass accretion. The internal average luminosity of such objects would thus be smaller than an object with a constant mass accretion rate (\citealt{Kenyon95,Dunham10b}). \\
\noindent While detailed studies of the physical properties of VeLLOs are rare, even less is known about the chemistry in sources of this type. VeLLOs are faint objects, they are usually studied with typical molecular tracers such as CO and its isotopologues, N$_2$H$^+$ or HCN to quantify their large-scale structures and system properties (e.g., \citealt{Hsieh18,Kim19,Kim21}). Species with a higher degree of complexity are rarely observed.
\noindent Complex organic molecules (COMs; carbon-bearing species consisting of at least 6 atoms; \citealt{Herbst09}) are readily detected in cores at the prestellar stage (e.g., \citealt{Bacmann12,JimenezSerra16,JimenezSerra21,Lattanzi20,Scibelli20,Megias22}), and are also commonly observed around low-mass protostars (e.g., \citealt{Jorgensen16,Yang21}). In order to follow the trail of chemical complexity in space, the study of the chemistry of VeLLOs is a crucial step to close the gap between pre- and protostellar objects. To date, only \cite{Favre20} have discovered emission of methanol, the simplest COM, at a distance of $\sim$~1~000~au of the VeLLO in L1521F. A search for additional COMs in that source has not been successful to date, due to a lack of sensitivity. To complete our understanding of the complex chemistry that is occurring in these types of objects, more VeLLOs like L1521F need to be studied in detail. \\
\noindent The study of these young objects is also important in terms of understanding the composition of forming planets. Substructures in the dust of more evolved Class II objects have been linked to the formation of gas giant planets (e.g., \citealt{Andrews18,vanderMarel19}), but signs of on-going planet formation such as rings and gaps have already been detected in a system with the age of $\sim$~500~kyr (\citealt{SeguraCox20}), and most recently \cite{Maureira24} have reported evidence for annular substructures in a Class 0 source, which suggests that planets can already form earlier in the protostellar evolution (\citealt{Tychoniec20}). The composition of planets is set by the composition of the gas and solids they are accreting, and thus the presence of molecules at their location in the disk. The temperature determines which molecules are confined to the solid phase, and which molecules are present in the gas interior to their snowlines (\citealt{Minissale22,Ligterink23}). A change in the temperature, for instance due to an outburst, can temporarily change the composition of the gas due to the shifting of the snowlines and the associated sublimation of icy species, and by the formation of new gas-phase molecules induced by the species sublimating from the ice. The chemical response time (for e.g., freeze-out) is slower than the shift of physical conditions from the outburst to the quiescent stage. The duration between outbursting events can be estimated from observations and is $\sim$ 5~000--50~000~yr (\citealt{Scholz13,Hsieh19}), however most observational tracers that are enhanced due to the outburst remain altered for only $\sim$ 10--100~yr into the quiescent phase (\citealt{Zwicky24}), exceptions being e.g., CO, which can stay enhanced for up to 10$^4$~yr before it has frozen out again at typical envelope densities (\citealt{Visser15,Frimann17}). The emission of the freshly sublimated molecules is expected to be brighter and farther extended than if the object would be quiescent, which makes them easier to detect. Another possibility is to look for species whose formation can be attributed to chemistry that has occurred due to the outburst, and that remain present in the gas phase throughout the quiescent phase.\\
\noindent The aim of this study is to search for signposts of active chemistry that have been induced by the change of physical conditions in the VeLLO in the DC3272+18 cloud during its most recent outburst. The detection of NO for the first time in a source of this type is potentially directly tied to the outburst, and it is proposed that it could act as a tracer of the water snowline in outbursting sources. In addition, this work utilizes molecular tracers for different components to gain a better understanding of the structure of the system.
\newline \noindent This paper is structured as follows: the targeted object and the observations are described in Section \ref{sect:obs}. The detected lines and their analysis are detailed in Section \ref{sect:results} and discussed in Section \ref{sect:discussion}. The conclusions are stated in Section \ref{sect:conclusions}. \\


\section{Observations}\label{sect:obs}

\subsection{Targeted VeLLO in DC3272+18}
This study investigates the chemistry toward the isolated VeLLO in the DC3272+18 region, which is located at a distance of 250 $\pm$ 50~pc (\citealt{Kim19}). There are no massive stars nearby, so external irradiation influences are minimal for this object. The VeLLO has an internal luminosity (L$_{\rm int}$) of 0.04~$\pm$ 0.02~L$_{\odot}$, and an envelope mass (M$_{\rm env}$) of 0.04~$\pm$~0.08~M$_{\odot}$ (\citealt{Hsieh18,Kim19}). Its bolometric luminosity (L$_{\rm bol}$) of 0.06~$\pm$~0.01~L$_{\odot}$ and its bolometric temperature (T$_{\rm bol}$) of 105~$\pm$~3~K classify it as a late Class 0 object (\citealt{Dunham08,Enoch09,Hsieh18,Kim19}). Observations with the Atacama Submillimeter Experiment (ASTE) have found proof of an outflow in the $J$=3$-$2 transitions of $^{12}$CO, $^{13}$CO, and C$^{18}$O, showing blue-shifted asymmetries and wings in the line profiles, mapping of an area of 120$^{\prime \prime}$~$\times$~140$^{\prime \prime}$ with the same facility has revealed outflow lobes as well (\citealt{Kim19}). The physical properties of the outflow, such as its dynamical time, were quantified from the bipolarity of the outflow and used to infer a current accretion rate from the outflow, $\dot{M} _{\rm acc}$, of 0.09~$\pm$~0.01~$\times$~10$^{-6}$~M$_{\odot}$~yr$^{-1}$ (\citealt{Kim19}). Over the estimated dynamical time of the outflow, the VeLLO will be able to accrete 0.02~$\pm$~0.01~M$_{\odot}$, which classifies it as a proto-brown dwarf according to \cite{Kim19}. The emission of the $J$=3$-$2 transition of $^{13}$CO peaks at a distance of 2.4$^{\prime \prime}$ from the continuum emission peak, which is likely due to the outflow reported in \cite{Kim19}. Observations with the Atacama Large Millimeter/submillimeter Array (ALMA) presented in \cite{Hsieh18} have inferred  from maps of $^{13}$CO, C$^{18}$O, and N$_2$H$^+$ that the VeLLO in DC3272+18 has recently undergone an outburst. The observations reveal that the center is devoid of N$_2$H$^+$ emission where CO evaporates, because gaseous CO destroys N$_2$H$^+$ via the reaction N$_2$H$^+$ + CO $\longrightarrow$ HCO$^+$ + N$_2$. Thus, the location of the N$_2$H$^+$ emission can be used to infer the snowline of CO, the location where it freezes out onto the grains. Chemical models by \cite{Hsieh18} have shown that the CO sublimation radius at T~=~20~K should be located at 102--164~au, based on its current internal and bolometric luminosity. However, the current sublimation radius of CO is found to be at a distance of 275--311~au. Moreover, the peak of the N$_2$H$^+$ emission is located at a distance of 775--1149~au. This is further evidence that the VeLLO has undergone a recent outburst, which has moved the CO snowline outwards, closer to the current peak abundance position of N$_2$H$^+$ (\citealt{Hsieh18}). The study of \cite{Hsieh18} also conducted modeling in order to constrain the outburst luminosity that is required to lead to the peak position of N$_2$H$^+$ and concluded that it can be explained if L$_{\rm burst}$ was 1--4~L$_{\odot}$, leading to a mass accretion rate of 6~$\pm$~4~$\times$~10$^{-6}$~M$_{\odot}$~yr$^{-1}$. This is one order of magnitude higher than what is derived from the outflow by \cite{Kim19}. The models of \cite{Hsieh18} put the CO sublimation radius during the outburst at a distance of 633--836~au.\\

\subsection{APEX observations}\label{sect:APEXobs}
Single-pointing observations with the Atacama Pathfinder EXperiment (APEX) were centered on the VeLLO in the DC3272+18 cloud ($\alpha _{\rm 2000}$ = 15$^h$42$^m$16$^s$.99, $\delta _{\rm 2000}$ = -52$^\circ$48$^\prime$02$^{\prime\prime}$.2) with the nFLASH230 instrument (project-ID O-0109.F-9305A-2022, PI: N. F. W. Ligterink). Two spectral settings were centered on frequencies of 219.6 and 254.4~GHz. The FFTS spectrometer was used as the backend, which provided a bandwidth of 32~GHz observed across 8~$\times$~65~536 channels with a spectral resolution of 61~kHz (0.07--0.08 km~s$^{-1}$). This resulted in spectral windows of 213.6--221.6~GHz and 229.6--237.6~GHz for setup 1, and 248.4--256.4 GHz and 264.4--272.4~GHz for setup 2. The observations were conducted on April 21-22 and April 28 - May 1 2022 with a precipitable water vapor (pwv) between 0.6 and 1.8~mm. Inspection of the data revealed the presence of prominent atmospheric features, but at frequencies that do not concern the transitions discussed in this work, the meteorological data point towards somewhat unstable conditions during the observations. The typical noise range of the observations is 3--10~mK (rms) for velocity resolutions of 0.07--0.08~km~s$^{-1}$. An outlier is a noise of 140~mK for the CO line at 230.538~GHz (Section \ref{sect:gridfitting}). Typical system temperatures during the observations ranged from 76 to 152~K. The main beam efficiency ($\eta _{\rm mb}$) for observations at 230~GHz is listed by the APEX website\footnote{\url{https://www.apex-telescope.org/telescope/efficiency/?yearBy=2022}} as 0.81, and the half power beam width (HPBW) as 26.2$^{\prime\prime}$ for April and May 2022. The observations were conducted in wobbler-switching mode during the first observing session. However, the wobbler position showed contamination of the CO $J$~=~2$-$1 transition at 230.538~GHz from the off position, as the off position was not emission-free from CO. The remaining observations were carried out in position-switching mode with the off position at $\alpha _{\rm 2000}$ = 15$^h$37$^m$00$^s$, $\delta _{\rm 2000}$ = -51$^\circ$10$^\prime$00$^{\prime\prime}$. Thus, observations of the CO line from April 21 were not taken into account for the data analysis. For all other lines, the wobbler-switched and position-switched data were examined, found to be free from contamination from the wobbler-switched off positions, and thus considered for the data analysis. Dedicated calibration uncertainties for nFLASH230 are not published, thus the calibration uncertainties of 10\% as published for the previous APEX-1 and APEX-2 receivers (\citealt{Dumke10}) are assumed. Adopting a distance of 250~pc (\citealt{Kim19}) to the source and at the beam size of 26.2$^{\prime \prime}$, the observations probe spatial scales of 6~550~au in DC3272+18.

\section{Results}\label{sect:results}

\subsection{Detected molecules}\label{sect:detections}

Line identification and baseline fitting were carried out with the \textsc{class} package of the \textsc{gildas}\footnote{\url{http://www.iram.fr/IRAMFR/GILDAS}} software. Line lists were taken from the Jet Propulsion Laboratory (JPL) catalog (\citealt{Pickett98}), and the Cologne Database for Molecular Spectroscopy (CDMS; \citealt{Muller05,Endres16}), the references to the spectroscopic parameters of the molecules that were fit in this work can be found in the Appendix in Table \ref{tab:specparams}. Identified spectral lines were then fitted with a Gaussian profile with the same program that calculates the peak temperature (T$_{\rm peak}$), the noise (T$_{\rm rms}$), the line width ($\delta$v), and the line velocity (v$_{\rm lsr}$). The securely and tentatively detected transitions and the results of the Gaussian fits are listed in Table \ref{tbl:obstransitions}, their spectra are shown in the appendix (Figs. \ref{fig:secure-transitions}, \ref{fig:tentative-transitions}). Calibration errors are assumed to be 10\% (Section \ref{sect:APEXobs}), thus the uncertainty of the peak intensities is derived as $\sqrt{(0.1 T_{\rm peak})^2 + (T_{\rm peak})^2}$. If the T$_{\rm peak}$ of a line is $\geq$~3~$\times$~rms, the line is identified as a detection, if it is $\sim$~3~$\times$~rms, it is identified as a tentative detection. The rms is calculated with the baseline routine in \textsc{class}, Gaussian profiles were fit with \textsc{class} and the \textsc{curve\_fit} package of \textsc{scipy} (\citealt{scipy}). \\
Most of the detected molecules (Table \ref{tbl:obstransitions}) are species that are commonly found around protostars. CO, and its $^{13}$C- and $^{18}$O- isotopologues are securely detected, its minor isotopologue of $^{13}$C$^{17}$O is not detected (Fig. \ref{fig:non-detections}, Table \ref{tbl:non-detections}). \cite{Kim19} found that CO traces the outflow in this VeLLO; in this case one could expect to detect multiple velocity components. While the CO emission is indeed double-peaked, which can hint that the material traces different velocity components, its line shape is consistent with self-absorption. It is to be noted that CO and also $^{13}$CO are optically thick, thus self-absorption can indeed contribute to the observed line shapes. This is supported by the finding that the CO absorption peaks at the systemic velocity of the cloud at $\sim$~-0.1 km/s (\citealt{Hsieh18}), and therefore one could conclude that the bulk of the emission traced by our observations stems from the on-source position covering the envelope enshrouding the VeLLO instead of the outflow. Fitting CO with two line components did not lead to a good match. The fit of CO in Fig. \ref{fig:secure-transitions} was obtained by masking the region that shows self-absorption between -1.7 and 0.75~km/s, and fitting the line with one component. On the other hand, $^{13}$CO can be fit well with two components (Fig. \ref{fig:secure-transitions}), their respective line widths of 0.8 and 0.5~km/s are slightly broadened compared to the rest of the molecules, which predominantly exhibit line widths between 0.35 and 0.5~km/s. This can be seen as a hint that some of $^{13}$CO that is observed stems from the outflow reported in \cite{Kim19}, which would explain the broader line widths. Alternatively, this could also be due to the lines being optically thick. The $J$~=~3$-$2 transitions of HCO$^+$, HCN, and HNC, are also securely detected. Moreover, species that are commonly found in UV-irradiated cavity walls carved out by the outflow (\citealt{Tychoniec21}), such as c-C$_3$H$_2$ are detected. SO$_2$, a species that is associated with shocks, outflows and sublimation products (\citealt{Tabone17}), is securely detected in one transition, and tentatively detected in a second transition. In addition, H$_2$CO (formaldehyde), CH$_3$OH (methanol) and SO are detected towards the VeLLO. Deuterated molecules, namely DCO$^+$, DCN, N$_2$D$^+$, one line of D$_2$CO, and C$_2$D are found as well. The latter is also a tracer of the cavity walls. Two CH$_2$DOH lines with favorable upper state energies ($<$~25~K) and Einstein A coefficients ($\sim$~10$^{-5}$~s$^{-1}$) are below the detection threshold for a tentative detection (Fig. \ref{fig:non-detections}, Table \ref{tbl:non-detections}). This is also true for the H$_2 ^{13}$CO transition at 219.909~GHz. H$_2$CO, CH$_3$OH and SO are species that are typically associated with the emission of warm gas close to a protostar and/or the envelope (e.g., \citealt{Ceccarelli07, Tychoniec21}). However, the upper energies of the securely detected lines are $\sim$ 20--44~K. Other H$_2$CO, CH$_3$OH and SO transitions covered by the observations with E$_{\rm up}$~$\sim$~50--90~K are not detected. This points toward their origin from the cold, extended envelope. With the exception of CO and $^{13}$CO that show multiple line components, the detected molecules have line widths of $\sim$~0.35--0.50~km/s, and there is no clear trend that can attribute different molecules to different structures of the system based on their line widths and the v$_{\rm lsr}$ of the different lines. All of the so far mentioned species are routinely observed towards protostars, but the molecular inventory probed by the APEX observations also reveals multiple transitions of nitric oxide (NO) at an intensity of 3--17~$\times$~rms, and line widths of $\sim$~0.4~km/s in the VeLLO, which adds this source to the small list of objects with reported NO detections in the literature (Section \ref{sect:NOchem}). \\
The SiO $J$~=~5$_{\rm 0}$~$-$~4$_{\rm 0}$ transition is targeted in the observations, but not detected (Fig. \ref{fig:non-detections}). SiO is a species that is commonly associated with shocks, but the transition at 217.1050~GHz (E$_{\rm up}$ = 31.3~K, A$_{\rm ij}$~=~5.2~$\times$~10$^{-4}$~s$^{-1}$; Fig. \ref{fig:non-detections}) is not detected in the spectrum. This could well be due to beam dilution effects, as SiO is a tracer of shocked regions that are more compact (\citealt{Tychoniec21}). It is also to note that no species with a higher degree of chemical complexity than methanol are detected in the data. This is also true for species with upper level energies that point toward an origin in the hot gas in close proximity to the VeLLO. This does not rule out their presence, it could also be that the low spatial resolution of the data is not sufficient to detect them due to beam dilution. The highest upper energy for a transition with a secure detection is 48~K, tentative detections have upper level energies of 41--131~K (Table \ref{tbl:obstransitions}).

\begin{table}[ht]
    \centering
    \caption{Column densities for the detected species.}
    \label{tab:Columndensities}
    \begin{tabular}{l|c|c|c}
    \hline \hline 
    && N$_{\rm obs}$ (10$^{11}$~cm$^{-2}$) & \\
    Species & T$_{\rm ex}$ = 10 K & T$_{\rm ex}$ = 30 K & T$_{\rm ex}$ = 50 K \\
    \hline 
    HCN & 3.02 $\substack{+0.53 \\ -0.51}$ & 1.51 $\substack{+0.25 \\ -0.22}$ & 1.74 $\substack{+0.29 \\ -0.30}$ \\
    DCN & 0.51 $\substack{+0.06 \\ -0.08}$ & 0.36 $\substack{+0.05 \\ - 0.04}$ & 0.44 $\substack{+0.06 \\ -0.07}$  \\
    HCO$^+$ & 18.62 $\substack{+3.26 \\ -2.77}$ & 7.59 $\substack{+0.83 \\ -0.93}$ & 8.71 $\substack{+0.95 \\ -1.07}$\\
    DCO$^+$ & 4.27 $\substack{+1.23 \\ -1.25}$ & 2.75 $\substack{+0.76 \\ -0.79}$  & 3.47 $\substack{+0.96 \\ - 0.90}$  \\
    HNC & 18.20 $\substack{+1.30 \\ 1.60}$ & 7.59 $\substack{+0.51 \\ -0.73 }$  & 8.71 $\substack{+0.58 \\ - 0.62}$ \\
    N$_2$D$^+$ & 1.32 $\substack{+0.34 \\ -0.32}$ & 0.83 $\substack{+0.22 \\ -0.20}$ & 0.10 $\substack{+0.03 \\ - 0.02}$ \\
    NO & 1000.00 $\substack{+96.48 \\ -160.11}$  & 707.95 $\substack{+86.38 \\ -76.99}$ & 912.01 $\substack{+111.28 \\ -99.18}$  \\
    D$_2$CO & 0.79 $\substack{+0.22 \\ - 0.20}$  & 0.54 $\substack{+0.15 \\ - 0.14}$ & 0.79 $\substack{+0.22 \\ -0.21}$ \\
    SO & 32.73 $\substack{+7.54 \\ -7.90}$ & 8.51 $\substack{+0.28 \\ -0.28}$  & 8.41 $\substack{+0.32 \\ - 0.31}$ \\
    C$_2$D & 4.67 $\substack{+1.05 \\ -1.08}$ & 3.24 $\substack{+0.72 \\ -0.75}$ & 4.07 $\substack{+0.91 \\ -0.94}$ \\
    CH$_3$OH & 97.72 $\substack{+28.48 \\ -31.14}$ & - & - \\
    H$_2$CO & 20.95 $\substack{+3.61 \\ -3.74}$  & - & -\\    
    \hline 
    \end{tabular}
    \tablefoot{The column densities were derived via the grid-fitting method for T$_{\rm ex}$ of 10, 30, and 50~K with the exception of CH$_3$OH and H$_2$CO, where only the values calculated at 10~K are shown (see text for discussion).}
\end{table}

\subsection{Constraints on the excitation temperature and column density}\label{sect:Tex}

Subsequent analysis of the detected lines to determine their excitation temperatures and column densities was carried out under the assumption of local thermal equilibrium (LTE). It is assumed that the beam size equals the source size and that both distributions are Gaussian, which leads to a beam filling factor of 0.5. This choice is made based on the assumption that the envelope material fills the beam. The beam-filling factor affects the column density, but as the detected lines are mainly optically thin, the derived column densities would be scaled by the beam-filling factor. If the molecules emit from the same region, the abundance ratios are independent of the choice of filling factor.

\subsubsection{Rotational diagram analysis of c-C$_3$H$_2$}

Under the assumption that LTE is applicable, the distribution of the molecular energy levels follows a Boltzmann distribution, and the upper state level population (N$_{\rm u}$) divided by the statistical weight (g$_{\rm u}$) of the molecular energy levels can be related to their upper level energies (E$_{\rm u}$) by the Boltzmann equation:
\begin{equation}\label{eq:boltzmanndistro}
     \frac{N_u}{g_u} =  \frac{N_{tot}}{Q(T_{rot})} e^{-E_u / kT_{rot}} 
\end{equation}
where k is the Boltzmann constant, N$_{\rm tot}$ is the total column density, and Q corresponds to the partition function value at the corresponding rotational temperature, T$_{\rm rot}$. In a conventional rotational diagram analysis (e.g., \citealt{Blake87,Goldsmith99}), taking the logarithm of Eq. \ref{eq:boltzmanndistro} allows for a linear least squares regression:
\begin{equation}\label{eq:rotdiag}
    ln  \frac{N_u}{g_u} = ln N_{tot} - lnQ(T_{rot}) - \frac{E_u}{kT_{rot}}.
\end{equation}
In LTE, the rotational temperature T$_{\rm rot}$ corresponds to the excitation temperature T$_{\rm ex}$. By constructing a semi-log plot of N$_{\rm u}$ / g$_{\rm u}$ against E$_{\rm u}$, the rotational temperature, T$_{\rm rot}$, and the total column density, N$_{\rm tot}$, can be derived from the best-fit slope and intercept, respectively. As the optical depth of the transitions is unknown, an optical depth correction factor, C$_{\tau}$, needs to be applied, to account for cases in which the optical depth is $\tau \NOTll$1
\begin{equation}\label{eq:Ct}
    C_ \tau = \frac{\tau}{1 - e^{- \tau}}.
\end{equation}
Therefore, the right-hand side of Eq. \ref{eq:rotdiag} is rewritten as:
\begin{equation}\label{eq:rotdiagtau}
     ln(N_{tot}) - lnQ(T_{rot}) - lnC_{\tau} -  \frac{E_u}{kT_{rot}}.
\end{equation}
The optical depth of each transition is calculated via
\begin{equation}\label{eq:tau}
     \tau _{ul} = \frac{A_{ul} c^3}{8\pi \nu ^3 \delta v} N_u (e^{\frac{h\nu }{kT_{rot}}} - 1).
\end{equation}
A$_{\rm ul}$ corresponds to the Einstein A coefficient, c to the speed of light, $\nu$ to the frequency, and $\delta$v to the line width. This allows to re-write C$_{\tau}$ as a function of N$_{\rm u}$, and substitute it into Eq. \ref{eq:rotdiagtau} to construct a likelihood function $\mathcal{L}$(N$_{\rm u}$, T$_{\rm rot}$) for $\chi ^2$ minimization.
This likelihood function is used with the affine-invariant Markov Chain Monte Carlo (MCMC) code $\mathtt{emcee}$ (\citealt{Foreman-Mackey_2013}) to determine N$_{\rm u}$ and T$_{\rm rot}$.\\
This method requires multiple transitions per molecule to ensure a meaningful fit, thus this method is only applied to c-C$_3$H$_2$. The result is obtained with two calculations. First, a broad parameter space for the total column density N$_{\rm tot}$ = 10$^7$--10$^{14}$~cm$^{-2}$ and T$_{\rm rot}$ = 10--500~K is explored with 50 walkers and 500 steps. Those results are used to narrow down the parameter space to N$_{\rm tot}$ = 10$^{11}$--10$^{13}$~cm$^{-2}$ and T$_{\rm rot}$ = 10--100~K. After 2000 steps with 50 walkers, a column density of N$_{\rm tot}$ = 3.78 $\substack{+0.37 \\ -0.40}$~$\times$~10$^{11}$~cm$^{-2}$ and a T$_{\rm rot}$ of 35.4 $\substack{+3.2\\-4.9}$~K are obtained. Parameters and uncertainties correspond to the 50th, 16th, and 84th percentiles from the marginalized posterior distributions. The resulting population diagram is plotted in Fig. \ref{fig:RTDs} and reveals that c-C$_3$H$_2$ traces a warm gas component in the VeLLO in DC3272+18. This species is commonly associated with the cavities carved by protostellar outflows (e.g., \citealt{Murillo18,Tychoniec21}), and it is likely that it indeed traces the cavities of the outflow that \cite{Kim19} have detected in CO. The opacities of the c-C$_3$H$_2$ transitions are in the range of 0.002-0.006.

\subsubsection{Constraining the excitation temperature of methanol and formaldehyde}\label{sect:Texmeth}

Methanol and formaldehyde emit from a cold gas reservoir in the envelope surrounding the VeLLO. In DC3272+18 one H$_2$CO transition at 218.2222~GHz with an upper level energy of 20.96~K is detected (Fig. \ref{fig:secure-transitions}). However, two additional transitions at 218.4756~GHz ($J$=3$_{\rm 2,2}-$2$_{\rm 2,1}$) and 218.7601~GHz ($J$=3$_{\rm 2,1}-$2$_{\rm 2,0}$), both with E$_{\rm up}$ of 68.1~K, are not detected. Assuming that intensities $>$~3~$\times$~rms correspond to a detection, an LTE model that is based on the formalism of the \textsc{cassis}\footnote{\url{http://cassis.irap.omp.eu/}} (\citealt{Vastel15}) software was utilized to explore column densities in combination with excitation temperatures of 10--50~K in 5~K steps that simultaneously match the T$_{\rm peak}$ value of the detection at 218.222~GHz and the non-detection of the two other transitions. First, the fixed excitation temperature was used to find a column density that can reproduce the line intensity of the observed H$_2$CO transition. In a second step, this column density and excitation temperature was used to obtain the peak intensity of the two other H$_2$CO transition. Only at T$_{\rm ex}$ = 10~K the two other transitions remain undetected. \\
This was also tested for 12 additional transitions of CH$_3$OH with E$_{\rm up}$ in the range of $\sim$ 50--90~K in addition to the two lines that are detected in DC3272+18. As for H$_2$CO, it was not possible to reproduce the detected lines without overproducing the intensities of the non-detections at T$_{\rm ex} >$~10~K. Thus, the H$_2$CO and CH$_3$OH detected in the APEX data stem from a cold gas component and not from a potential hot corino, {which are regions close to the protostar that have T $>$~100~K and are thus associated with the sublimation of COMs (\citealt{Ceccarelli07}). If a second, warmer gas reservoir is present, the spatial resolution of the observations is not sufficient to detect this emission.

\subsubsection{Grid-fitting of species tracing the cold gas}\label{sect:gridfitting}

Species such as N$_2$D$^+$, HNC, HCN, DCN, HCO$^+$, and DCO$^+$ are commonly associated with the cold envelope (e.g., \citealt{Tychoniec21}), and thus likely emit from a region with conditions similar to H$_2$CO and CH$_3$OH. In order to determine the column densities of these species, the excitation temperature is set to a fixed value of 10~K, which corresponds to T$_{\rm ex}$ derived from H$_2$CO and CH$_3$OH, but it is also the value used by \cite{Hsieh18} to fit the column densities of N$_2$H$^+$ and the CO isotopologues. \textsc{cassis} is used to fit over a grid of N$_{\rm tot}$ assuming LTE conditions. The position of the lines is set to the source v$_{\rm LSR}$ of -0.1~km/s, the full width half maximum (FWHM) are set to the values given in Table \ref{tbl:obstransitions}, and the source size is set to match the beam size. Model spectra are computed with a step size of 0.01 in logarithmic space to explore column densities in the range of 10$^{10}$--10$^{15}$~cm$^{-2}$. The best-fit column densities and their 2$\sigma$ uncertainties for each of the above-mentioned molecules, as well as NO, SO, and D$_2$CO are listed in Table \ref{tab:Columndensities}. CO and its isotopologues were excluded from this process. CO and $^{13}$CO do suffer from optical depth issues. C$^{18}$O is potentially affected as well, while the rms at the position of the covered $^{13}$C$^{17}$O transition is only $\sim$~2~$\times$~rms, and therefore counted as a non-detection. Optical depth is not an issue for the remaining molecules, as $\tau$ is found to be $<$~1.

\begin{figure}
    \centering
    \includegraphics[width=0.95\columnwidth]{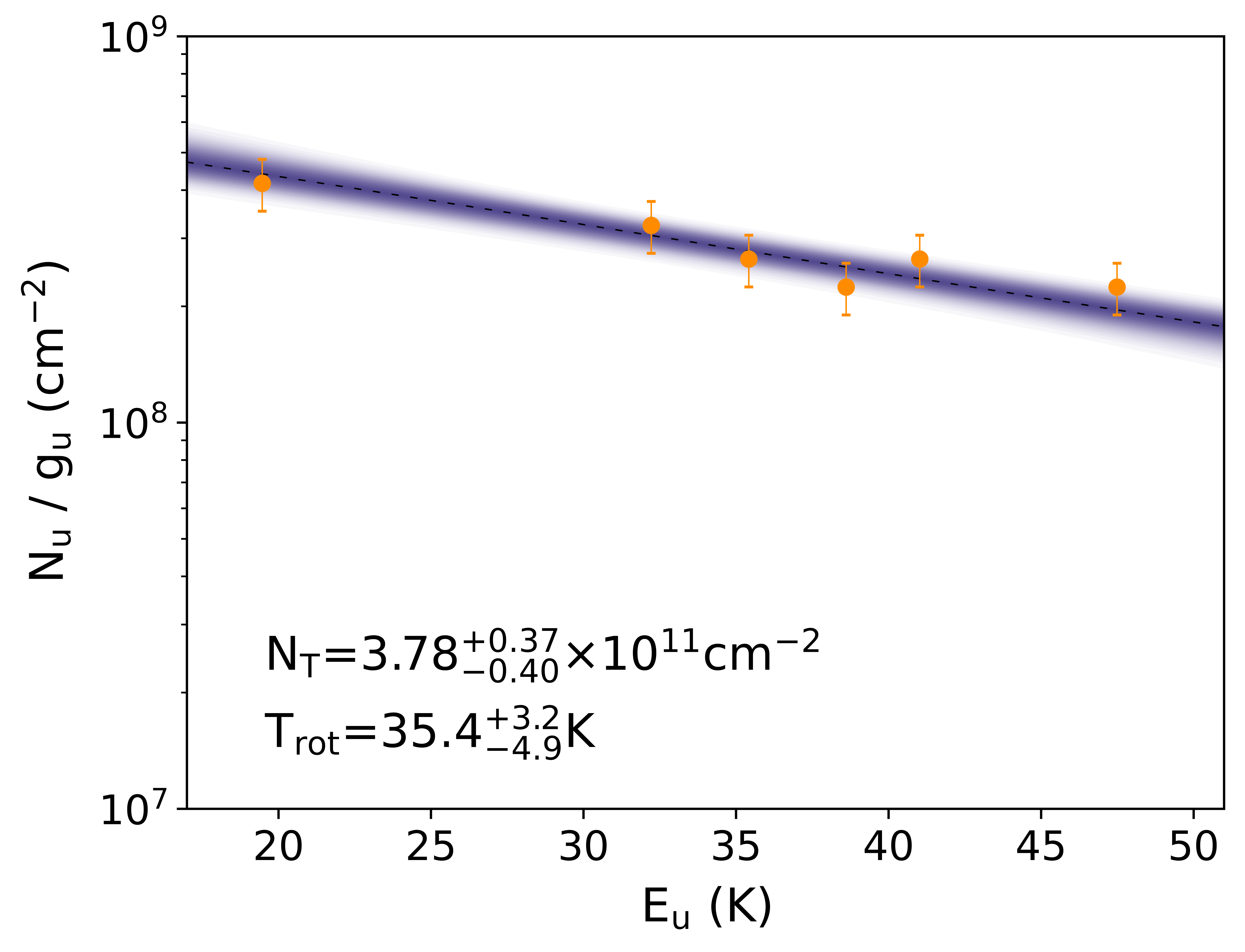}
    \caption{Population diagram for c-C$_3$H$_2$. The shaded region in purple marks the region between the 16th and 84th percentile. Here, N$_{\rm T}$ corresponds to the total column density.}
    \label{fig:RTDs}
\end{figure}

\subsection{Detection of NO}\label{sect:NOchem}
Seven transitions of NO (six secure, one tentative) are detected around the DC3272+18 VeLLO, which adds it to a small list of different types of sources that have reported detections of this molecule. It has been detected in the dark clouds TMC-1 and L134N (\citealt{McGonagle90,Gerin93}), around the low-mass protostars NGC 1333 IRAS 4A, SVS13-A, and IRAS 16293-2422 B, in the shock of L1157-B1 (\citealt{Yildiz13,Codella18,Ligterink18}), in the high-mass star-forming region Sgr B2 toward Sgr B2(M) and Sgr B2(N) (\citealt{Liszt78,Ziurys94,Halfen2001}), and most recently in the dust trap of the Oph-IRS 48 disk (\citealt{Brunken22,Leemker23}). For Oph-IRS 48, models have shown that its abundance can be increased when ice sublimation boosts the abundance of H$_2$O and NH$_3$ (\citealt{Leemker23}). The main formation pathway of NO is the gas-phase neutral-neutral reaction 
\begin{equation}\label{eq:NO_1}
    \rm N + OH \longrightarrow NO + H
\end{equation} in oxygen-enriched gas, where OH is a product of the photodissociation of water. \\
\noindent It can also form via a second neutral-neutral reaction, namely 
\begin{equation}\label{eq:NO_2}
    \rm NH + O \longrightarrow NO + H
\end{equation}
(\citealt{Millar91,Baulch05,Wakelam12}). In addition, photodissociation of species such as HNO, N$_2$O, and NO$_2$ can also produce NO (\citealt{vanDishoeck06,Heays17}), but these species are usually neither abundant nor commonly detected around protostars, so their contribution to the abundance of NO is presumably small. After its formation NO can deplete from the gas by accreting onto dust grains, where it gets hydrogenated to form species such as NH$_2$OH (\citealt{Congiu12,Fedoseev12}). The formation of N$_2$O is a by-product of the hydrogenation process of NO to NH$_2$OH. A ratio $<$1 for NO/N$_2$O would indicate that some of the NO is lost to this formation channel. However, the observed frequency range only covers one N$_2$O transition at 251.2216~GHz that is not detected, likely due to its high E$_{\rm up}$ (63.3~K) and low A$_{\rm ij}$ coefficient (2.3~$\times$~10$^{-6}$~s$^{-1}$), so it is not possible to assess whether some NO has already been frozen out and has been incorporated in grain surface processes via this pathway. If returned back into the gas phase, the photodissociation of NH$_2$OH can also form NO, but this destruction channel is not efficient (\citealt{Gericke94,Fedoseev16}). The predominant product channel from photodissociated NH$_2$OH leads to NH$_2$ + OH, which is not a direct parent species for NO (\citealt{Betts65}). In addition to freezing out, NO can be directly destroyed in the gas phase via photodissociation reactions, or the reaction N + NO $\longrightarrow$ N$_2$ + O (\citealt{Millar91}). However, photodissociation induced by internal UV photons from the VeLLO during its quiescent phase is less effective than during the outburst, so while increased UV flux during the outburst can indeed boost the photodissociation of species such as H$_2$O and NH$_3$, which would deliver the reactants to form NO from the two neutral-neutral reactions in Eq. \ref{eq:NO_1} and \ref{eq:NO_2}, the photodissociation of NO after the VeLLO has returned to its quiescent phase does play a smaller role. NO has a low binding energy ($\sim$~1~600~K; \citealt{Wakelam17}), and thus does not freeze out rapidly after the gas in the envelope cools down to its pre-outburst temperature. The sublimation temperature of NO is $\sim$30--50~K, so in current envelope conditions its presence in the gas phase cannot be explained by thermal desorption. In addition, internal UV photons from the quiescent VeLLO do not lead to significant photodissociation, while its isolated position in the DC3272+18 region distances it from external UV flux. On the other hand, H$_2$O has a binding energy of $\sim$~5~640~K, which requires temperatures between 90 and 140~K in the interstellar medium (ISM) to sublimate, and already freezes out after 10$^2$--10$^3$~yr at typical envelope densities (\citealt{Minissale22}). \\
\noindent As for the other species, the beam size of the observations does not allow the location of the NO emission to be spatially resolved. Under the assumption that the molecular emission is co-spatial, the column density of NO is e.g., 10 times higher than that of CH$_3$OH, which is a species that freezes out on similar time scales as water (e.g., \citealt{Collings04}), so this indicates that the freeze-out process of NO has not progressed as far as for e.g., CH$_3$OH, yet.

\subsection{D/H ratios}\label{sect:DHratios}
The average ratio of deuterium over hydrogen (D/H ratio) in the local ISM is $\sim$~2$\times$~10$^{-5}$ (\citealt{Linsky06,Prodanovic10}). Molecules with an enrichment in deuterium by up to 4 orders of magnitude compared to the interstellar average are often linked to a formation before the protostellar stage, as temperatures $<$~20~K boost deuterium fractionation (\citealt{Caselli12}). The D/H ratios for the deuterated molecules in Table \ref{tab:DHratios} have been calculated based on the column densities in Table \ref{tab:Columndensities}, but it is to note that all ratios have been derived based on one transition per molecule, and therefore, the presented values should be interpreted with caution. All D/H ratios are in the percent range (17--23\%), thus pointing to an origin from a cold environment, potentially before the formation of the VeLLO. DCO$^+$ and DCN formation occurs in the gas phase (\citealt{Willacy07}), while D$_2$CO forms via grain-surface reactions at temperatures $<$~20~K (\citealt{Nagaoka07,Hidaka09}). The derived D/H ratios in the VeLLO are comparable to findings in prestellar cores and around low-mass protostars (\citealt{Bizzocchi14,ChaconTanarro19,Ambrose21,Drozdovskaya22,Lin23L}) and thus favor their prestellar origin and subsequent inheritance to the protostellar stage. 

\begin{table}[h]
    \centering
    \caption{D/H ratios calculated with a fixed T$_{\rm ex}$ of 10~K.}
    \small{\begin{tabular}{r|c}
    \hline \hline 
    Ratio  & D/H ratio w. stat. corr. \\
    \hline 
    DCO$^+$/HCO$^+$ & 0.23 $\substack{+0.10 \\ -0.11}$ \\
    DCN/HCN & 0.17 $\pm$ 0.05 \\
    D$_2$CO /H$_2$CO & 0.19 $\pm$ 0.04 \\
    \hline          
    \end{tabular}}
    \label{tab:DHratios}
\end{table}

\subsection{Physical conditions implied by line intensity ratios}\label{sect:physicalcond}

The line intensity ratios of emission lines can be utilized to probe the physical conditions in the ISM. \cite{Hacar20} demonstrated that the intensity ratio of the $J$=1$-$0 transitions of HCN and HNC can be used to infer the kinetic gas temperature and derived a two-component scaling relation from large-scale observations in Orion. This scaling relation is attributed to different HNC destruction pathways whose efficiency depends on the temperature. \cite{Hacar20} deduce the scaling relation from the $J$=1$-$0 transitions of HCN and HNC. As a test in this work, it is assumed that the relation also holds true for the $J$=3$-$2 transitions that are detected in DC3272+18. The peak intensities listed in Table \ref{tbl:obstransitions} assume a flux calibration uncertainty of 10\%. However, both HCN and HNC are observed in the same sideband, thus a relative calibration uncertainty within the band of 1\% is assumed, resulting in an error of $\sqrt{(0.01 T_{\rm peak})^2 + (T_{\rm rms})^2}$. The line intensity ratio $I_{\rm HCN}$/$I_{\rm HNC}$ derived from the T$_{\rm peak}$ values in DC3272+18 is 0.27~$\pm$~0.07. Thus, the scaling relation for $I_{\rm HCN}$/$I_{\rm HNC}$ $<4$ given in Eq. 3 in \cite{Hacar20} would apply, which yields T$_{\rm kin}$ (K) = 10 $\times$ $I_{\rm HCN}$/$I_{\rm HNC}$ = 2.68 $\pm$ 0.7~K, which would put the kinetic temperature below the cosmic background temperature if this scaling law holds true. However, \cite{Hacar20} point out that this relation is only valid down to kinetic temperatures of 15~K, below the errors become large, e.g., errors in T$_{\rm kin}$ are $>$5~K for intensity ratios $<1$. Therefore, this method cannot be used to derive the kinetic temperature, but it reveals that it must be cold, likely below the 15~K threshold, where the scaling law is not valid anymore. This is also supported by additional line ratios that offer tighter constraints.\\
\noindent \cite{Murillo18} conducted a survey of twelve low-mass protostellar systems in Perseus with APEX. The frequency range of their observations partially overlaps with the range observed in this work, thus the line intensity ratios of the two works can be compared. The first intensity ratio that is considered is $I_{\rm DCN}$/$I_{\rm DCO^+}$. This is due to the D/H ratio of both molecules, which is tied to chemical pathways leading to their deuteration at different temperatures. Both transitions discussed here were observed in the same sideband, thus a flux calibration uncertainty of 1\% is assumed. This ratio can be used as a proxy for the gas temperature. While $\sim$~2~$\times$~10$^{-5}$ corresponds to the average D/H ratio in the local ISM (\citealt{Linsky06,Prodanovic10}), species get enriched in deuterium when the deuterium fractionation becomes efficient once temperatures drop below 20~K and gas densities reach 10$^4$~cm$^{-3}$ via the reaction
\begin{equation}
    \rm H_3^+ + HD \rightleftharpoons H_2D^+ + H_2 + \Delta E,
\end{equation}
where $\Delta$E~=~232~K (\citealt{Watson74}). DCO$^+$ subsequently forms from the reaction of H$_2$D$^+$ + CO, thus it has been deemed a suitable tracer for the cold gas (\citealt{Jorgensen05,Murillo15}). Then, DCO$^+$ can propagate its deuteration to DCN (\citealt{Willacy07}) via
\begin{equation}
    \rm DCO^+ + HNC \longrightarrow HNCD^+ + CO
\end{equation}
and 
\begin{equation}
    \rm HNCD^+ + e^- \longrightarrow DCN + H.
\end{equation}
Once the temperature increases to $>$30~K, a second pathway leading to deuteration fractionation starts to dominate, namely via the reactions 
\begin{equation}
    \rm CH_3 ^+ + HD \rightleftharpoons CH_2D^+ + H_2 + \Delta E
\end{equation}
and 
\begin{equation}
    \rm C_2H_2 ^+ + HD \rightleftharpoons C_2HD^+ + H_2 + \Delta E
\end{equation}
(\citealt{Millar89,Roueff13}), where $\Delta$E is $\sim$ 390~K for the first reaction (\citealt{Asvany04}), and $\sim$~550~K for the second reaction (\citealt{Herbst87}).\\
\noindent CH$_2$D$^+$ is the starting point for DCO$^+$ (via CH$_2$D$^+$ + CO $\longrightarrow$ DCO$^+$ + CH$_2$; \citealt{Favre15}) and DCN formation (via CH$_2$D$^+$ + e$^-$ $\longrightarrow$ CHD + H followed by CHD + N $\longrightarrow$ DCN + H; \citealt{Millar89}). Both routes are most efficient for temperatures of $\sim$~70--100~K (\citealt{Favre15}). The intensity ratio of $I_{\rm DCN}$/$I_{\rm DCO^+}$ in DC3272+18 is 0.050~$\pm$~0.005, which is lower by a factor of $\geq$~2 than the values that \cite{Murillo18} derive for their sample of low-mass protostars. The low intensity of DCN does not suggest significant contribution from the warm formation pathway. It is also most likely that DCO$^+$ forms solely from the cold formation pathway, and has not proceeded to form a substantial amount of DCN, yet. Thus, the combination of the low value derived for $I_{\rm HCN}$/$I_{\rm HNC}$ and the high intensity of DCO$^+$ strongly hints that the probed material around the VeLLO is predominately cold. \\
\noindent This is also concluded from two additional sets of line ratios that are calculated to probe the gas temperature from c-C$_3$H$_2$ transitions. The 6$_{\rm 0,6}$ - 5$_{\rm 1,5}$ and 3$_{\rm 3,0}$ - 2$_{\rm 2,1}$ transitions of c-C$_3$H$_2$ have E$_{\rm up}$ of 38.6 and 19.5~K, respectively, thus their intensity ratio can give hints about the gas temperature, too. This also applies to the ratio of the 5$_{\rm 1,4}$ - 4$_{\rm 2,3}$ transition  (E$_{\rm up}$ = 35.4~K) over the 3$_{\rm 3,0}$ - 2$_{\rm 2,1}$ (E$_{\rm up}$ = 19.5~K) transition. As for the other two ratios discussed above, all transitions were observed in the same sideband, a flux uncertainty of 1\% is therefore assumed. The derived intensity ratios are 0.297~$\pm$~0.040, and 0.173~$\pm$~0.014, respectively, which yet again indicates that the warmer transitions are less excited as the gas probed by the observations is cold. The line ratios obtained for the VeLLO are also on the lower end of the range of ratios that \cite{Murillo18} derive for their sample of low-mass protostars, which are $\sim$ 0.38--1.15 and 0.3--0.92 for the two c-C$_3$H$_2$ ratios, respectively. The ratios do not seem to depend on the mass of the envelope, but they do scale with the bolometric luminosity (Fig. \ref{fig:cC3H2_protostars}), as lower ratios are found for the faintest sources. \\
\noindent Thus, even though it is not possible with the current data to derive an exact value for the excitation temperature of the detected molecules, the line intensity ratios also strongly suggest that the probed material is predominantly cold. 
The low ratio of $I_{\rm DCN}$/$I_{\rm DCO^+}$ suggests a kinetic gas temperature of $<$30~K, as otherwise DCN should be more abundant, and the ratio of $I_{\rm HCN}$/$I_{\rm HNC}$ suggests an even lower value of $\leq$15~K. These findings are in line with the excitation temperature that was derived for formaldehyde and methanol (Section \ref{sect:Texmeth}).

\begin{figure}
    \centering
    \includegraphics[width=0.9\columnwidth]{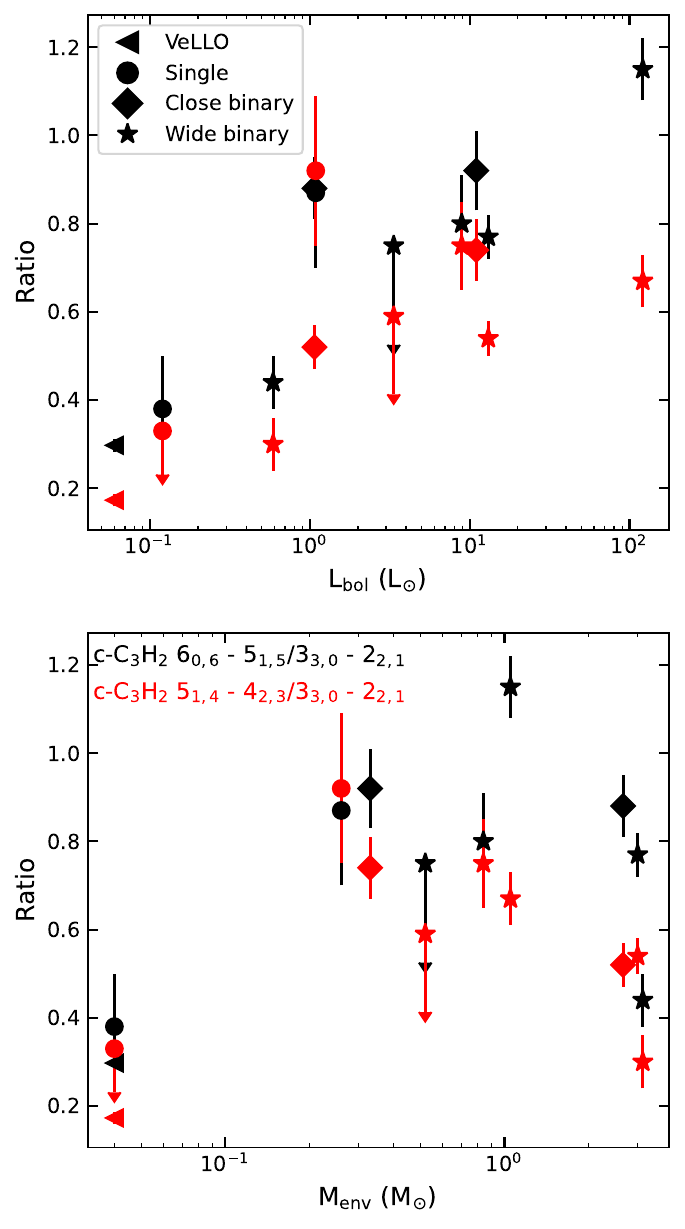}
    \caption{Line intensity ratios of c-C$_3$H$_2$ transitions for the VeLLO, and the sample of \cite{Murillo18} that consists of single protostars, close binary systems, and wide binaries. Close binaries in \cite{Murillo18} are separated by $<$2$^{\prime \prime}$, the wide binaries in this work have separations in the range of 7--46$^{\prime \prime}$. The ratios are plotted against the bolometric luminosity, L$_{\rm bol}$ in the upper panel, and against the envelope mass, M$_{\rm env}$ in the lower panel.}
    \label{fig:cC3H2_protostars}
\end{figure}

\subsection{Line emission and system parameters}\label{sect:systemparams}

The large beam of the observations (26.2$^{\prime \prime}$) does not allow to spatially resolve different physical components of the system. However, the variety of the detected species and specifically their detected transitions can constrain the physical conditions and structures around the VeLLO. The discussed line ratios in Section \ref{sect:physicalcond}, and the constraint on T$_{\rm ex}$ of 10~K for species such as H$_2$CO and CH$_3$OH imply that most of the gas around the VeLLO is cold. This is supported by the detection of molecular tracers that are commonly associated with the cold envelope, such as DCO$^+$ and N$_2$D$^+$ (\citealt{Tychoniec21}). In addition to their excitation temperature, their narrow line width of $\sim$~0.5~km/s points to its origin from the cold envelope as well. This is also likely for the species NO, SO, H$_2$CO, and CH$_3$OH. In contrast, c-C$_3$H$_2$ is a molecule commonly associated with the cavity walls carved out by the outflow. This is strongly supported by the population diagram fit that derives a rotational temperature of $\sim$~35~K for c-C$_3$H$_2$, indicating that it traces a second, warm gas reservoir around the VeLLO. The presence of an outflow cavity is also supported by the detection of multiple C$_2$D lines (\citealt{Tychoniec21}). SiO is commonly used to trace shocks around protostars, as it is linked to the destruction of Si-rich grains due to the high temperatures induced by shocks. The non-detection of SiO in the DC3272+18 cloud suggests that either no shock is present in this source, or that it is too weak to destroy dust grains. The presence of SO$_2$ does not clearly support the detection of a shock either. While it can be a tracer of shocked gas, it is also often associated with sublimated grain-surface species (\citealt{Tychoniec21}). The 4$_{\rm 2,2}$ - 3$_{\rm 1,3}$ transition with an upper energy of 19.0~K is securely detected, and most likely traces the same gas reservoir as species like NO and CH$_3$OH, namely the cold envelope. A second SO$_2$ transition with E$_{\rm up}$ of 130.7~K is only tentatively detected, a warmer gas reservoir, e.g., one that is heated by a shock, would be necessary to excite this line. However, no other shock tracers such as H$_2$CO, CO and SO transitions with high E$_{\rm up}$ are detected (\citealt{Tabone20,Tychoniec21}), so it is unlikely that the APEX observations have picked up signatures of shocked material. In contrast, tracers of the outflow from the VeLLO are detected. Previous observations of \cite{Kim19} had already confirmed the presence of an outflow in $^{12}$CO, $^{13}$CO, and C$^{18}$O, and the line widths of those molecules are broadened by $\sim$~20\% compared to species such as C$_2$D or CH$_3$OH in the APEX data.  \\
\noindent The here presented observations indicate that the majority of the probed material stems from the cold envelope. Molecular tracers attributed to the outflow, and the cavity walls have also been detected. 

\section{Discussion}\label{sect:discussion}

\subsection{Does NO trace the fossilized water snowline?}

The high column density of NO relative to CH$_3$OH can be interpreted as a chemical signpost of the past outburst. If NO and methanol indeed probe the same region, with all evidence pointing to it being the cold envelope, their column density ratio at T$_{\rm ex}$ of 10~K comes to $\sim$~10 (Table \ref{tab:Columndensities}). Thus, NO is estimated to be more abundant than CH$_3$OH by one order of magnitude. This is in contrast to all NO detections that have been reported so far. NO is 1--3 orders of magnitudes less abundant in the low-mass protostars IRAS 16293-2422 B (\citealt{Jorgensen18,Ligterink18}) and SVS-13A (\citealt{Bianchi17}), and the shock in L1157-B1 (\citealt{Codella18}). Grain-surface formation of NO and CH$_3$OH and subsequent thermal sublimation during the quiescent phase can be ruled out. NO has a binding energy of 1~600~K and thermally desorbs at $\sim$~30--50~K (\citealt{Collings04,Wakelam17}), CH$_3$OH has a binding energy of 5~500--6~600~K, which leads to a desorption temperature of $\sim$~100~K (\citealt{Collings04,Sakai13,Minissale22}). Thus, the current temperature in the envelope would not allow for efficient thermal desorption of either of the two species. In addition, NO is hydrogenated to form NH$_2$OH once it freezes out onto the grain surfaces (\citealt{Fedoseev16}), therefore it is not likely that abundant NO is sublimated directly from the grain surfaces.
It is more likely that the high NO column density is linked to its efficient formation after the outburst in the gas phase. Its main formation pathway in the gas phase (Eq. \ref{eq:NO_1}) requires N and OH. The increased temperature during the heating event leads to the sublimation of icy material containing species such as CH$_3$OH, but more importantly H$_2$O and NH$_3$, which all desorb at temperatures of $\sim$~100~K (\citealt{Collings04,Wakelam17,Minissale22}). Photodissociated products of the major ice constituents H$_2$O and NH$_3$ (\citealt{Boogert15}) are then consumed to form NO at the location where the water snowline got shifted to during the outburst. Due to the increased UV flux from the VeLLO during the outburst the efficiency of photodissociation is elevated, which provides the necessary ingredients to form abundant NO in the gas phase. Thus, the presence of NO at a high column density relative to methanol could be a direct consequence of the outburst of the VeLLO. After the end of the accretion outburst the temperature will decrease again as the VeLLO reaches its quiescent phase. Subsequently, species that have been evaporated during the outburst will start to freeze out again, that order depends on the binding energies of the molecules. Chemical models by \cite{Visser15} have shown that the water snowline moves outwards by a factor of $\sim$~10 during the outburst and that H$_2$O abundances remain enhanced for 10$^2$--10$^3$~yr after the outburst. In the case of CO the enhancement is seen for 10$^3$--10$^4$~yr, so while it is not possible to pinpoint the exact point in time of the outburst, the extended CO snowline compared to its quiescent luminosity (\citealt{Hsieh18}) is evidence that the outburst happened less than 10$^4$~yr ago. The results presented in this paper show that NO could be another molecule that serves as a tracer that is available for longer timescales than the typical 10$^2$--10$^3$~yr of most direct tracers of the water snowline. Its longevity as a tracer is reflected in the elevated NO/CH$_3$OH ratio. Due to its higher binding energy, CH$_3$OH starts its freeze-out in a point in time closer to the outburst, while NO still remains in the gas phase. If NO indeed forms from the reaction in Eq. \ref{eq:NO_1}, it can be used to expose the location of the position of the shifted water snowline during the outburst, as this proposed formation pathway is tightly linked to species that sublimate/freeze out at the water snowline. This would add NO to the list of tracers of the water snowline that is more commonly probed via HCO$^+$, CH$_3$OH, HDO and H$_2 ^{18}$O (\citealt{Hsieh19,Tobin23}). Observations with a better spatial resolution are required to confirm the suspected co-spatial origin of the NO and CH$_3$OH emission.

\subsection{Post-outburst chemistry, non-thermal desorption, or outflows?}

While an outburst that leads to sublimation of grain-surface species due to its accompanied temperature increase in the envelope followed by boosted photodissociation due to its higher UV flux during its active phase is a possibility to explain the peculiarly high NO abundance in the VeLLO, other options also need to be considered. \\
Another mechanism that can release grain surface species into the gas phase and that consequently has to be considered as an alternative explanation is reactive desorption. Reactive desorption is a non-thermal desorption process where the energy of chemical reactions on the grains is not fully absorbed by the grain, and the remaining energy causes the ejection of a percentage of the newly formed product in the gas-phase (e.g., \citealt{Vasyunin17,Pantaleone20}). This process is responsible for the presence of COMs such as CH$_3$OH in the gas in prestellar cores (e.g., \citealt{JimenezSerra16}), and could also contribute to the molecular abundances in the envelope around the VeLLO. Chemical models find that 1--10\% of the formed grain surface product gets ejected into the gas phase via this desorption process (e.g., \citealt{Garrod08}). Thus, reactive desorption processes could have released CH$_3$OH into the gas phase, which would be an alternative explanation to thermal desorption induced by the outburst. This is also true for H$_2$O and NH$_3$. However, without an outburst that increases the internal UV flux and photodissociates H$_2$O and NH$_3$ it is unlikely that the reactants required to form NO via Eq. \ref{eq:NO_1} and \ref{eq:NO_2} are abundant enough to result in a column density ratio of 10 for NO/CH$_3$OH. \\
Another possibility is that the outflow emanating from the VeLLO sublimates grain-surface species locally. Yet again, this option could therefore explain the presence of CH$_3$OH in the gas. However, there are two caveats coming with this option. First, lines of molecules in the outflow are usually broadened, as it is seen for CO and $^{13}$CO, but not for CH$_3$OH. Second, as for the explanation with reactive desorption, this option does not present an effective path towards photodissociation of H$_2$O and NH$_3$. Thus, it is unlikely that the outflow played a major role in the formation of NO, and the presence of gaseous CH$_3$OH in this source. If it contributes marginally can only be assessed with observations with a higher spatial resolution. If those species are originating from the outflow cavity wall, a shift in their v$_{\rm vlsr}$ would be expected. \\
Thus, an outburst is the most likely explanation for the high abundance of NO in the VeLLO, but based on the available data it is not possible to fully rule out other options or a combination of multiple scenarios.

\subsection{Limits on Complex Organic Molecules}

COMs are precursors of prebiotic molecules, thus revealing their formation pathways in space is a central interest of astrochemical studies. Around protostars they are often associated with hot cores or hot corinos, which is a region in close proximity to the protostar that has warmed up sufficiently to sublimate COMs from icy grains and form them actively in the gas phase (\citealt{Ceccarelli07,Herbst09}). Nevertheless, they are also readily detected in cold, prestellar cores, where their presence in the gas phase is attributed to non-thermal desorption processes such as reactive desorption (\citealt{Garrod06,Vasyunin17}), or cosmic ray-induced desorption (\citealt{Sipilä21}). The most complex species detected in the VeLLO is CH$_3$OH, the question now becomes whether the VeLLO is COM-poor, or if the current sensitivity is not sufficient to detect more complex species, or if the beam dilution effects hamper the chance of a detection. This has been tested for three COMs that are routinely detected around protostars and in cores, namely methyl formate (CH$_3$OCHO), dimethyl ether (CH$_3$OCH$_3$), and acetaldehyde (CH$_3$CHO). Their abundance relative to methanol is found to be in the range of 1--40\% for low-mass protostars (\citealt{Jorgensen18,Belloche20,Yang21}). In the well-studied prestellar core L1544, \cite{JimenezSerra16} find an abundance of 2.0--2.5\% for dimethyl ether relative to methanol, 5.9--7.4\% for methyl formate, and 2.0--8.2\% for acetaldehyde for two positions in this core. In addition, \cite{Scibelli20} detect acetaldehyde in 22 prestellar cores, with a ratio of CH$_3$CHO/CH$_3$OH of 2--26\%. As the majority of the probed material in the VeLLO is cold, and it is likely that potential COM emission from a warm gas component close to the VeLLO will not be detected due to the large beam of the observations, an excitation temperature of 10~K was chosen to test which ratio of COM/CH$_3$OH would have led to a detection in the APEX data. The rms of the data is typically 3--5~mK (Table \ref{tbl:obstransitions}), and the same LTE model that was used to constrain the excitation temperature of H$_2$CO and CH$_3$OH was used to derive column densities that would lead to T$_{\rm peak}$ intensities of 3~$\times$~rms for transitions covered in the data that have E$_{\rm up}$ of $\leq$~50~K. In the case of acetaldehyde, a ratio of 30--40\% compared to methanol would be required for a detection in the APEX observations. For acetaldeyde and methyl formate this value is $>$~60\%. Thus, it is very unlikely that they could have been detected in the current observational data.

\section{Conclusions}\label{sect:conclusions}

\noindent This work presents APEX observations of the VeLLO in the isolated DC3272+18 cloud that has undergone an outburst that likely occurred less than 10$^4$~yr ago. The presence of molecules such as CH$_3$OH, H$_2$CO, SO, SO$_2$, and for the first time in a source of this type, NO, is likely tied to the past heating event. Moreover, typical chemical tracers of multiple physical components of the protostellar system are detected. Line intensity ratios and column densities are utilized to constrain the kinetic gas temperature of the system and the excitation temperature of the detected species. The main findings are listed below:
\begin{itemize}
    \item For the first time, NO is detected in a VeLLO. The most likely explanation for its high column density is its formation after species such as H$_2$O and NH$_3$ have been sublimated and subsequently photodissociated after the outburst. If this proposition holds true, it could be used to trace the position of the extended water snowline during the outburst. Due to its high volatility it remains enhanced in the gas phase long after the central object has returned to its quiescent stage. Its potential as a tracer for past outbursts has to be tested with observations with a higher spatial resolution for this source, and for a larger sample of sources as well. This is especially interesting to test for Class I and II objects when planet formation is already an on-going process.
    \item The detection of CH$_3$OH and H$_2$CO with an excitation temperature of $\sim$~10~K suggests that they stem from the cold envelope and have sublimated from the grains during the outburst. The securely detected transitions of SO with E$_{\rm up}$~$<$~44~K most likely originate from the same gas reservoir. The high column density ratio of NO/CH$_3$OH is attributed to different freeze-out timescales of molecules related to their binding energies. Observations with a higher spatial resolution are required to confirm that those molecules are indeed co-spatial.
    \item The low line intensity ratio of $I_{\rm HCN}$/$I_{\rm HNC}$ indicates that the gas kinetic temperature is $\leq$15~K. A low kinetic temperature is also supported by the low line intensity of the $I_{\rm DCN}$/$I_{\rm DCO^+}$ ratio. When compared with APEX observations of a sample of low-mass protostars in Perseus (\citealt{Murillo18}), the ratios of the VeLLO are low compared to the other objects, in line with its low bolometric luminosity and envelope mass compared to the rest of the sample.
    \item The detections of c-C$_3$H$_2$ and C$_2$D are consistent with the presence of cavity walls carved out by the outflow driven by the VeLLO. The population diagram analysis of the c-C$_3$H$_2$ transitions calculates a rotational temperature of 35$\substack{+3 \\ -5}$~K, and thus reveals a second, warmer layer of gas around the VeLLO. However, ratios of the 6-5/3-2 and 5-4/3-2 transitions of c-C$_3$H$_2$ are also lower than what is found in the sample by \cite{Murillo18}, so while c-C$_3$H$_2$ traces a reservoir that is warm compared to the rest of the probed material in the VeLLO, it is still colder compared to the material that is found around low-mass protostars.
    \item The detections of three deuterated species, namely DCO$^+$, DCN, D$_2$CO, and their respective D/H ratios reveal that deuteration has been effective in the past and potentially also in the present. C$_2$D is also detected in the data, but C$_2$H is not covered, so its D/H ratio cannot be calculated. D$_2$CO is a product of grain-surface formation and has sublimated into the gas phase during the outburst. DCO$^+$, DCN, and C$_2$D are products of gas-phase chemistry, their deuterium enhancement could still be on-going in the cold layers of the envelope. 
\end{itemize}

\noindent The study of outbursting objects is crucial to understand to which extent bursts sublimate and reset icy grain mantles and influence the ice and gas-phase chemical inventory of future planet-forming material by sublimating species that would otherwise remain on the icy grains. These kind of bursts may completely alter the balance between inheritance and reconstitution of the volatile reservoirs during the formation of protoplanetary systems. Investigating the chemical composition of VeLLOs allows the study of objects at the low luminosity/low mass end of star formation, which has been understudied so far. The APEX data have revealed that VeLLOs, at least the one in DC3272+18, is rich in molecules. To fully understand their chemical composition it is necessary to obtain observations with a higher spatial resolution with facilities like ALMA. This will allow to spatially resolve emission of molecules, and also to probe for the existence of a hot corino, where COMs are expected to be in the gas phase, if the chemical composition in VeLLOs is in line with what is found around low-mass protostars, just at closer distances to the central star due to its lower luminosity. Complementary observations with the \textit{James Webb} Space Telescope would offer the opportunity to study the level of reprocessing of the ices induced by the past outburst.

\begin{acknowledgements}
      This publication is based on data acquired with the Atacama Pathfinder Experiment (APEX) under programme ID O-0109.F-9305A-2022. APEX is a collaboration between the Max-Planck-Institut fur Radioastronomie, the European Southern Observatory, and the Onsala Space Observatory. We want to thank the APEX staff for support with these observations. B.M.K acknowledges the SNSF Postdoc.Mobility stipend P500PT\_214459. B.M.K and M.N.D acknowledge the Swiss National Science Foundation (SNSF) Ambizione grant no. 180079. M.N.D. acknowledges the Holcim Foundation Stipend. S.F.W. acknowledges the financial support of the SNSF Eccellenza Professorial Fellowship (PCEFP2\_181150). N.F.W.L. and K.A.K. acknowledge support from the Swiss National Science Foundation (SNSF) Ambizione grant 193453. T.-H.H. acknowledges the support by the Max Planck Society. P.B. acknowledges the support of the Swedish Research Council (VR) through contract 2017-0492. M.K.M. acknowledges financial support from the Dutch Research Council (NWO; grant VI.Veni.192.241).
\end{acknowledgements}

\bibliographystyle{aa.bst} 
\bibliography{Manuscript_revised2.bib}

\begin{appendix}

\section{Spectra and parameters of the detected lines}

\begin{table*}[!h]
\caption{Observed molecular transitions in this work.}
\label{tbl:obstransitions}
\centering
\begin{tabular}{l|c|c|c|c|c|c|c|c}
\hline \hline 
Molecule & Transition & Frequency & E$_{\rm up}$ & A$_{\rm ij}$ & T$_{\rm mb, peak}$ $^{\rm (b)}$ & $\Delta$v$^{\rm (b)}$ & v$_{\rm LSR}$ $^{\rm (b)}$ & T$_{\rm rms}$ \\
& & (GHz) & (K) & (s$^{-1}$) & (K) & (km~s$^{-1}$) & (km~s$^{-1}$) & (mK) \\
\hline 
CO & 2-1 & 230.53800 & 16.60 & 6.91~$\times$~10$^{-7}$ & 4.656 & 2.14~$\pm~$~0.03& -0.30~$\pm$~0.01 & 65.9 \\
$^{\rm 13}$CO$^{\rm (a)}$ & 2-1 & 220.39863 & 15.87 & 6.08~$\times$~10$^{-7}$ & 4.089 & 0.75~$\pm$~0.01 & -0.16~$\pm$~0.01 & 8.3 \\ 
& & & & & 1.529 & 0.53~$\pm$~0.02 & 0.38~$\pm$~0.01 & \\
C$^{\rm 18}$O & 2-1 &  219.56035 & 15.81 & 6.01~$\times$~10$^{-7}$ & 2.227 & 0.54 $\pm$ 0.00 & -0.01 $\pm$ 0.00 &  3.6 \\
N$_2$D$^+$ $^{\rm (c)}$ & 3-2 & 231.32186  & 22.20 & 7.14~$\times$~10$^{-4}$ & 0.142 & 0.42 $\pm$ 0.02 & -0.24 $\pm$ 0.01 & 4.6  \\
 & 3-2 & 231.31996 & 22.20 & 6.65~$\times$~10$^{-5}$ & 0.016 & 0.31 $\pm$ 0.06 & 2.29 $\pm$ 0.03$^{\rm (d)}$ & 4.6 \\ 
 & 3-2 & 231.32145 & 22.20 & 6.00~$\times$~10$^{-4}$ & 0.034 & 0.49 $\pm$ 0.08 & 0.28 $\pm$ 0.03$^{\rm (d)}$ & 4.6 \\
 & 3-2 & 231.32444 & 22.20 & 9.87~$\times$~10$^{-5}$ & 0.009 & 0.28 $\pm$ 0.11 & -3.48 $\pm$ 0.05$^{\rm (d)}$ & 4.6 \\
HCO$^+$ & 3-2 & 267.55753 &  25.68 & 1.45~$\times$~10$^{-3}$ & 1.480 & 0.65 $\pm$ 0.00 & -0.32 $\pm$ 0.00 & 6.2 \\
DCO$^+$ & 3-2 & 216.11258 & 20.74 & 7.66~$\times$~10$^{-4}$ & 0.559 & 0.43 $\pm$ 0.00 & 0.0 $\pm$ 0.0 & 2.7 \\
HCN$^{\rm (c)}$ & 3-2 & 265.88618 & 25.52 & 8.36~$\times$~10$^{-4}$ & 0.259  & 0.67 $\pm$ 0.02 & -0.50 $\pm$ 0.01 & 5.4 \\
 & 3-2 & 265.88852 & 25.52 & 3.10~$\times$~10$^{-5}$ & 0.040 & 0.41 $\pm$ 0.08 & -2.83 $\pm$ 0.03 &  5.4 \\
 & 3-2 & 265.88489 & 25.52 & 3.09~$\times$~10$^{-5}$ & 0.138 & 0.41 $\pm$ 0.02 &  1.29~$\pm$ 0.01 & 7.6 \\
DCN$^{\rm (b)}$ & 3-2 & 217.23863 & 20.85 & 4.58~$\times$~10$^{-4}$ & 0.033 & 0.40 $\pm$ 0.06 & 0.08 $\pm$ 0.04 &  3.2 \\
& 3-2 & 217.23823 & 20.85 & 9.15~$\times$~10$^{-5}$ & 0.010 & 0.49 $\pm$ 0.28 & 0.53 $\pm$ 0.16 & 3.2 \\
HNC & 3-2 & 271.98114 & 26.61 & 9.34~$\times$~10$^{-4}$ & 0.970 & 0.57 $\pm$ 0.01 & -0.13 $\pm$ 0.00 &  5.9 \\
H$_2$CO & 3$_{\rm 0,3}$ - 2$_{\rm 0,2}$  & 218.22219 & 20.96 & 2.82~$\times$~10$^{-4}$ & 0.373 & 0.46 $\pm$ 0.00 & 0.03 $\pm$ 0.00 & 3.2 \\
\textbf{*}CH$_3$OH & 4$_{2,3}$ - 3$_{1,2}$& 218.44006 & 45.46 & 4.69~$\times$~10$^{-5}$ & 0.013 & 0.26 $\pm$ 0.06 & 0.1 $\pm$ 0.03 &  3.8 \\
CH$_3$OH & 2$_{\rm 0,2}$ - 1$_{\rm 1,1}$  & 254.01538  & 20.09 & 1.90~$\times$~10$^{-5}$ &  0.081 & 0.38 $\pm$ 0.02 & 0.01 $\pm$ 0.01 & 6.5  \\
NO & 5/2-3/2 & 250.43685 &  19.23 & 1.84~$\times$~10$^{-6}$  & 0.089  & 0.38 $\pm$ 0.02 & -0.00 $\pm$ 0.01 & 5.3 \\
NO & 5/2-3/2 & 250.44066 &  19.23 & 1.55~$\times$~10$^{-6}$ & 0.055 & 0.41 $\pm$ 0.08 & -0.02 $\pm$ 0.02 & 5.1 \\
NO & 5/2-3/2 &  250.44853 & 19.23 & 1.38~$\times$~10$^{-6}$ & 0.034 & 0.38 $\pm$ 0.06 & -0.01 $\pm$ 0.03 & 5.1 \\
\textbf{*}NO & 5/2-3/2 & 250.47541 & 19.23 & 4.42~$\times$~10$^{-7}$ & 0.016 & 0.37 $\pm$ 0.09 & -0.01 $\pm$ 0.04 & 5.0 \\
NO & 5/2-3/2 & 250.79644 &  19.28 & 1.85~$\times$~10$^{-6}$ & 0.094 & 0.41 $\pm$ 0.01 & -0.06 $\pm$ 0.01 &  5.3 \\
NO & 5/2-3/2 & 250.81559 &  19.28 & 1.55~$\times$~10$^{-6}$ & 0.055 & 0.42 $\pm$ 0.03 & -0.49 $\pm$ 0.01 &  5.1 \\
NO & 5/2-3/2 &  250.81695 & 19.27 & 1.39~$\times$~10$^{-6}$  & 0.032 & 0.45 $\pm$ 0.06 &  0.03 $\pm$ 0.00 & 5.1 \\
c-C$_3$H$_2$ & 3$_{\rm 3,0}$ - 2$_{\rm 2,1}$ & 216.27876 & 19.47 & 2.56~$\times$~10$^{-4}$ & 0.111 & 0.45 $\pm$ 0.01 & 0.00 $\pm$ 0.01 & 3.4 \\
c-C$_3$H$_2$ & 6$_{\rm 0,6}$ - 5$_{\rm 1,5}$  & 217.82215 & 38.61 & 5.40~$\times$~10$^{-4}$ & 0.034 & 0.53 $\pm$ 0.04 & 0.00 $\pm$ 0.02 & 4.2  \\
c-C$_3$H$_2$  & 5$_{\rm 1,4}$ - 4$_{\rm 2,3}$ & 217.94005 &  35.41 & 4.03~$\times$~10$^{-4}$ & 0.017 & 0.51 $\pm$ 0.07 & 0.10 $\pm$ 0.03 & 3.6 \\
\textbf{*}c-C$_3$H$_2$ & 5$_{2,3}$ - 4$_{3,2}$ & 249.05437 & 41.02 & 4.16~$\times$~10$^{-4}$ & 0.012 & 0.45 $\pm$ 0.16 & -0.02 $\pm$ 0.05 & 5.2 \\
c-C$_3$H$_2$ & 6$_{\rm 2,5}$ - 5$_{\rm 1,4}$ & 251.52731  & 47.49 & 6.75~$\times$~10$^{-4}$ &  0.015 & 0.37 $\pm$ 0.08 & 0.07 $\pm$ 0.05 & 4.5 \\
c-C$_3$H$_2$ & 4$_{\rm 4,1}$ - 3$_{\rm 3,0}$ & 265.75948  & 32.22 & 7.27~$\times$~10$^{-4}$ & 0.022 & 0.45 $\pm$ 0.06 & -0.10 $\pm$ 0.03 & 4.8 \\
C$_2$D & 3$_{\rm 4,5}$ - 2$_{\rm 3,4}$ & 216.37284 &  20.77 & 2.99~$\times$~10$^{-5}$ & 0.014 & 0.27 $\pm$ 0.06 & 0.07 $\pm$ 0.04 & 2.8 \\
C$_2$D & 3$_{\rm 4,3}$ - 2$_{\rm 3,2}$ & 216.37331 &  20.77 & 2.67~$\times$~10$^{-5}$ & 0.013 & 0.27 $\pm$ 0.07 & 0.04 $\pm$ 0.03 & 2.8 \\
C$_2$D & 3$_{\rm 4,4}$ - 2$_{\rm 3,3}$& 216.37332 &  20.77 & 2.76~$\times$~10$^{-5}$ & 0.009 & 0.28 $\pm$ 0.06 & -0.04 $\pm$ 0.02 & 2.8 \\
C$_2$D & 3$_{\rm 3,4}$ - 2$_{\rm 2,3}$ & 216.42825 &  20.77 & 2.77~$\times$~10$^{-5}$ & 0.010 & 0.38 $\pm$ 0.08 & -0.02 $\pm$ 0.04 & 2.8 \\
C$_2$D & 3$_{\rm 3,3}$ - 2$_{\rm 2,2}$  & 216.42843 &  20.77 & 2.33~$\times$~10$^{-5}$ & 0.010 & 0.38 $\pm$ 0.08 & 0.22 $\pm$ 0.03 &  2.8\\
C$_2$D & 3$_{\rm 3,2}$ - 2$_{\rm 2,1}$ & 216.42888 & 20.77 & 2.09~$\times$~10$^{-5}$ & 0.010 & 0.38 $\pm$ 0.09 & 0.75 $\pm$ 0.04 & 2.8 \\
D$_2$CO & 4$_{\rm 0,4}$ - 3$_{\rm 0,3}$ & 231.41023 & 27.88 & 3.47~$\times$~10$^{-4}$  &  0.012 & 0.36 $\pm$ 0.10 & -0.24 $\pm$ 0.04 & 3.4 \\
SO$_2$ & 4$_{\rm 2,2}$ - 3$_{\rm 1,3}$ & 235.15172 & 19.03 & 7.69~$\times$~10$^{-5}$ & 0.012 & 0.64 $\pm$ 0.13 & -0.26 $\pm$ 0.05 & 3.5 \\
\textbf{*}SO$_2$ & 16$_{1,15}$ - 15$_{2,14}$ & 236.21669 & 130.67 & 7.50~$\times$~10$^{-5}$ & 0.011 & 0.27 $\pm$ 0.13 & 0.04$\pm$0.04 & 3.0 \\
SO & 5,5 - 4,4 & 215.22065 &  44.10 & 1.19~$\times$~10$^{-4}$ & 0.018 & 0.58 $\pm$ 0.05 & 0.06 $\pm$ 0.02 &  2.6 \\
SO & 5,6 - 4,5 &  219.94944 & 34.98 & 1.34~$\times$~10$^{-4}$ &  0.119 & 0.46 $\pm$ 0.01 & 0.01 $\pm$ 0.00 &  3.2 \\
\textbf{*}SO &  6,5 - 5,4 & 251.82577 & 50.66 & 1.92~$\times$~10$^{-4}$ & 0.017  & 0.31 $\pm$ 0.11 & -0.07 $\pm$ 0.04 &  5.6 \\
\hline 
\end{tabular}
\tablefoot{Tentative detections are marked with an asterisk in front of the molecule name. The spectra and their fits are plotted in Figs. \ref{fig:secure-transitions} and \ref{fig:tentative-transitions}. $^{\rm (a)}$ $^{13}$CO was fit with two Gaussians $^{\rm (b)}$ The peak temperature, T$_{\rm mb, peak}$, the line width of the spectral lines, $\Delta$v, and the v$_{\rm LSR}$, and the T$_{\rm rms}$ of the data of the spectral lines were determined with a Gaussian fit with the \textsc{class} package of the \textsc{gildas} software and the \textsc{curve\_fit} module of \textsc{scipy} (\citealt{scipy}); $^{\rm (c)}$ hyperfine components were detected; $^{\rm (d)}$ the position of the hyperfine transitions is given relative to the strongest transition (Fig. \ref{fig:secure-transitions}).}
\end{table*}

The spectra of all securely and tentatively detected lines (Table \ref{tbl:obstransitions}) are shown in Figs. \ref{fig:secure-transitions} and \ref{fig:tentative-transitions} and are overlaid with Gaussian fits in red. All fits were obtained in \textsc{class} and with the \textsc{curve\_fit} module in \textsc{scipy} (\citealt{scipy}, the rms displayed in the plots are taken from Table \ref{tbl:obstransitions}. In Fig. \ref{fig:secure-transitions}, velocity binning over two channels was applied to the D$_2$CO line, and the SO$_2$ line at 235.151~GHz, and for the NO transition at 250.475~GHz and the SO$_2$ transition at 236.217~GHz in Fig. \ref{fig:tentative-transitions}.

\begin{figure*}[!h]
    \centering
   \includegraphics[width=0.33\textwidth]{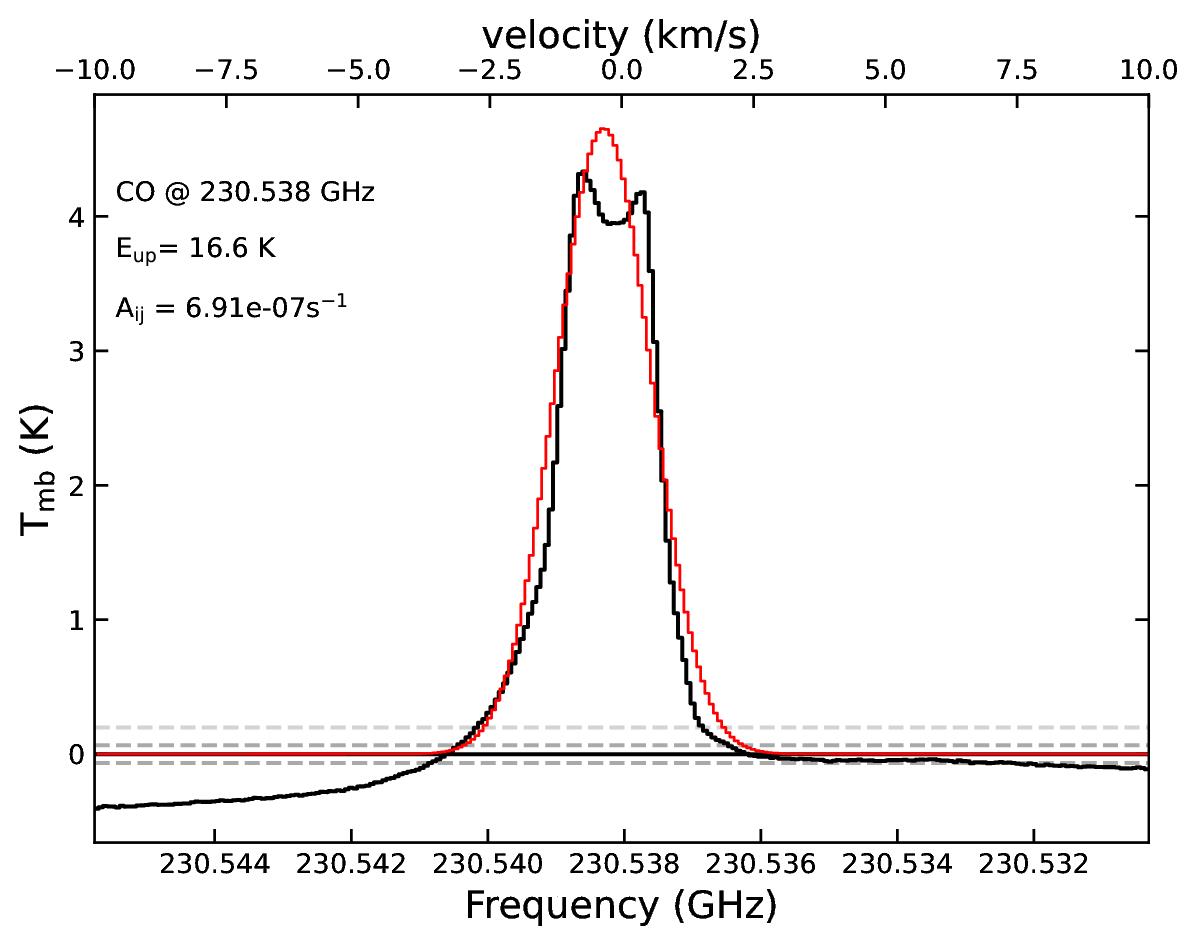}\includegraphics[width=0.33\textwidth]{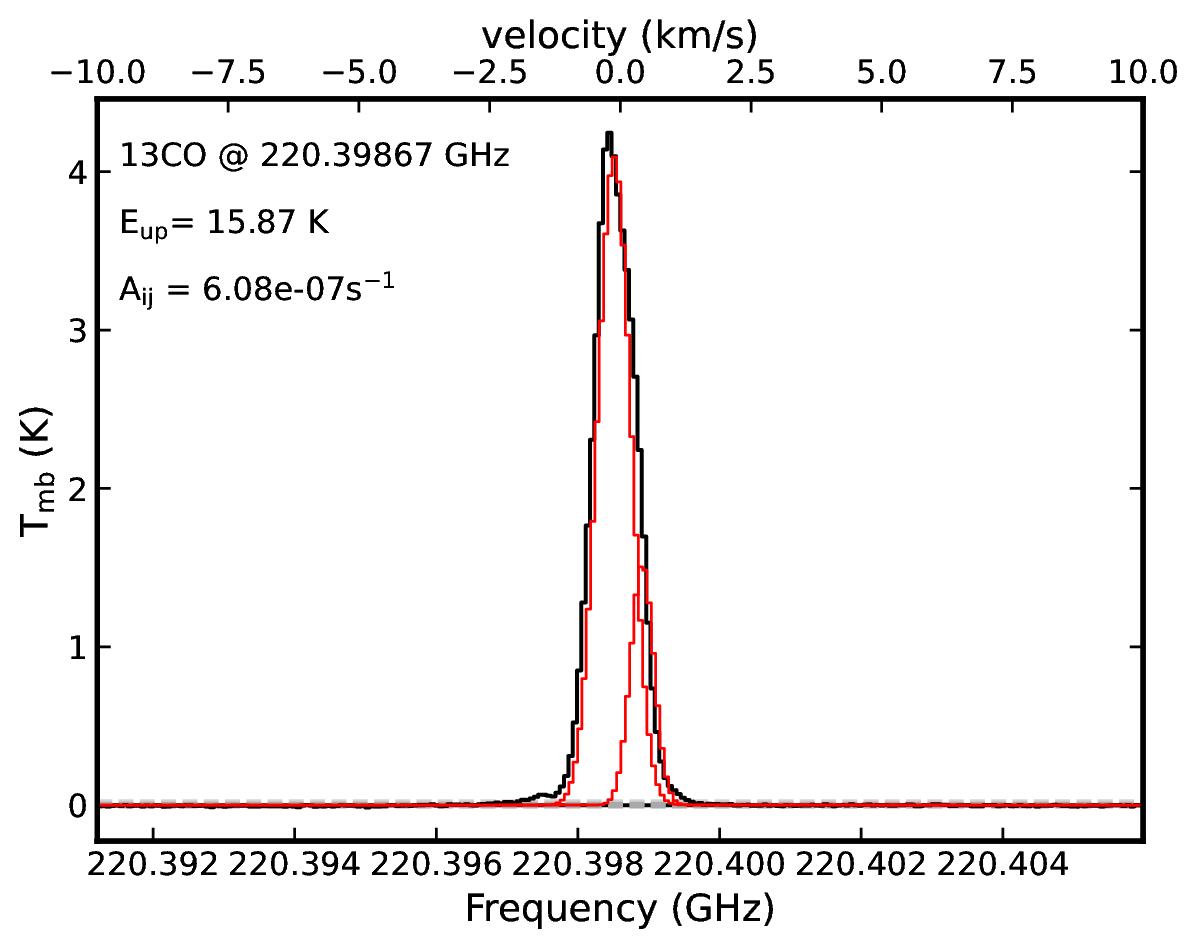}\includegraphics[width=0.33\textwidth]{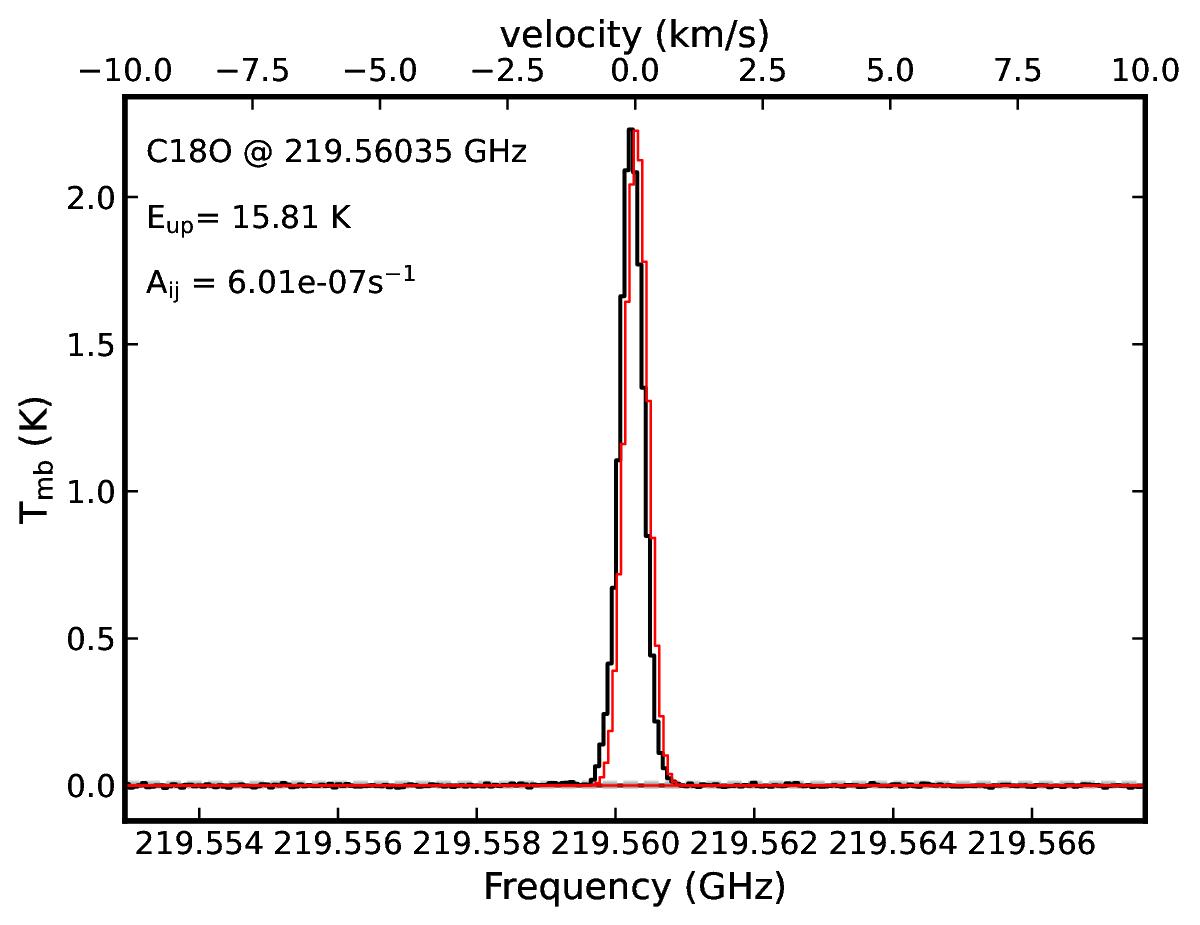}
   \includegraphics[width=0.33\textwidth]{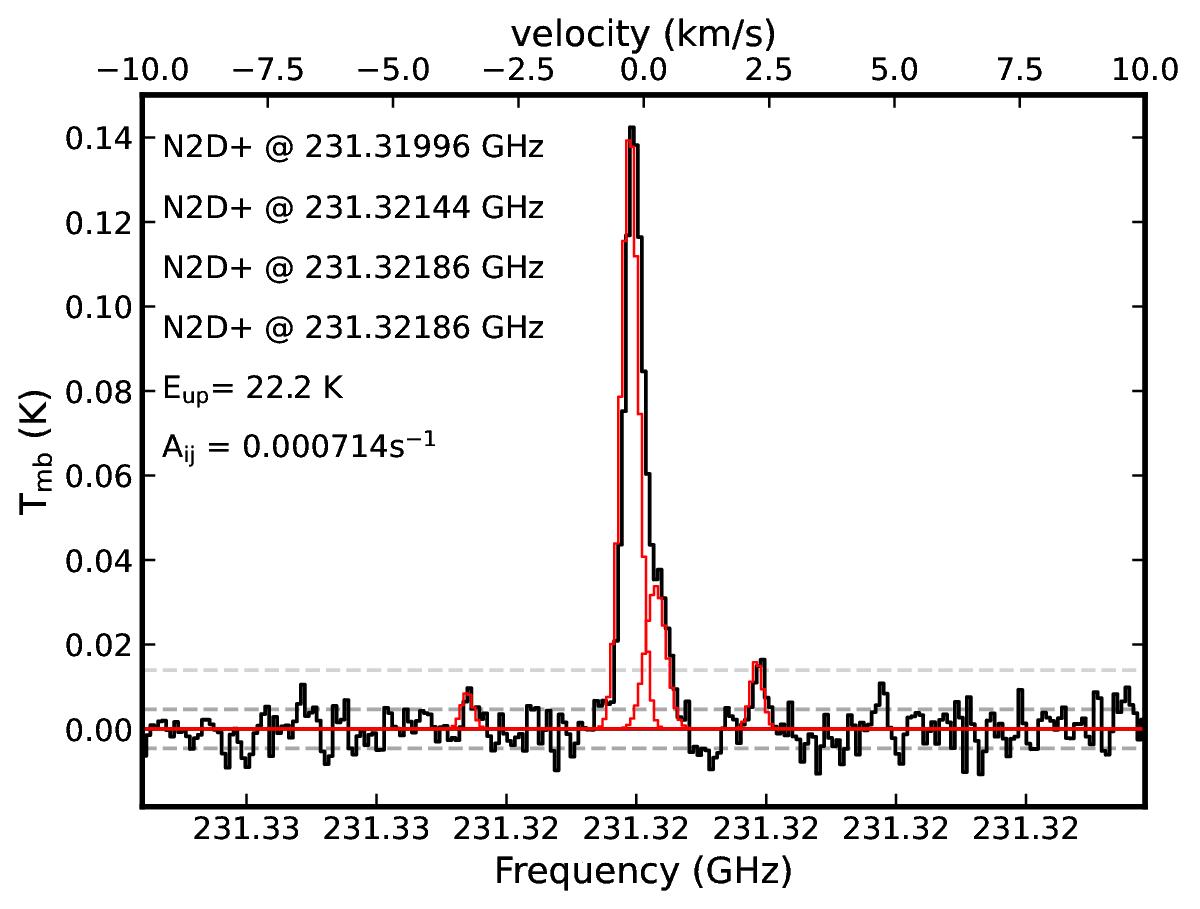}\includegraphics[width=0.33\textwidth]{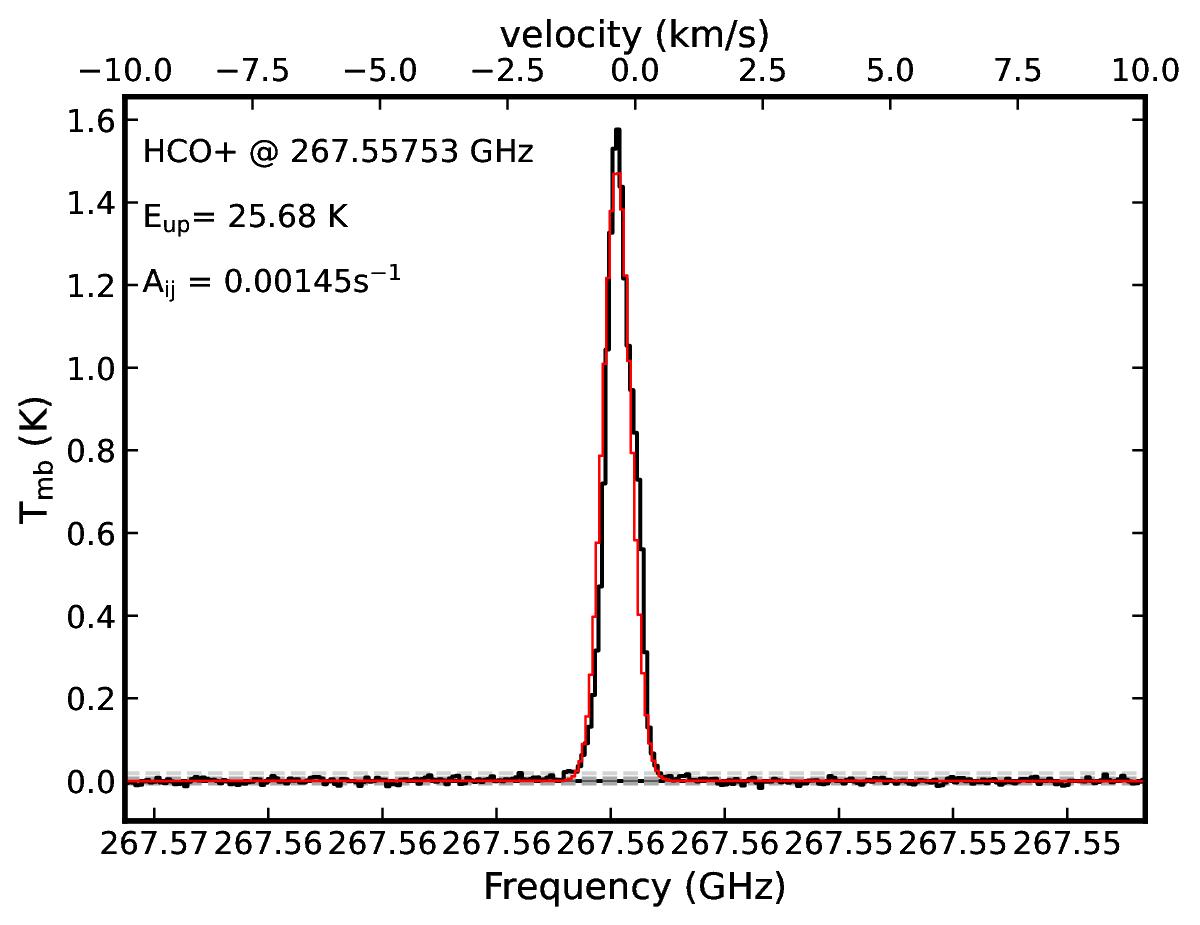}\includegraphics[width=0.33\textwidth]{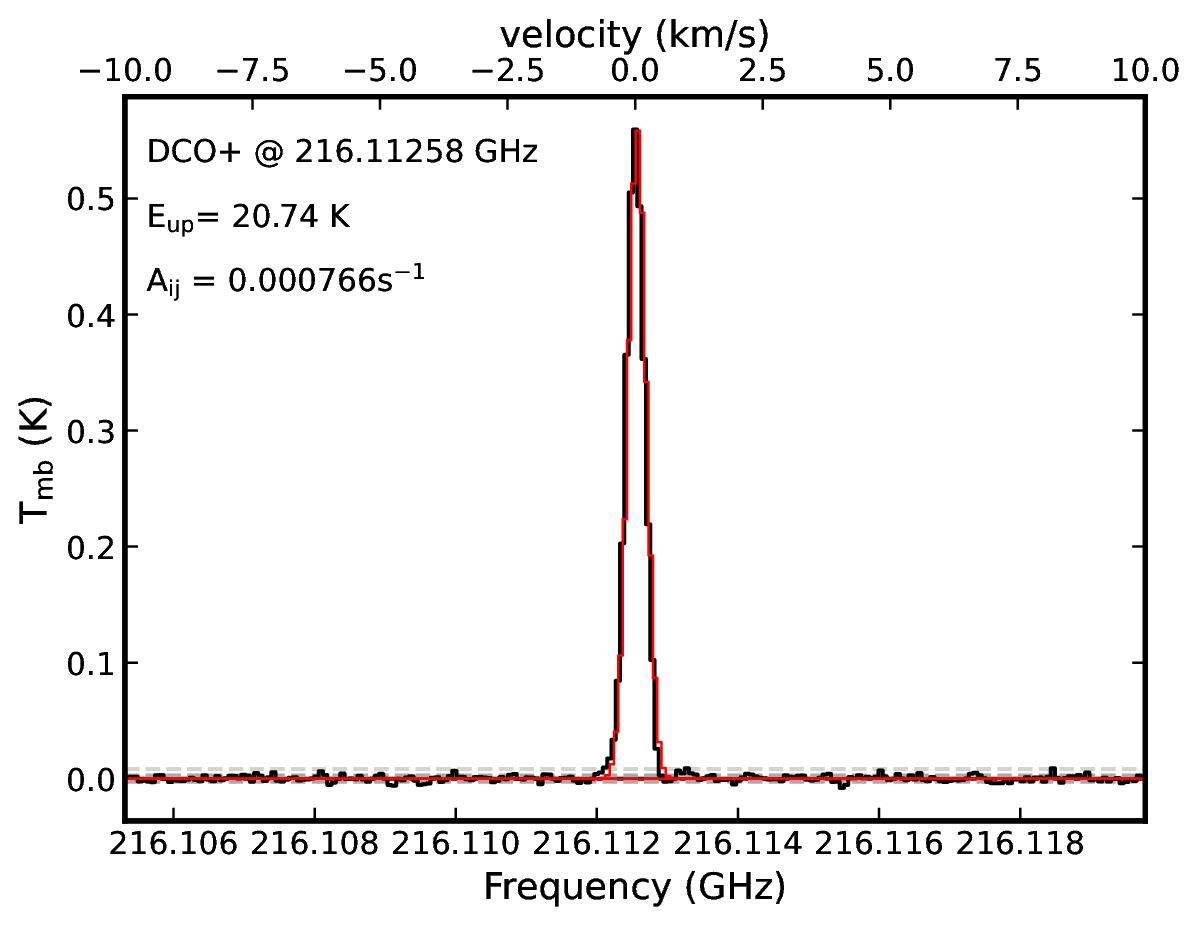} \includegraphics[width=0.33\textwidth]{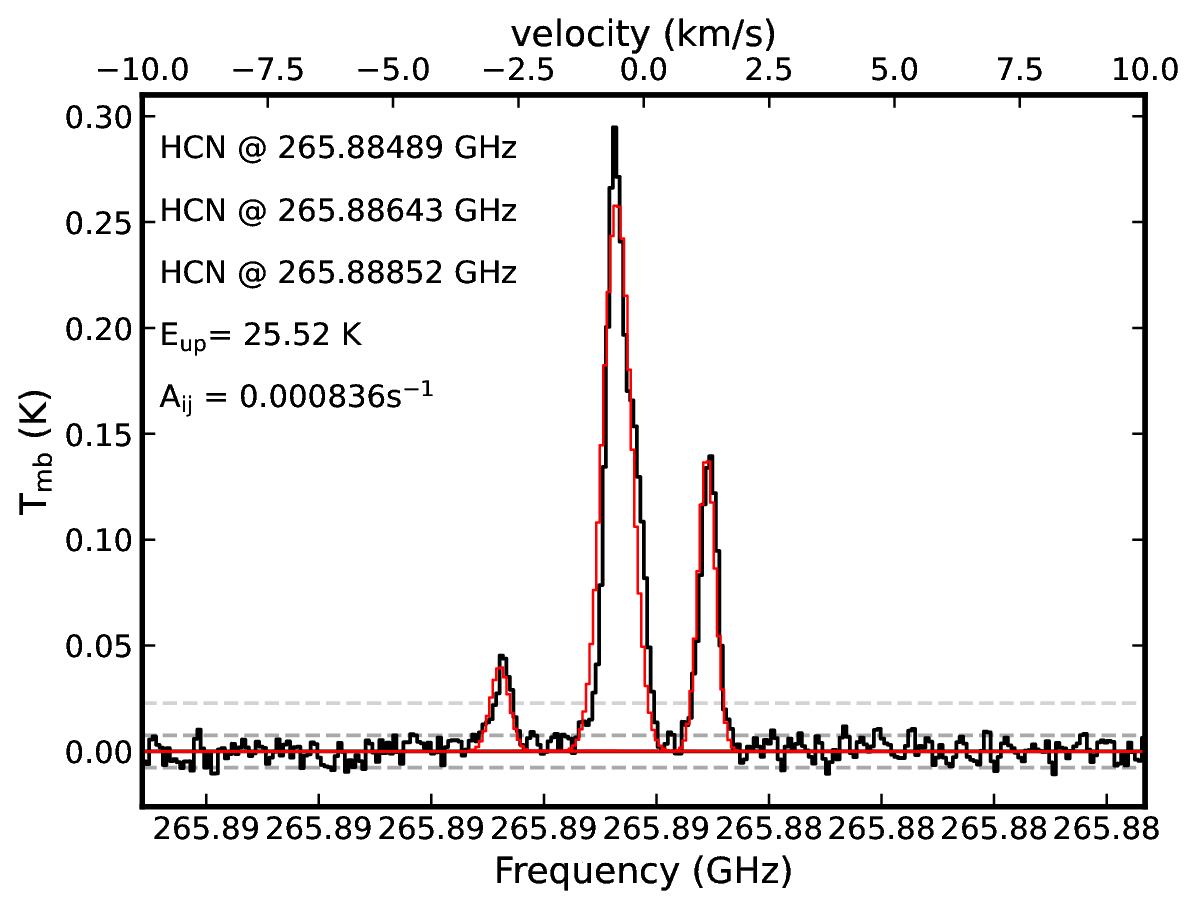}\includegraphics[width=0.33\textwidth]{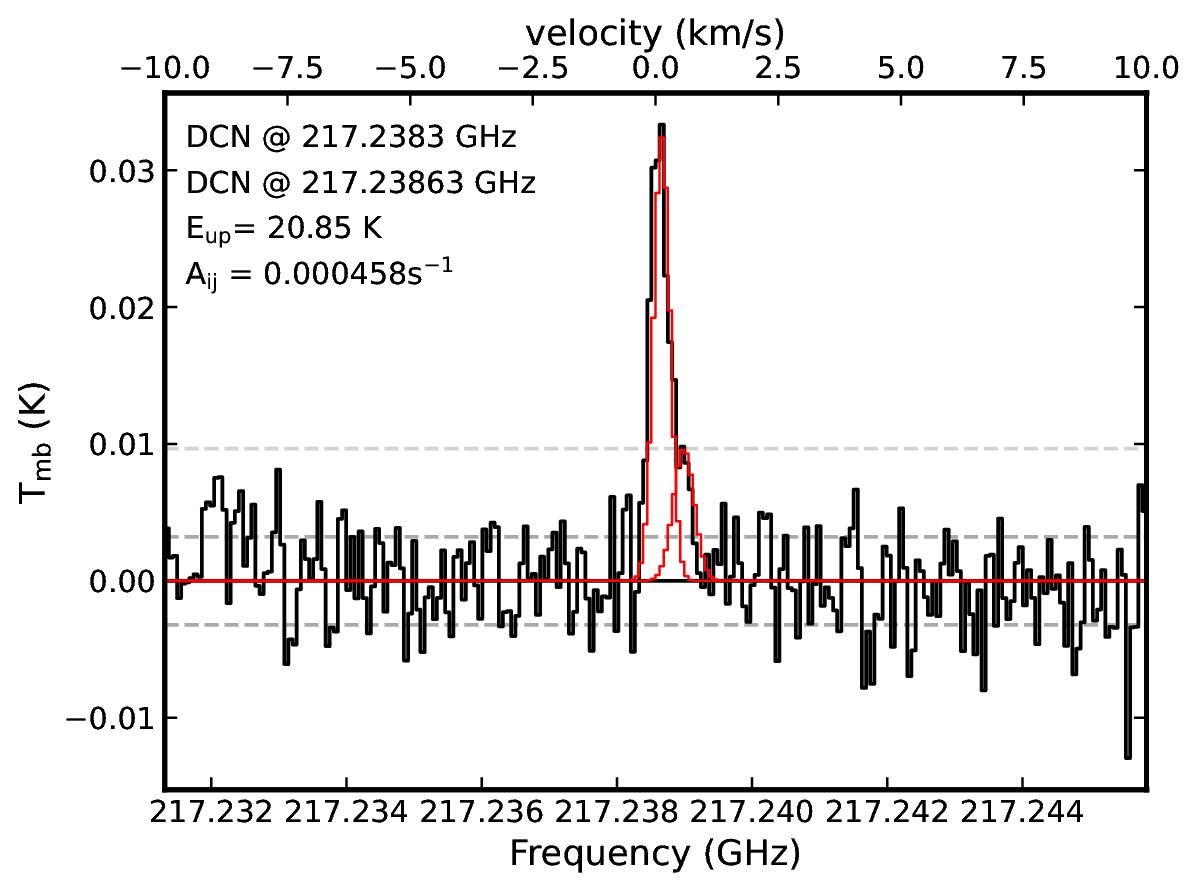}\includegraphics[width=0.33\textwidth]{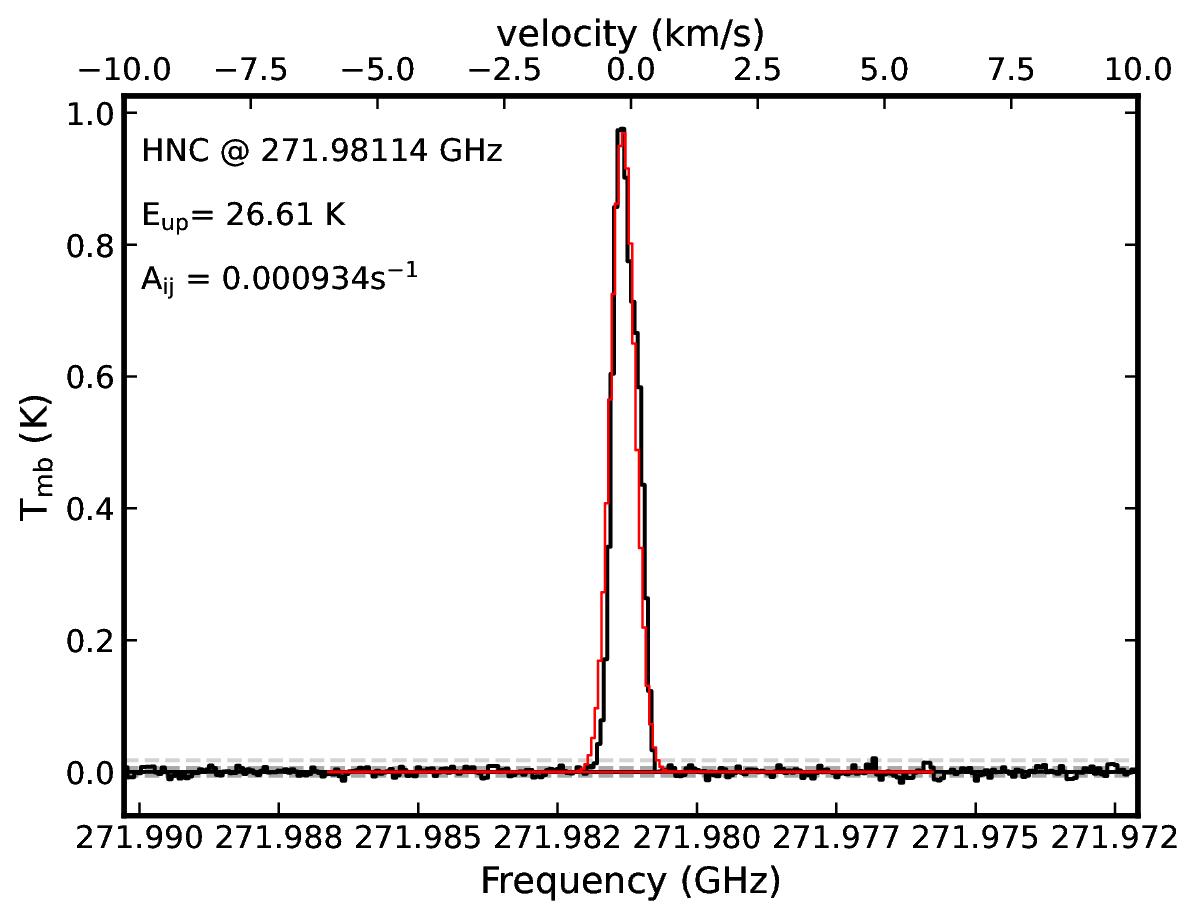} \includegraphics[width=0.33\textwidth]{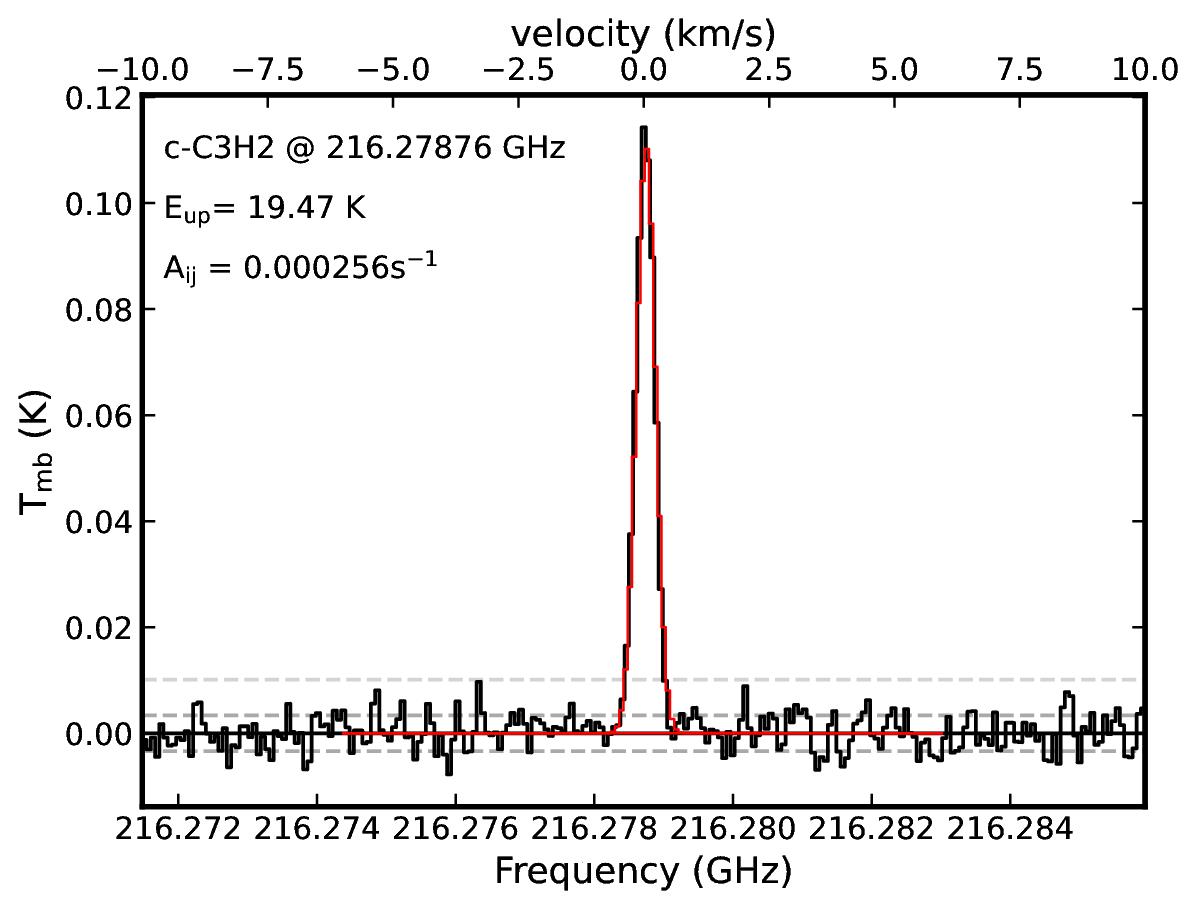}\includegraphics[width=0.33\textwidth]{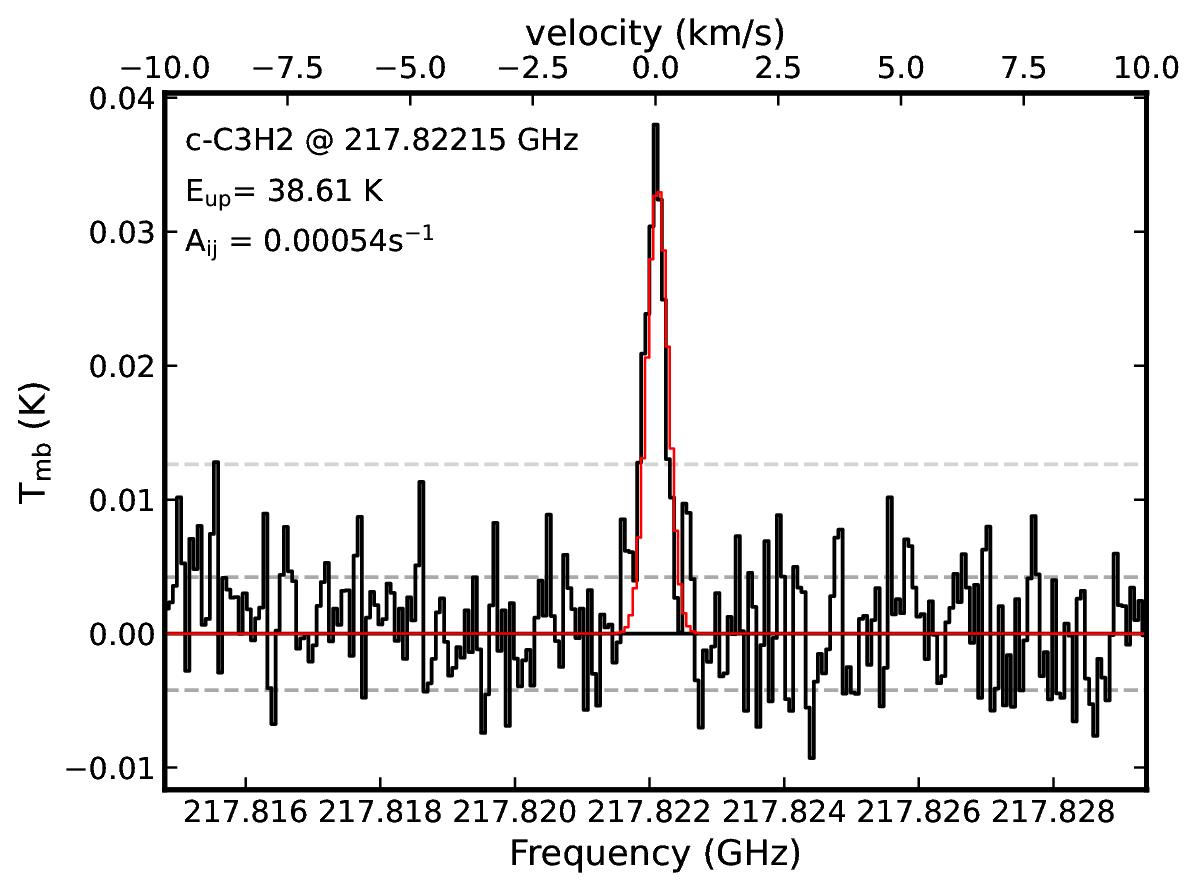}    \includegraphics[width=0.33\textwidth]{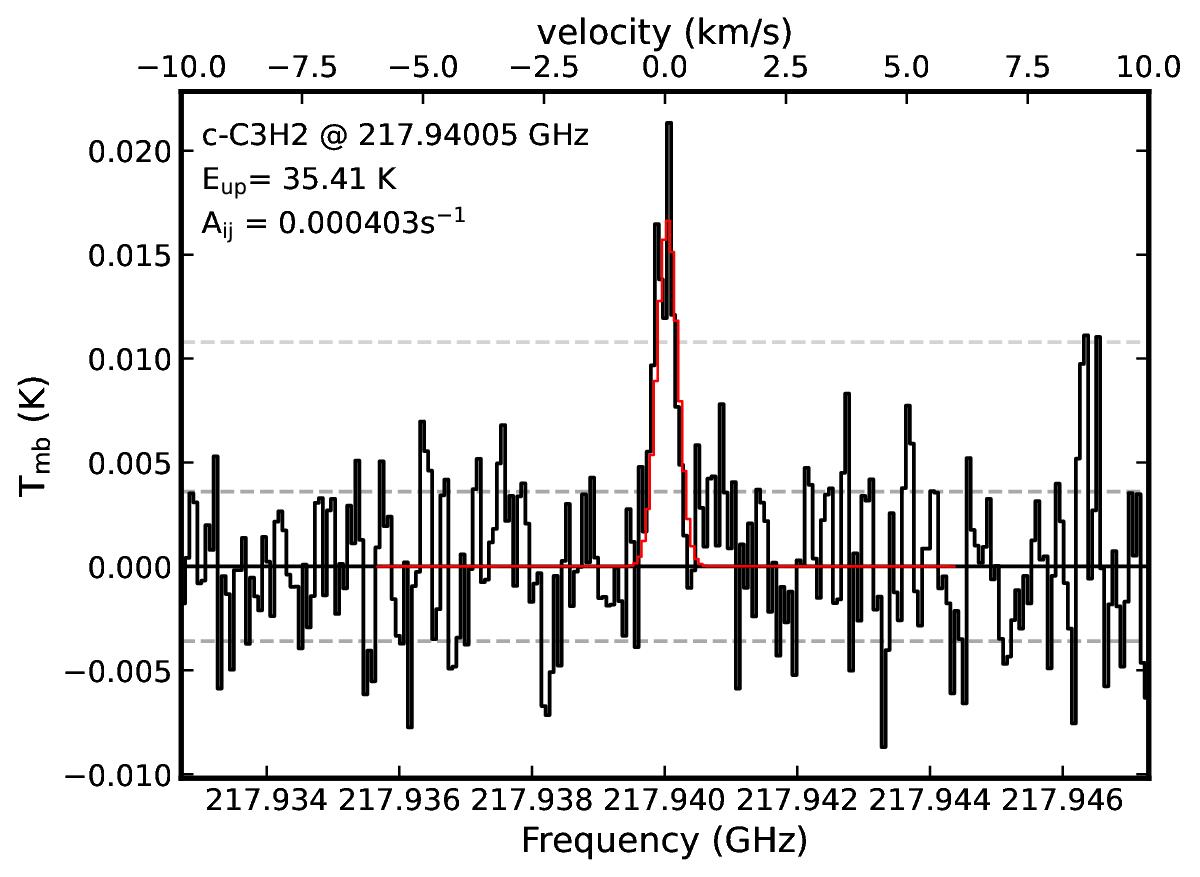}   
    \caption{Spectra of the securely detected transitions (Table \ref{tbl:obstransitions}) overlaid with the Gaussian best-fit model in red (Table \ref{tbl:obstransitions}). Not all lines are perfectly fit by a Gaussian, kinematics or line broadening are factors that can influence the line shape, in the case of CO and $^{13}$CO two components are used to fit the line. It is also to note that due to the large beam size of 26.2$^{\prime \prime}$ different physical components are not resolved, and emission is picked up from multiple components. The dashed, dark gray line indicates the 3$\times$rms value, the dashed, light gray line indicates the rms value.}
\end{figure*}
\begin{figure*}\ContinuedFloat
    \includegraphics[width=0.33\textwidth]{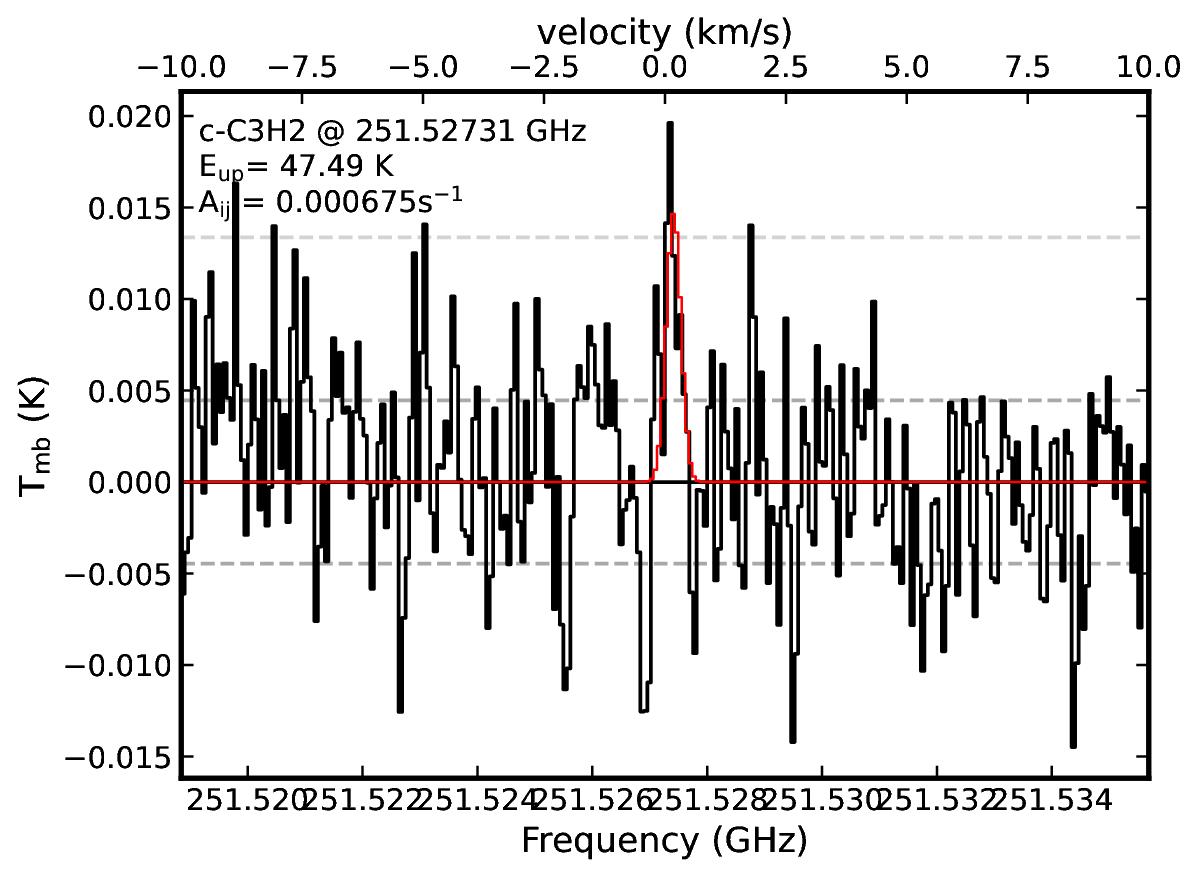}\includegraphics[width=0.33\textwidth]{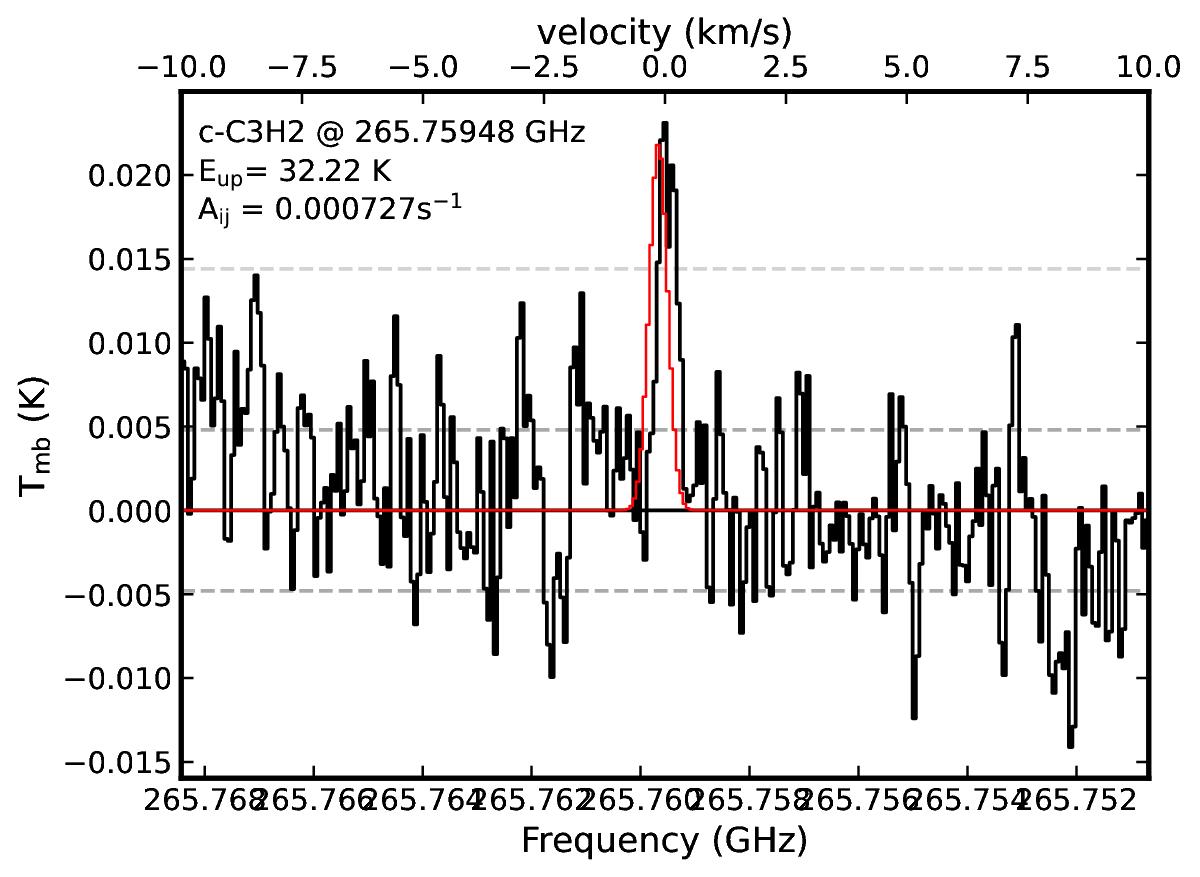}\includegraphics[width=0.33\textwidth]{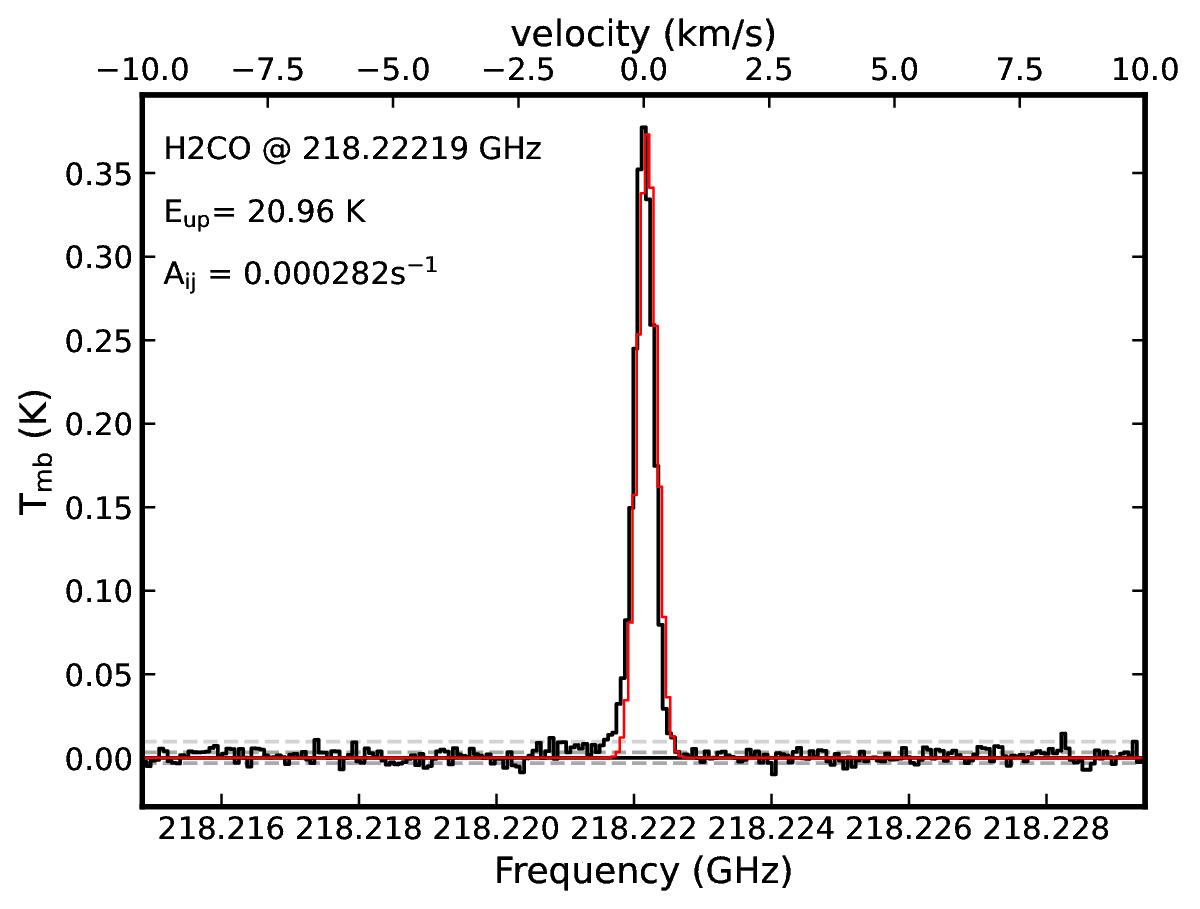} \includegraphics[width=0.33\textwidth]{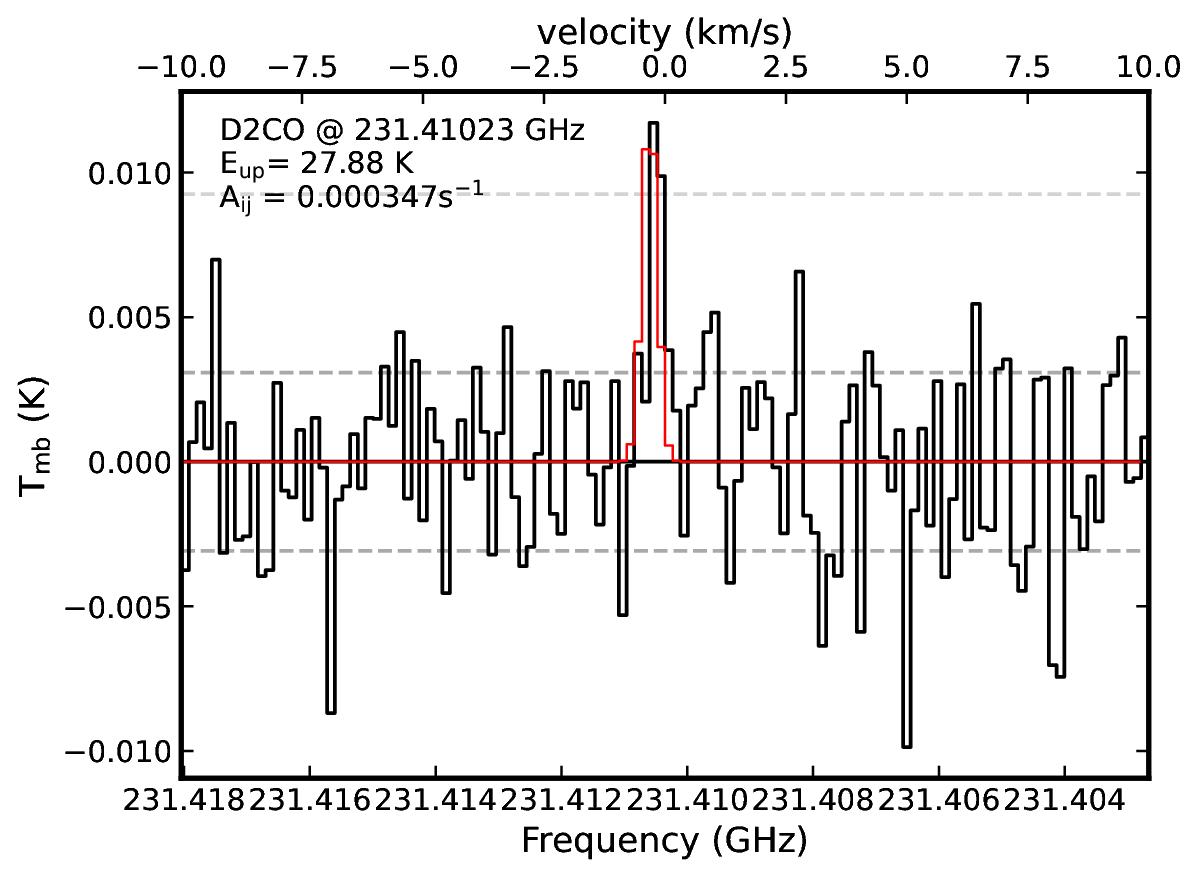}\includegraphics[width=0.33\textwidth]{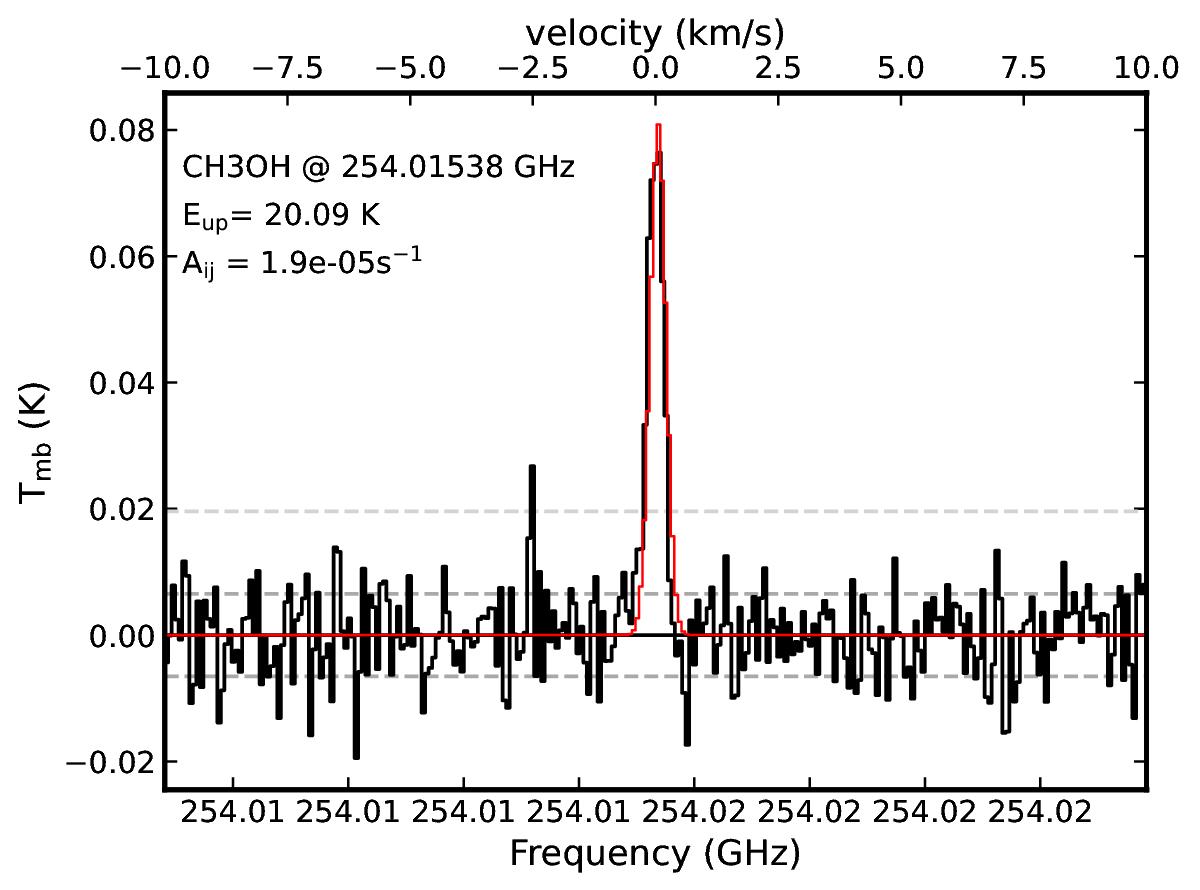}\includegraphics[width=0.33\textwidth]{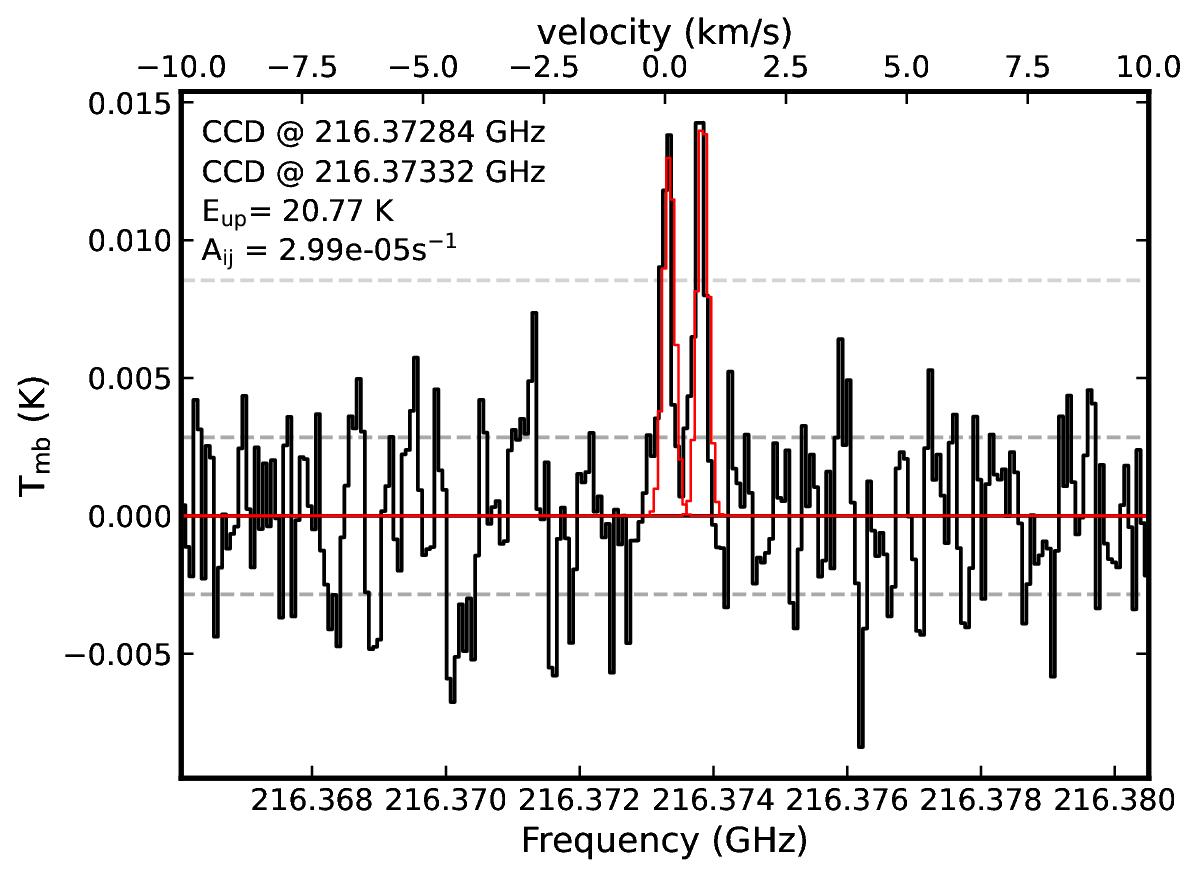} \includegraphics[width=0.33\textwidth]{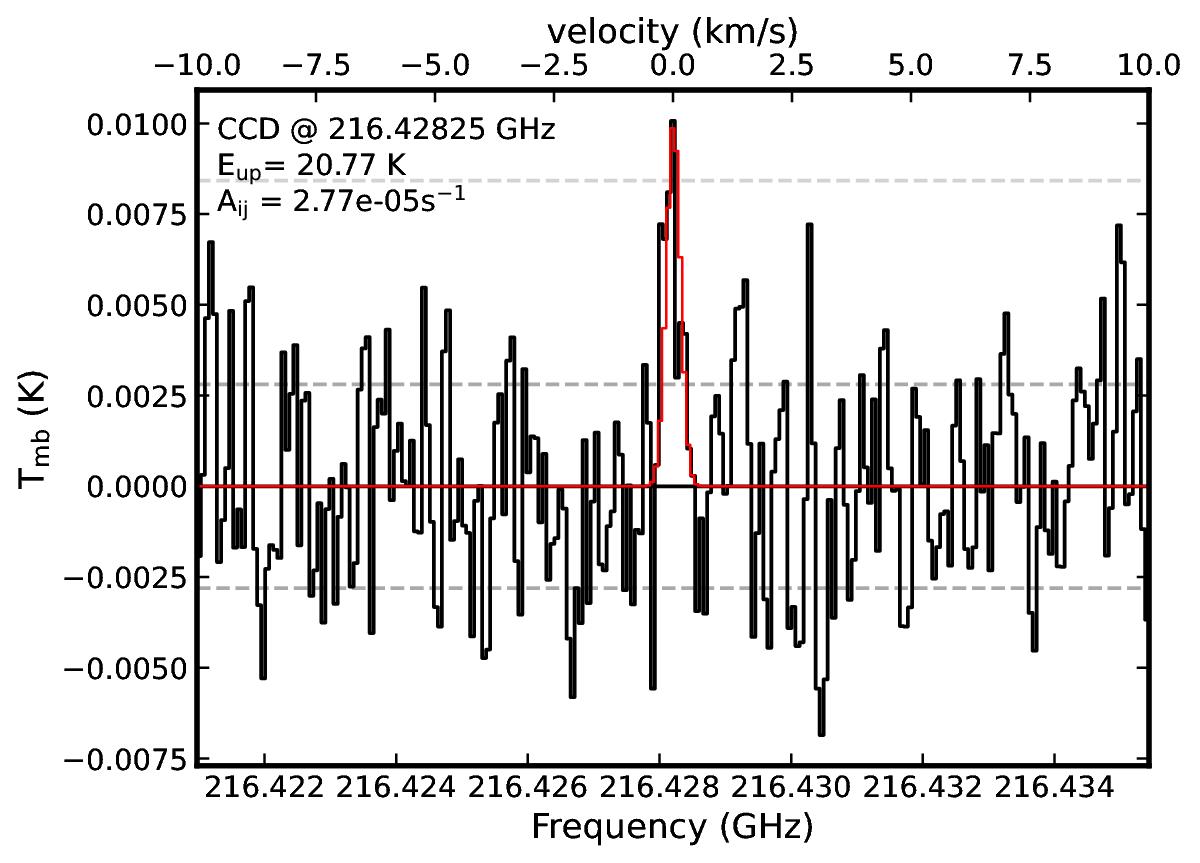}\includegraphics[width=0.33\textwidth]{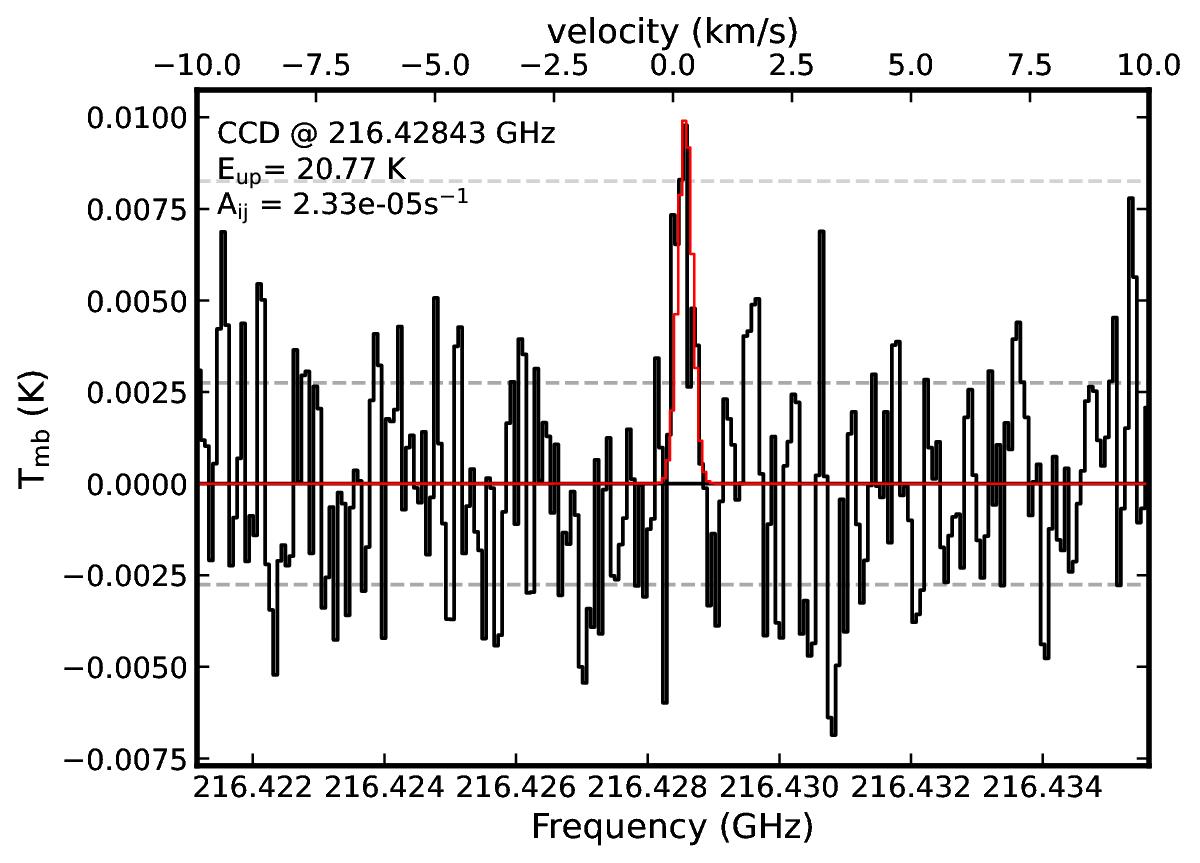}\includegraphics[width=0.33\textwidth]{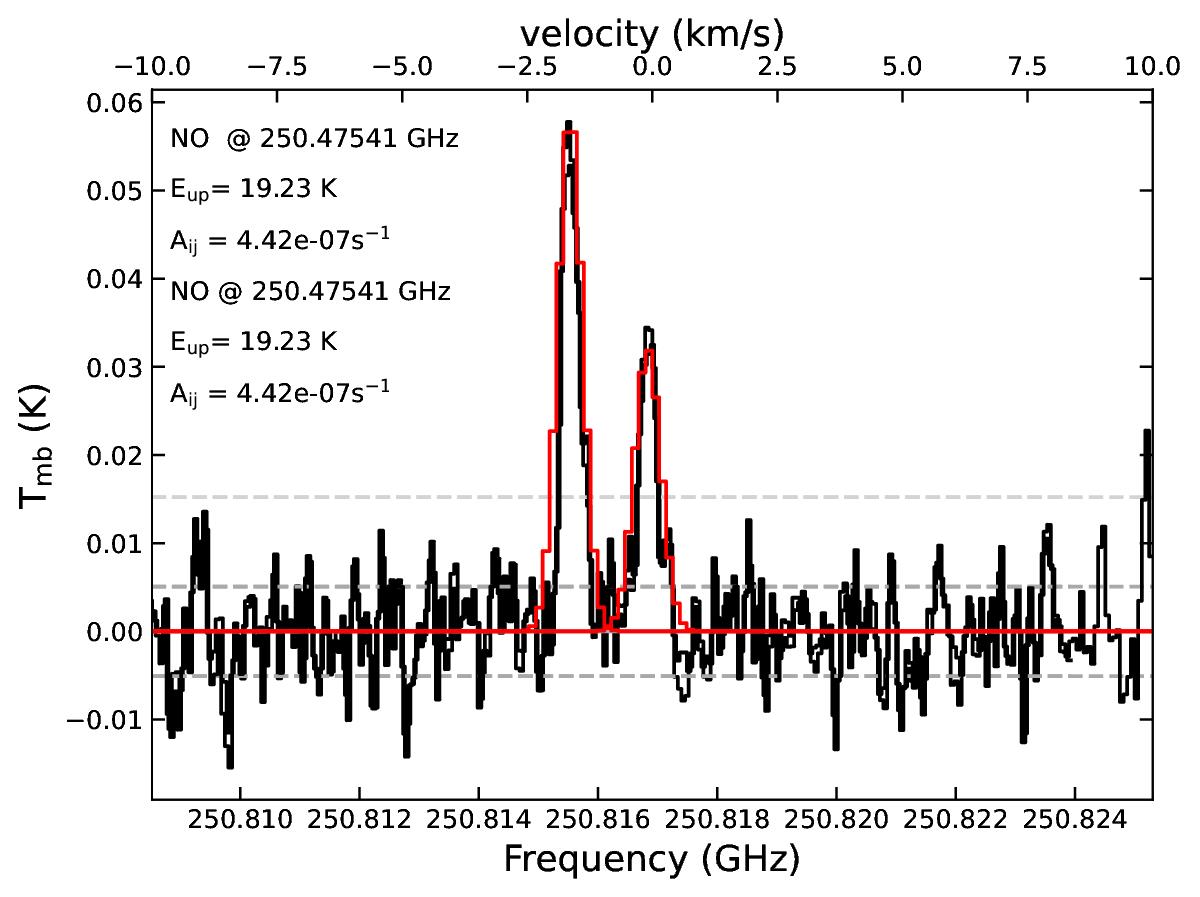}
    \includegraphics[width=0.33\textwidth]{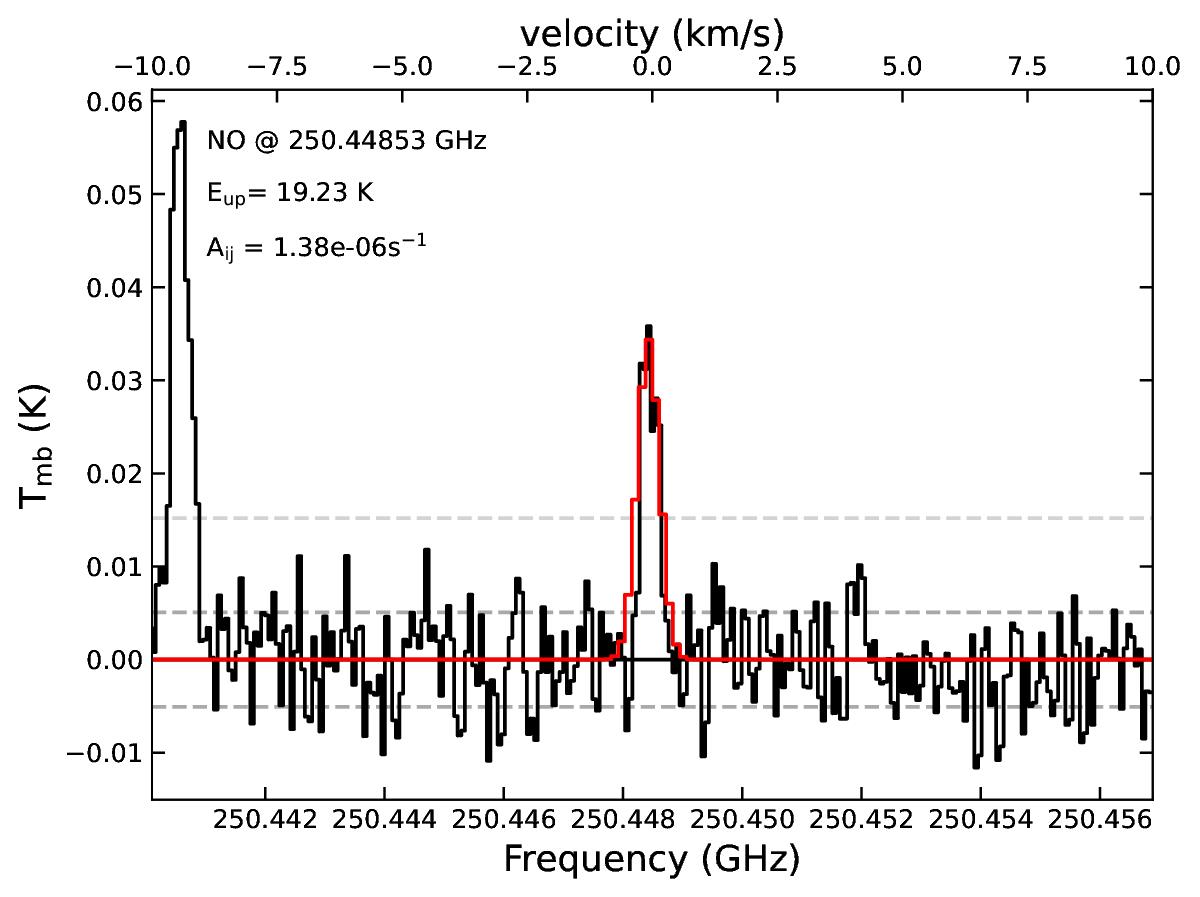}    \includegraphics[width=0.33\textwidth]{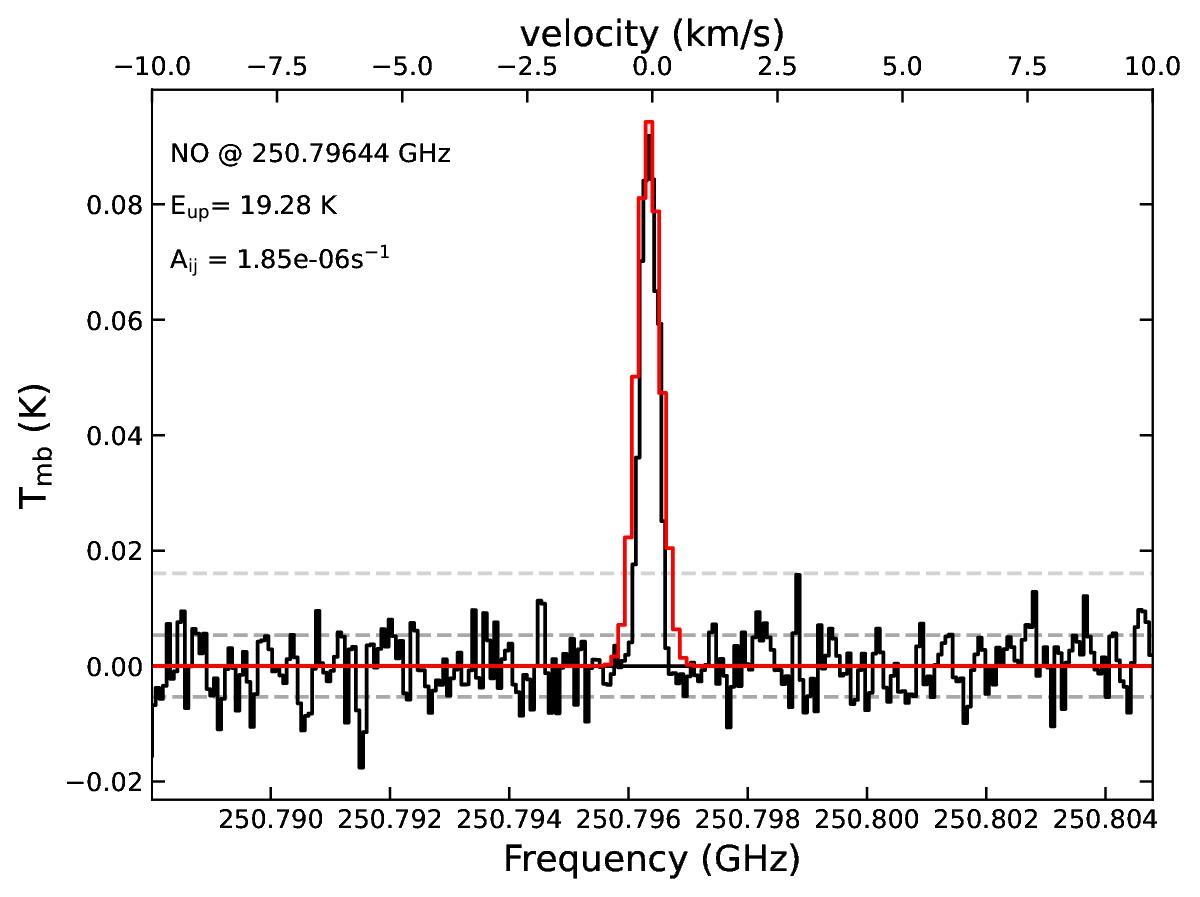}\includegraphics[width=0.33\textwidth]{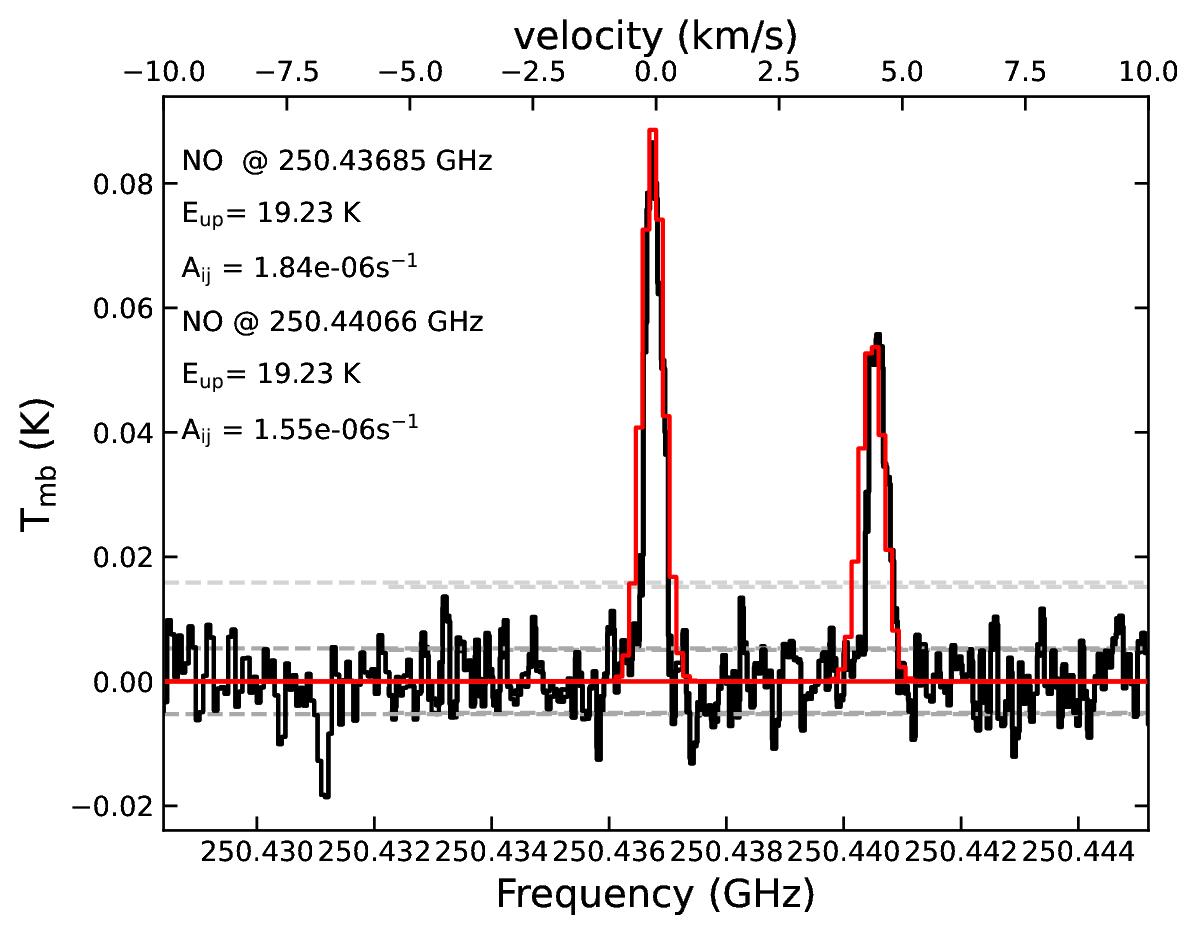}  
    \caption{(continued)}
\end{figure*}
\begin{figure*}\ContinuedFloat
 \includegraphics[width=0.33\textwidth]{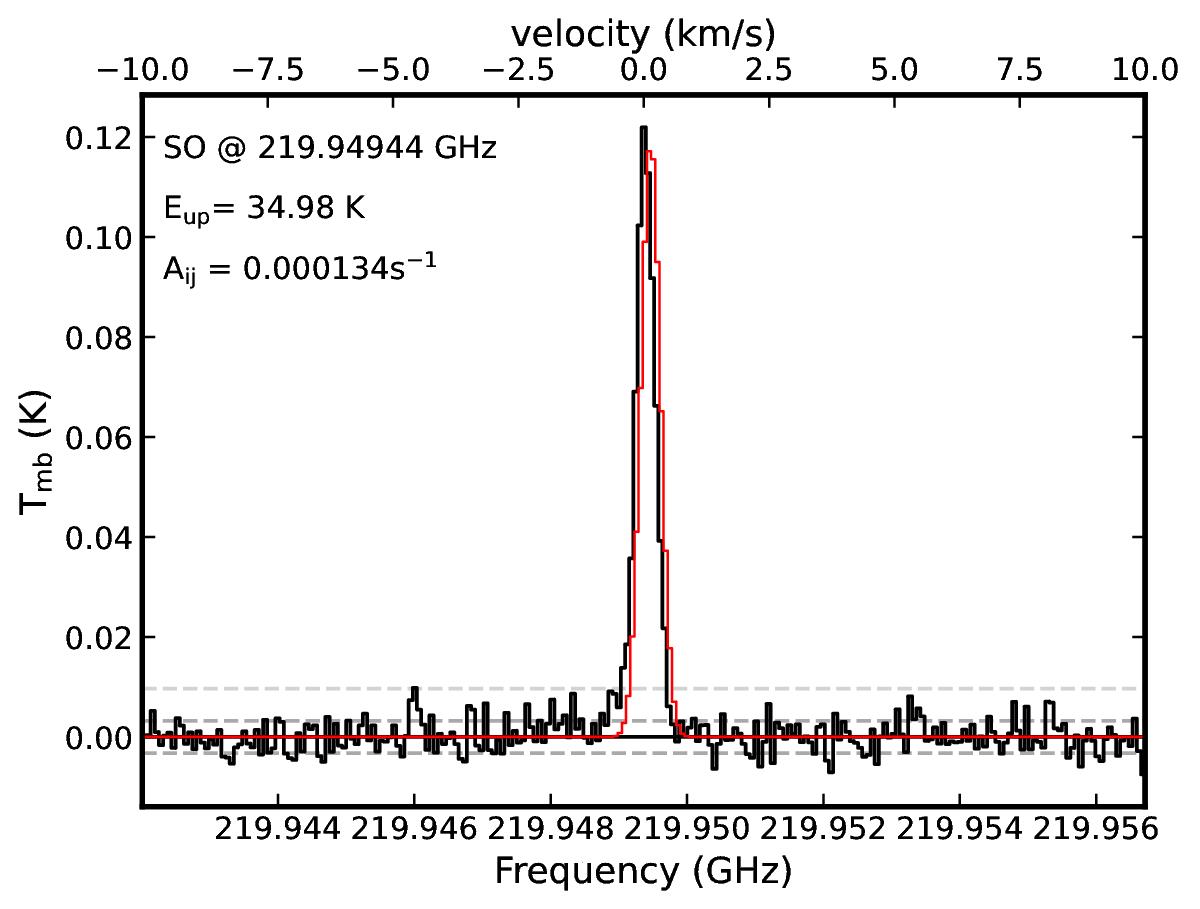}\includegraphics[width=0.33\textwidth]{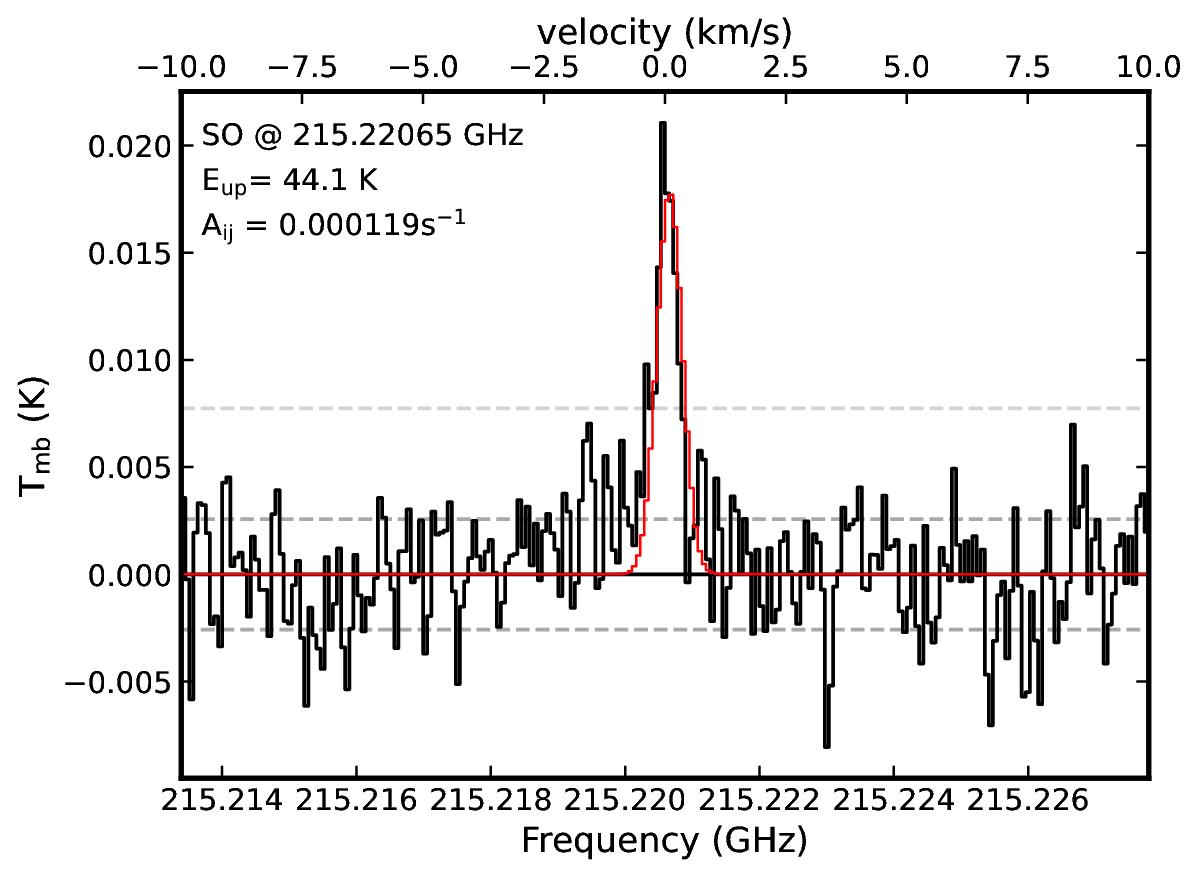}\includegraphics[width=0.33\textwidth]{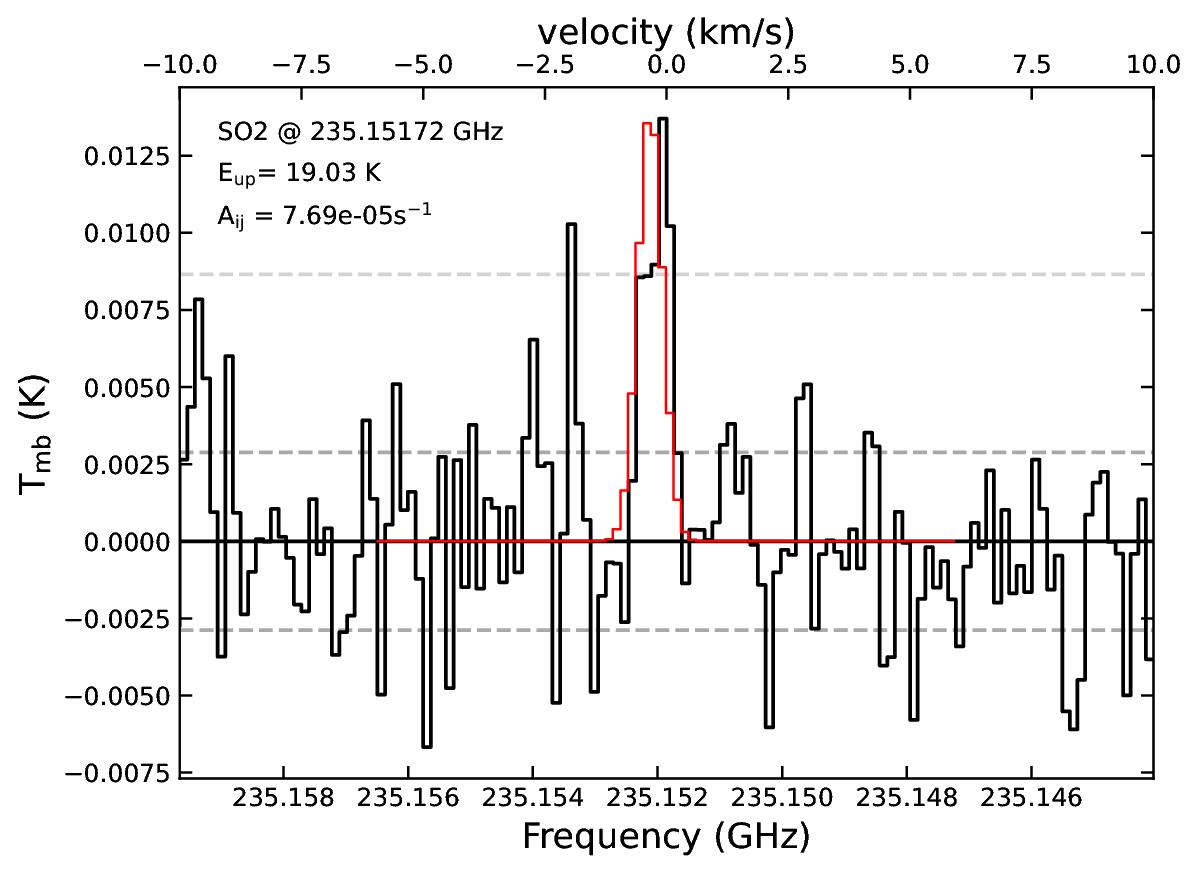}
    \caption{(continued)}
    \label{fig:secure-transitions}
\end{figure*}

\begin{figure*}
    \centering
    \includegraphics[width=0.33\textwidth]{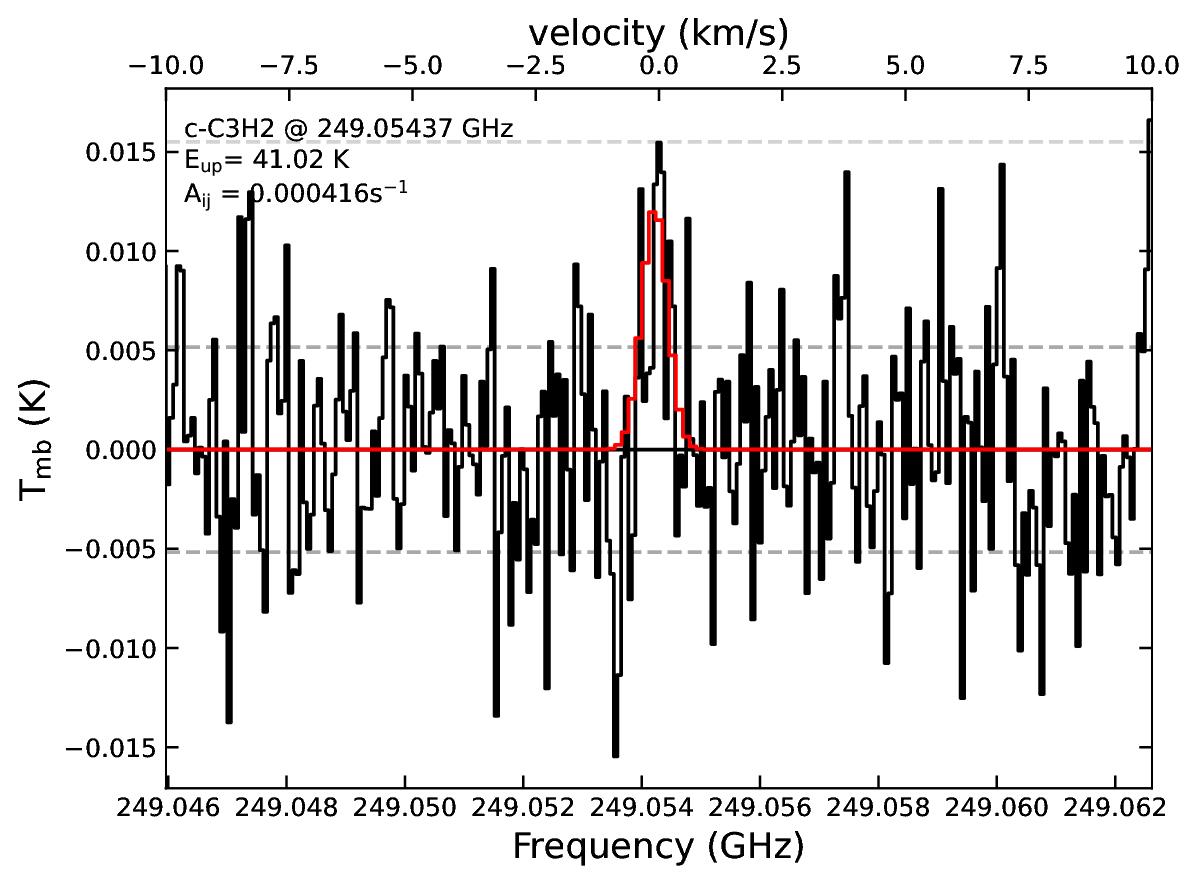}\includegraphics[width=0.33\textwidth]{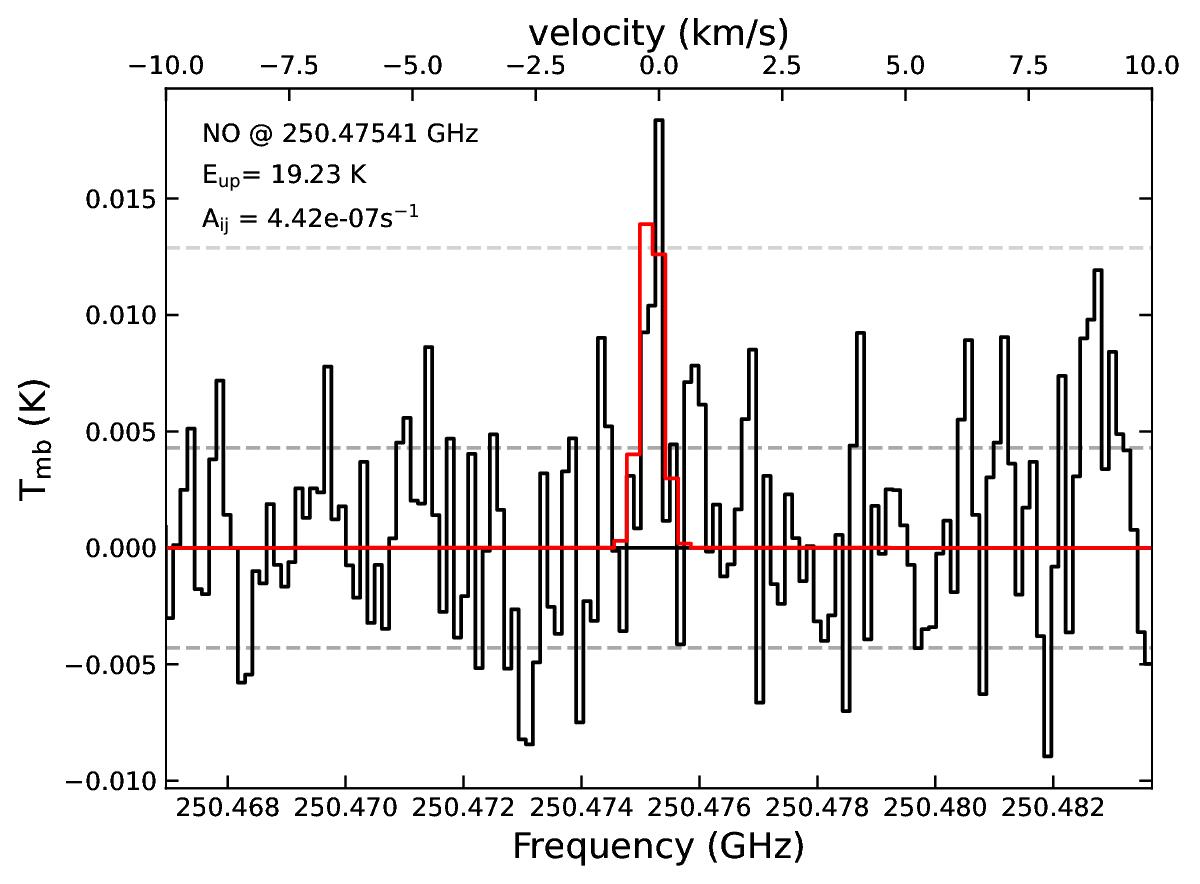}\includegraphics[width=0.33\textwidth]{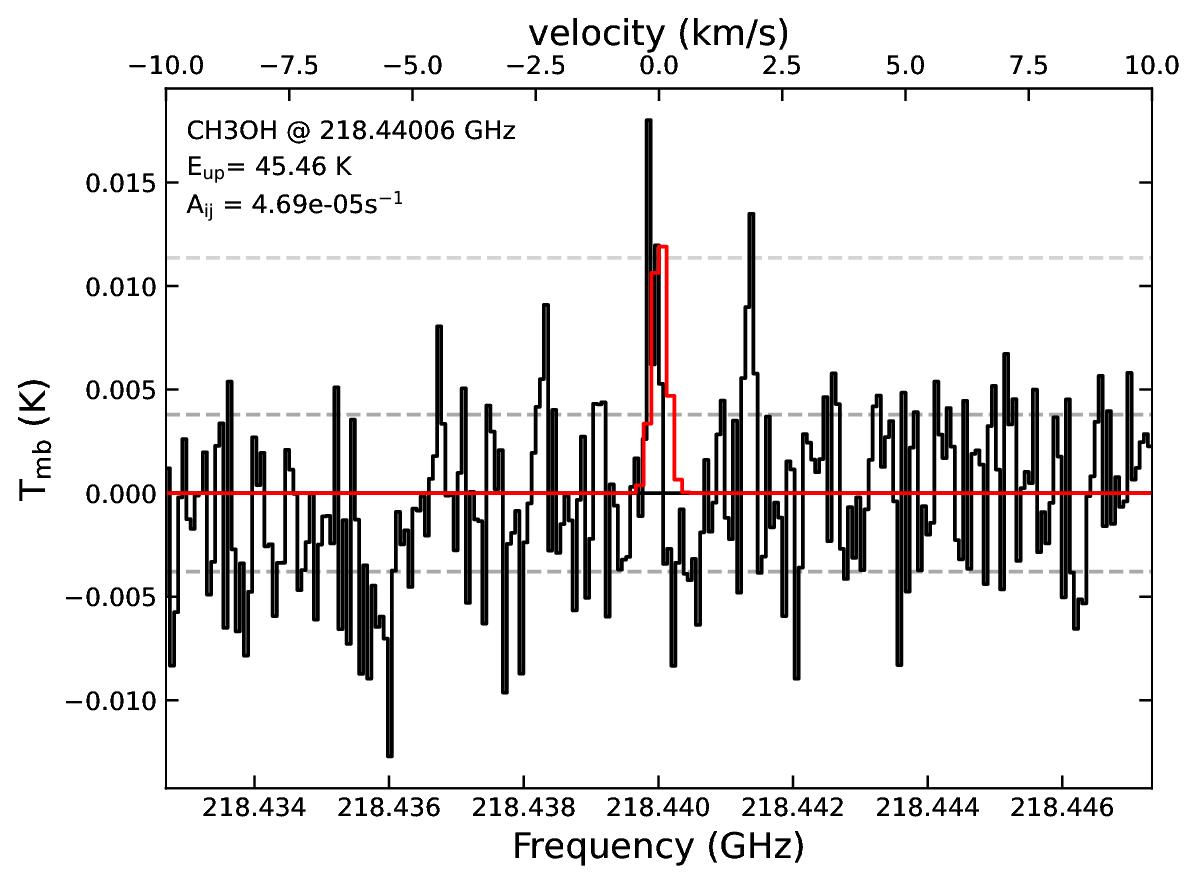} 
    \includegraphics[width=0.33\textwidth]{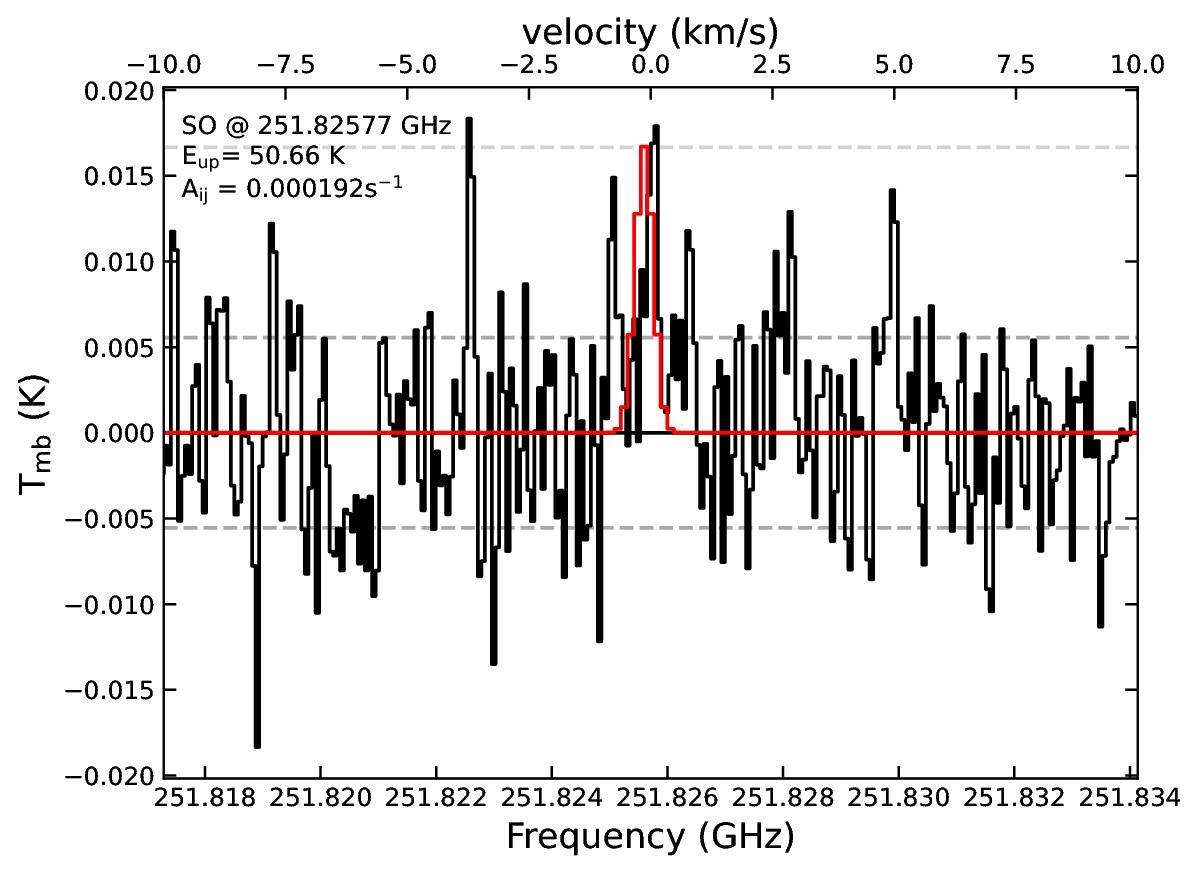} \includegraphics[width=0.33\textwidth]{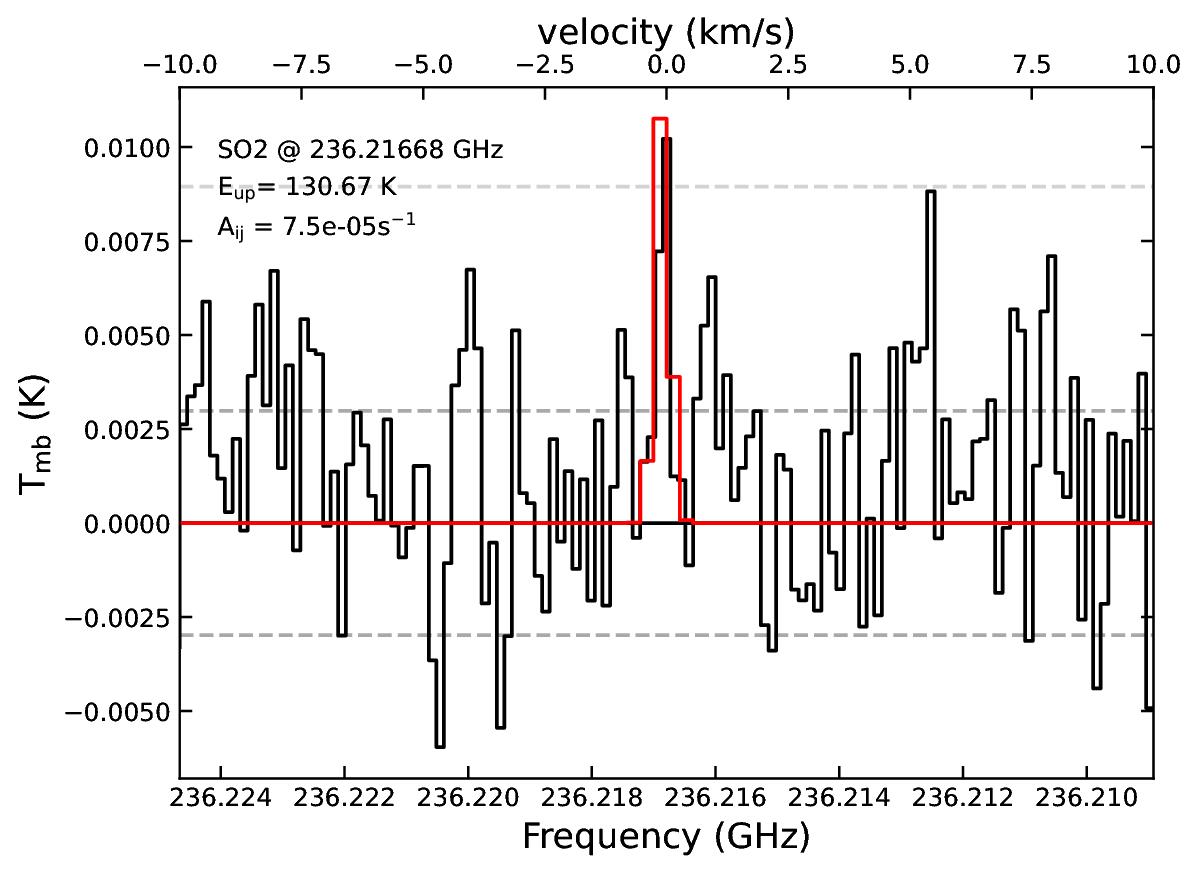}
    \caption{Spectra of the tentatively detected transitions (Table \ref{tbl:obstransitions}) overlaid with the Gaussian fit in red (Table \ref{tbl:obstransitions}). The dashed, dark gray line indicates the 3$\times$rms value, the dashed, light gray line indicates the rms value.}
    \label{fig:tentative-transitions}
\end{figure*}

\section{Spectra of notable non-detections}

Fig. \ref{fig:non-detections} shows the spectral window with the non-detections of the SiO line at 217.105~GHz, the $^{13}$C$^{17}$O line at 214.574~GHz, the H$_2 ^{13}$CO line at 219.909 GHz, and two CH$_2$DOH transitions at 214.702 and 253.629~GHz that were searched for in the data based on their similar upper state energies, and line strengths compared to detected species. Velocity-binning over two channels was applied to both CH$_2$DOH transitions and the H$_2 ^{13}$CO line.

\begin{figure*}[!h]
    \centering
    \includegraphics[width=0.33\textwidth]{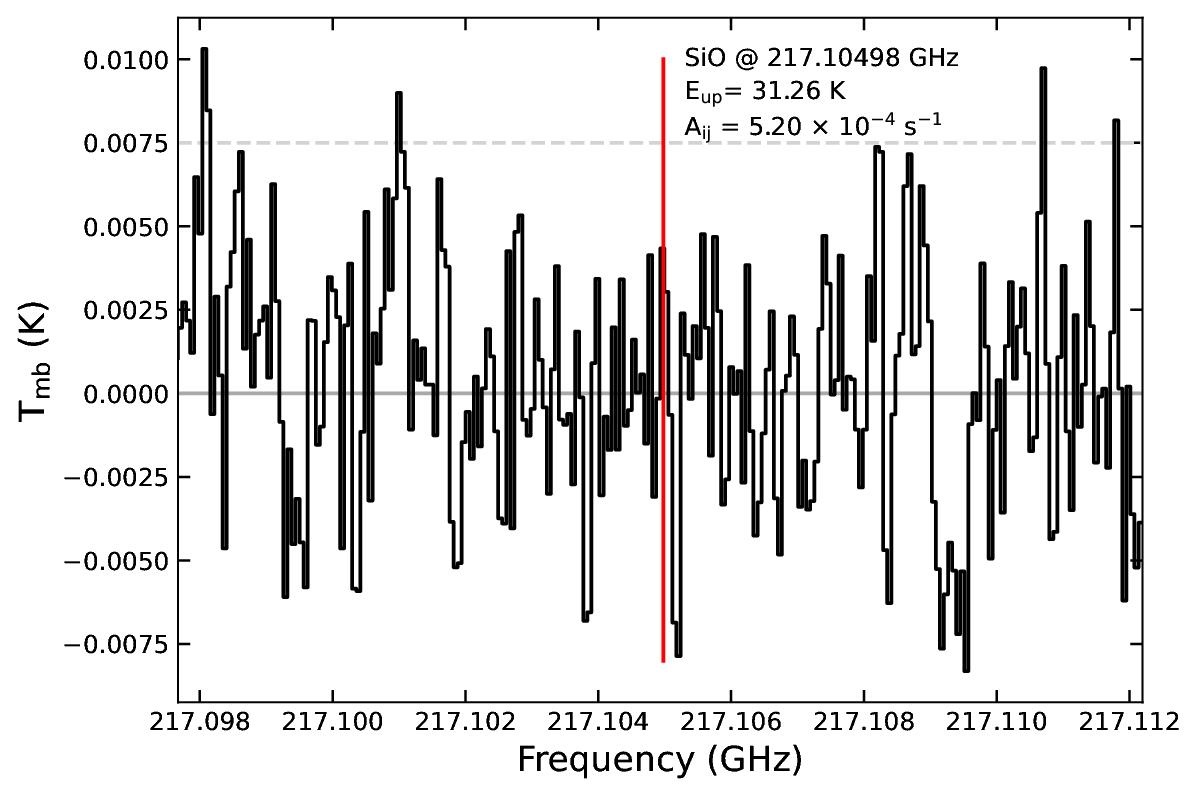}\includegraphics[width=0.33\textwidth]{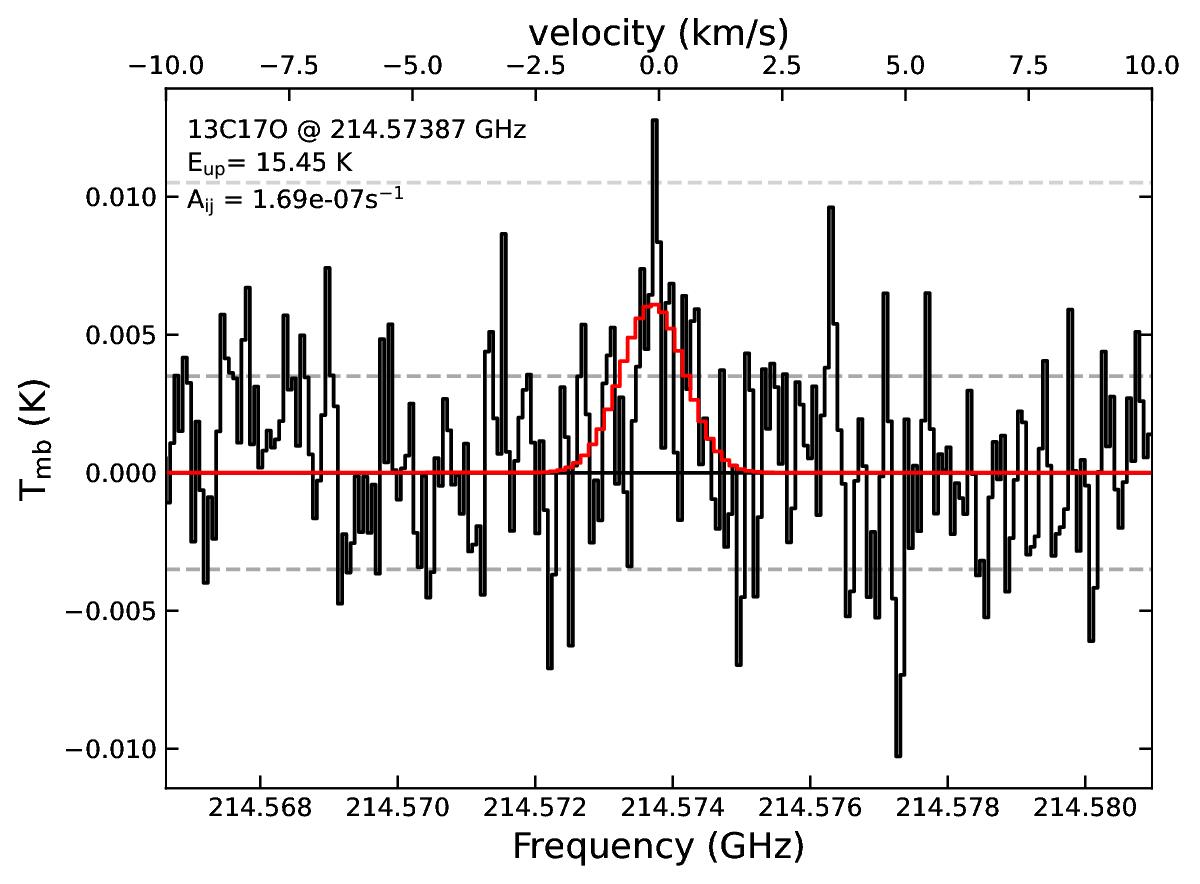}\includegraphics[width=0.33\textwidth]{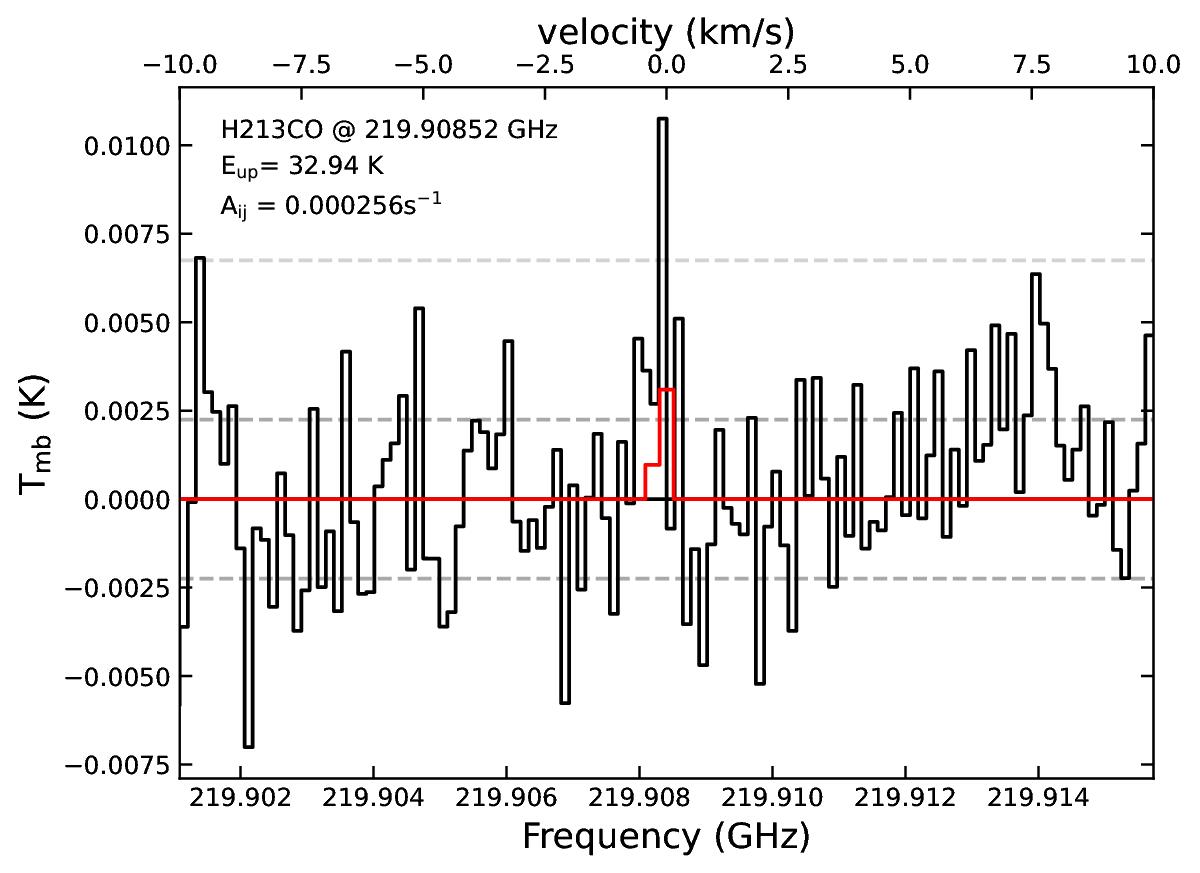}
    \includegraphics[width=0.33\textwidth]{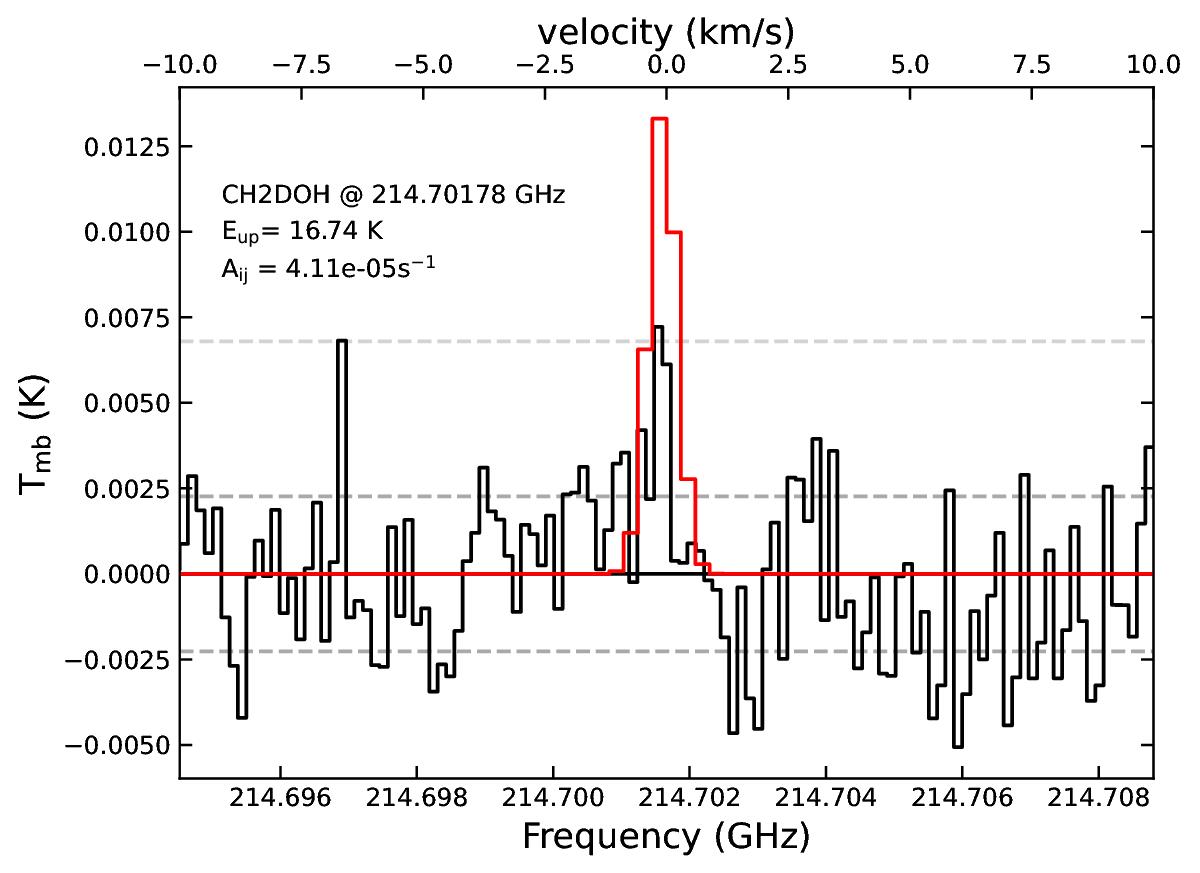}\includegraphics[width=0.33\textwidth]{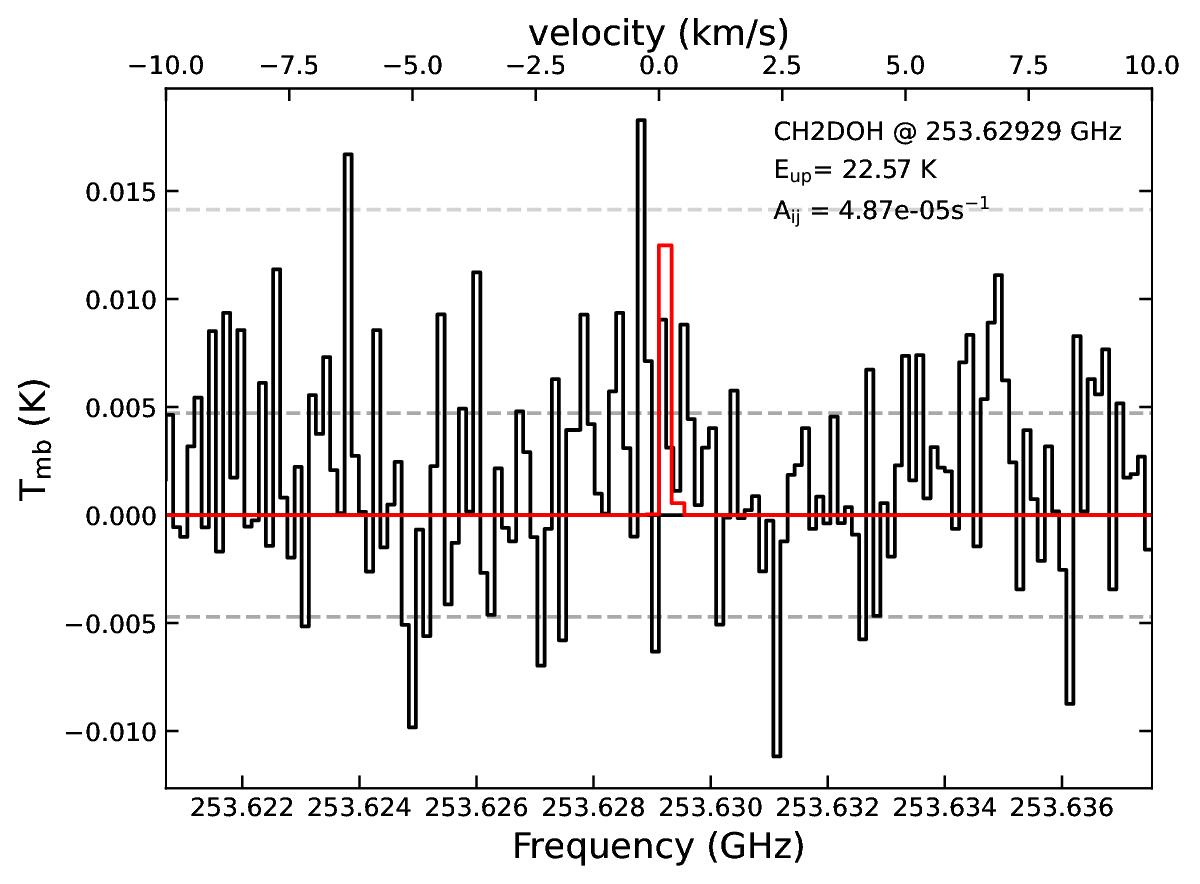}
    \caption{Notable non-detections in the data. The 3$\times$rms value is indicated by the dashed, light gray line, the rms value is indicated by the dashed, dark gray line. The position of the targeted SiO transition is indicated by the red line, for the remaining lines the attempt to fit a Gaussian to the data is displayed in red.}
    \label{fig:non-detections}
\end{figure*}

\begin{table}[!h]
\caption{Notable non-detections.}
\label{tbl:non-detections}
\centering
\begin{tabular}{l|c|c|c|c}
\hline \hline 
Molecule & Transition & Frequency & E$_{\rm up}$ & A$_{\rm ij}$  \\
& & (GHz) & (K) & (s$^{-1}$)  \\
\hline
SiO & 5$_{\rm 0}$~$-$~4$_{\rm 0}$ & 217.10492 & 31.26 & 5.20~$\times$~10$^{-4}$ \\
$^{13}$C$^{17}$O & 2-1 & 214.57387 & 15.45 & 1.69~$\times$~10$^{-7}$ \\
H$_2 ^{13}$CO & 3$_{\rm 1,2}$~$-$~2$_{\rm 1,1}$ & 219.90853 & 32.98 & 2.56~$\times$~10$^{-4}$ \\
CH$_2$DOH & 3$_{1,2}$ - 2$_{1,1}$ & 214.70178 &  16.74 & 4.11~$\times$~10$^{-5}$  \\
CH$_2$DOH & 2$_{2,1}$ - 2$_{1,2}$ &  253.62929 & 22.57 & 4.87~$\times$~10$^{-5}$ \\
\hline 
\end{tabular}
\end{table}

\section{Spectroscopic parameters}
Table \ref{tab:specparams} lists whether the spectroscopic data was taken from JPL or CDMS, and lists the papers that the catalog entries are based on.

\begin{table*}[]
    \caption{References to the spectroscopic data of all the molecules that were detected or searched for in the VeLLO in the DC3272+18 cloud and discussed in this work.}
    \centering
    \begin{tabular}{l|l|l}
    \hline \hline 
    Molecule     &  Database & References \\
    \hline 
     CO & CDMS & \cite{Goorvitch94,Winnewisser97}\\
     $^{13}$CO & CDMS & \cite{Zink90,Goorvitch94,Klapper00,Cazzoli04} \\
     C$^{18}$O & CDMS & \cite{Winnewisser85,Goorvitch94,Klapper01} \\
     N$_2$D$^+$ & CDMS & \cite{Dore04,Amano05,Pagani09} \\
     HCO$^+$ & CDMS & \cite{Botschwina93,Lattanzi07,Tinti07} \\
     DCO$^+$ & CDMS & \cite{Caselli05,Lattanzi07,vandertak09} \\
     HCN & CDMS & \cite{Ebenstein84,Ahrens02,Thorwirth03} \\
     DCN & CDMS & \cite{Deleon84,Möllmann02,Brünken04} \\
     HNC & CDMS & \cite{Blackman76,Saykally76} \\
     H$_2$CO & CDMS & \cite{Fabricant77,Muller17} \\
     CH$_3$OH & CDMS & \cite{Xu08} \\
     NO & CDMS & \cite{Muller15} \\
     c-C$_3$H$_2$ & CDMS & \cite{Bogey86,Bogey87,Vrtilek87,Lovas92,Spezzano12} \\
     C$_2$D & CDMS & \cite{Yoshida19,Cabezas21} \\
     D$_2$CO & CDMS & \cite{Fabricant77,Bocquet99} \\
     SO$_2$ & CDMS & \cite{Patel79,Muller05b} \\
     SO & CDMS & \cite{Lovas92,Bogey97} \\
     SiO & CDMS & \cite{Raymonda70,Muller13} \\
     $^{13}$C$^{17}$O & CDMS & \cite{Goorvitch94,Klapper03} \\
     H$_2 ^{13}$CO & CDMS & \cite{Fabricant77,Muller00} \\
     CH$_2$DOH & JPL & \cite{Pearson12} \\
    \hline \hline 
    \end{tabular}
    \label{tab:specparams}
\end{table*}

\end{appendix}

\end{document}